\documentclass[10pt,conference,singlecolumn]{IEEEtran}
\usepackage{amstext,amssymb}
\usepackage{graphicx}
\usepackage{times,color}
\usepackage{subfigure}
\usepackage{latexsym}
\usepackage{amsmath}
\usepackage{algorithm}
\usepackage{algorithmic}
\usepackage[final,ulem=normalem]{changes}
\usepackage{flushend}
\newtheorem{theorem}{Theorem}
\newtheorem{lemma}{Lemma}
\newtheorem{corollary}{Corollary}

\newtheorem{definition}{Definition}
\newtheorem{remark}{Remark}

\newcommand{\comment}[1]{}

\definecolor{dgreen}{rgb}{0.1,0.6,0.3}
\long\def\comment#1{}

\IEEEoverridecommandlockouts
\begin{document}
\title{The Cycle Consistency Matrix Approach to Absorbing Sets in  Separable Circulant-Based LDPC Codes 
\thanks{This research was supported by a gift from the Inphi Corporation, UC Discovery Grant 192837 and in part by grants CCF-1029030, 1161798 and 1150212  from NSF.  Portions of this paper were presented at the {IEEE} 6th International Symposium on Turbo Codes and Iterative Information Processing in Brest, France, Sept. 2010, the Information Theory and its Applications (ITA) Workshop in San Diego CA,  {Feb. 2010} and Feb. 2011,   the  {IEEE} International Conf. on Communications (ICC) in Kyoto, Japan, June, 2011, and the  {IEEE} International Symposium, on Information Theory (ISIT) in Saint Petersburg, Russia, July 2011.}}

\author{Jiadong Wang, Lara Dolecek, and Richard Wesel\\
wjd@ee.ucla.edu,~dolecek@ee.ucla.edu,~wesel@ee.ucla.edu}
\maketitle

\begin{abstract}
For LDPC codes operating over additive white Gaussian noise channels and decoded using message-passing decoders with limited precision, absorbing sets have been shown to be a key factor in error floor behavior.
Focusing on this scenario, this paper introduces the cycle consistency matrix (CCM) as  a powerful analytical tool for characterizing and avoiding absorbing sets in separable circulant-based (SCB) LDPC codes.   SCB codes include a wide variety of regular LDPC codes such as array-based LDPC codes as well as many common quasi-cyclic codes.   As a consequence of its cycle structure, each potential absorbing set in an SCB LDPC code has a CCM, and an absorbing set can be present in an SCB LDPC code only if the associated CCM {has a nontrivial null space.}

{CCM-based analysis can determine the multiplicity of an absorbing set in an SCB code and CCM-based constructions avoid certain small absorbing sets completely.  {While these techniques can be applied to an SCB code of any rate,} lower-rate SCB codes can usually avoid small absorbing sets because of their higher variable node degree.}  This paper focuses attention on the high-rate scenario in which the CCM constructions provide the most benefit.  {Simulation results demonstrate that {under limited-precision decoding} the new codes have steeper error-floor slopes and can provide one order of magnitude of improvement in the low FER region.}

\end{abstract}

\begin{IEEEkeywords}
Cyclic consistency matrix, absorbing set, error floor, quasi-cyclic LDPC codes.
\end{IEEEkeywords}

\section{Introduction}

Low-density parity-check (LDPC) codes were introduced by Gallager 
and are well-known for approaching capacity with iterative decoding
.  However, in the low FER region a flattening of the frame error rate (FER) curve called the error floor usually occurs for LDPC codes with moderate block lengths and high code rates.  This performance degradation is due at least in part to the sub-optimality of message passing in a graph with cycles.

This paper focuses on one class of regular LDPC codes: separable, circulant-based  (SCB) codes.    SCB codes include array-based LDPC codes as well as many common quasi-cyclic codes.  SCB codes retain standard properties of quasi-cyclic LDPC codes such as girth, code structure, and compatibility with existing high-throughput hardware implementations.  {These codes might be applicable for high-throughput data storage applications when message passing decoding is used with limited precision.}

This paper introduces the cycle consistency matrix (CCM) as  a powerful analytical tool for characterizing and avoiding any graphical structure in SCB LDPC codes that contains a well-defined cycle structure.  {For LDPC codes operating over additive white Gaussian noise channels and decoded using message-passing decoders with limited precision, absorbing sets have been shown in \cite{globecom06,globecom08,zhangTCOM} to be a key factor in error floor behavior.}  As a consequence of its cycle structure, each potential absorbing set in an SCB LDPC code has a CCM, and an absorbing set can be present in an SCB LDPC code only if the associated CCM is not full column-rank.   

For a specified circulant matrix, all SCB codes share a common mother matrix.  For a specified absorbing set, the CCM approach  yields with relative ease (because testing for a nontrivial nullspace is straightforward) a complete characterization (as to which codes have the absorbing set) for the entire family of SCB codes of a specified rate obtained by selecting a specified number of rows from a specified SCB mother matrix.   {The CCM-based analysis shows that quasi-cyclic code families described in \cite{tanner04}, \cite{FossorierIT04} and \cite{Fan00} include codes with good absorbing set spectra with a proper choice of parameters.} This paper also identifies absorbing-set-spectrum equivalence classes within a family of SCB codes, significantly simplifying this characterization.

Larger absorbing sets are more difficult to avoid, and our characterizations reveal that absorbing sets of large enough size occur in every code in the family.  A specific code avoiding as many small absorbing sets as possible can be selected for further improvement.   {This improvement can be accomplished by existing construction techniques such as \cite{AsvadiIT11} \cite{AsvadiArxiv} described in Section \ref{background:trapping}.   Alternatively, a CCM-based code shortening technique that we introduce in this paper can remove variable nodes so as to force the null space of the CCM to have zero dimension  thereby avoiding the target absorbing set while still precluding the smaller absorbing sets already absent before shortening.} \added{Compared to other algorithm-based code construction methods, the CCM approach is a systematic way to analyze a broad family of codes. Once the CCM is formulated, one can identify the best SCB code in the SCB family with the parameters of interest without requiring additional trial and error steps.}

Section~\ref{background:trapping} reviews trapping sets, absorbing sets, and prior work on designing LDPC codes that avoid these structures. Section~\ref{background} introduces separable circulant-based (SCB) codes and the cycle consistency matrix (CCM). 

{For SCB codes with column weights $r=5$, Section~\ref{analysis5} considers selected-row (SR) SCB codes, which are constructed by selecting $r=5$ appropriate row groups from the SCB mother matrix.  This section identifies the CCMs for the smallest absorbing sets for $r=5$ SR-SCB codes and uses these CCMs to select rows that avoid all (4,8), (5,9), and (6,8) absorbing sets.  Recognizing equivalence classes of SR-SCB codes that have the same absorbing sets greatly simplifies analysis and permits a complete characterization of $r=5$ SR-SCB codes with respect to the smallest absorbing sets.  Section~\ref{analysis5} concludes with an example characterization of all $r=5$ SR-SCB codes with respect to (4,8), (5,9), and (6,8) absorbing sets for code length $n=67^2=4489$.} 

{Section \ref{analysis4} identifies the CCMs for the smallest absorbing sets for $r=4$ SCB codes. These configurations are (6,4) absorbing sets.   In this section, the ability to characterize an entire SR-SCB family is utilized to show that one of the three configurations is necessarily present in every $r=4$ SR-SCB code regardless of $p$.  The other two configurations can be avoided by proper row selection.  This section concludes by showing how the CCM approach can guide the removal of specific columns from an SR-SCB parity-check matrix to eliminate the troublesome (6,4) configuration, producing a {\em shortened} SR-SCB code with no (6,4) absorbing sets.} 

{Section~\ref{section3} briefly remarks on the $r=3$ case, completing an analysis of small absorbing sets in SCB codes with $3\le r \le5$.  It is important to recognize that the CCM approach itself easily generalizes to any $r$.  This paper focuses on $3\le r \le5$ to provide illustrative examples and because  $r=4$ and $r=5$ are important for high-rate codes that are of interest to the authors. Section~\ref{results} provides simulation results demonstrating performance improvement in the error floor position and slope obtained by the newly identified codes.  Section~\ref{conclusion} delivers the conclusions.}

Conference presentations of parts of this paper appear in \cite{dolecekITA10} \cite{DOLECEKISTC10}, \cite{JWANGICC11}, \cite{WangITA2011}, and  \cite{WangISIT2011}.

 \section{Trapping sets and absorbing sets}\label{background:trapping}
Prior work indicates that certain sub-graphs called trapping sets~\cite{richardson,Bani07}, and, in particular, a subset of trapping sets called absorbing sets~\cite{dolecekIT10} are a primary cause of the error floor {in practical implementations}. Absorbing sets are trapping sets that are stable under bit-flipping decoding. {For LDPC codes operating over additive white Gaussian noise channels and decoded using message-passing decoders, some absorbing sets can be successfully overcome with sufficient precision.  However, in the practically important scenario of limited precision, absorbing sets have been shown in \cite{globecom06,globecom08,zhangTCOM} to be a key factor in error floor behavior.} 

{We refer to the smallest absorbing sets (in terms of the number of variable nodes in the set) as dominant because an absorbing set is activated when all of its variable nodes are in error and this activation is more likely for a smaller number of variable nodes than a larger number of variable nodes. } {Our simulation results also show that when precision is sufficiently limited, error performance is dominated by the smallest absorbing sets in the cases that we studied.}  However, there may be situations in which a relatively large absorbing set structure with high multiplicity could be more important to performance than an extremely rare absorbing set that is, nonetheless, smaller. \added{Even in the cases where the larger absorbing sets are important, the avoidance of smaller absorbing sets is often sufficient to eliminate certain larger absorbing sets. }

{Since the performance degradation due to trapping sets (including absorbing sets) is partially due to the sub-optimality of the iterative decoding algorithm {(especially under limited precision)}, one possible direction to {mitigate the effect of} trapping sets is to improve the decoding algorithm.} More effective message-passing algorithms \cite{CasadoTCOM2011, commletters09}, better iteration averaging schemes \cite{milenkovic}, more efficient quantizations \cite{quant2, zhangTCOM} and post-processing for absorbing sets \cite{globecom08} all can improve the error floor.

A complementary direction to improving the error floor, and the focus of this paper,  is to design a parity check matrix to avoid the trapping sets and absorbing sets.  Numerous previous papers have taken this approach.  The Approximate Cycle EMD\footnote{EMD is the acronym for Extrinsic Message Degree.} (ACE) algorithm \cite{TianTC2004} or the Progressive Edge Growth (PEG) algorithm \cite{XiaoCL2004} can be used for a column-by-column constructions. Algebraic methods for constructing LDPC codes also demonstrate good error floors as compared to randomly constructed codes. Results in \cite{LanIT2007}, \cite{ChenTCOM04}, \cite{ZhangTCOM2010}, \cite{HuangIT2012}, \cite{NguyenITW10},and \cite{NguyenIT11} present some notable work in this area.

Recent papers have proposed methods to improve the absorbing set spectrum. Introducing additional check nodes~\cite{milenkovicGlobecom06} or increasing the girth{~\cite{MilenkovicIT06}} eliminates small trapping sets for some codes. The algorithm in \cite{NguyenIT11} and \cite{NguyenITW10} constructs quasi-cyclic codes from Latin squares so that the Tanner graph does not contain certain trapping sets.  {In \cite{AsvadiIT11} \cite{AsvadiArxiv}, a deterministic construction of structured LDPC codes free of dominant trapping sets was developed by carefully swapping edges  in the lifted graph of the original code to provably eliminate the presence of detrimental trapping sets.}

{Several of these approaches have taken advantage of the cycle structure of the target objects.  For example, the ACE algorithm avoids stopping sets (and incidentally trapping sets and absorbing sets) by ensuring that all small cycles have sufficient extrinsic message degree.   \added{The cycle structure  of trapping sets is utilized in \cite{AsvadiIT11} \cite{AsvadiArxiv} to guide the edge swapping in the lifted graph}.   In \cite{dolecekIT10}, Dolecek \textit{et al.} used the cycle structure to study absorbing sets in array-based LDPC codes. }

{Our work builds on \cite{dolecekIT10} by capturing the relevant cycle structure of a specific absorbing set with a cycle consistency matrix and ultimately by {whether the nullspace of that matrix is zero or not}.   A unique feature of  the CCM approach is that it allows the characterization of an entire family of SCB codes rather than focus on constructing a single code by a column-by-column construction as in ACE or PEG or by a specific lifting as in \cite{AsvadiIT11} \cite{AsvadiArxiv}.   On the other hand, in this paper the CCM approach restricts attention to SCB codes while ACE, PEG, and the approach of  \cite{AsvadiIT11} \cite{AsvadiArxiv} can be applied in principle to any LDPC code.}  While the rest of the paper applies the CCM approach to absorbing sets, it can as well be applied to any trapping set or more generally any graphical object that is comprised of specific cycles connected in a specific way.

\section{SCB Codes and the Cycle Consistency Matrix }\label{background}
{Section~\ref{ba} introduces separable, circulant-based (SCB) codes. Section~\ref{mtrxrep} introduces absorbing sets and then puts forth the main concept of this paper: the cycle consistency matrix (CCM). Section \ref{sec:ECCM} uses graph theory to identify efficient 
CCMs that have the smallest possible number of rows.  Section \ref{sec:VNgraph} shows how the VN-graph approach simplifies the identification of efficient CCMs when the check node degree of the graphical structure is constrained to be two or less. Lastly, Section \ref{sec:NASC} 
presents a theorem providing necessary and sufficient conditions for the existence of a given absorbing set in SCB codes in terms of its CCM.}
\subsection{Separable, circulant-based LDPC codes}\label{ba}
Circulant-{based} LDPC codes are composed of circulant matrices and form a subset of $ (r,c) $ regular LDPC codes, where $r$ is the variable-node degree and $c$ is the check-node degree. {Each circulant matrix is a $ p \times p $ matrix, where $ p $ is a prime number.} {Parameters $r$ and $c$ are positive integers and are at most $p$}. The structure of these codes is compatible with high-throughput hardware implementations \cite{zhangTCOM}.

The parity-check matrix of circulant-based LDPC codes can be described as follows:
\begin{equation*}
\small
H_{p,f}^{r,c}=
\begin{bmatrix}
\sigma^{f(0,0)} & \sigma^{f(0,1)} & \sigma^{f(0,2)} & \ldots & \sigma^{f(0,c-1)}\\
\sigma^{f(1,0)} & \sigma^{f(1,1)} & \sigma^{f(1,2)} & \ldots & \sigma^{f(1,c-1)}\\
\sigma^{f(2,0)} & \sigma^{f(2,1)} & \sigma^{f(2,2)} & \ldots & \sigma^{f(2,c-1)}\\
\vdots & \vdots & \vdots & \ldots & \vdots \\
\sigma^{f(r-1,0)} & \sigma^{f(r-1,1)} & \sigma^{f(r-1,2)} & \ldots & \sigma^{f(r-1,c-1)}
\end{bmatrix},
\end{equation*}
where $ \sigma $ is {the following} $ p \times p $ circulant matrix:
\begin{equation*}
\sigma=\begin{bmatrix}
0 & 0 & \ldots & 0 & 1\\
1 & 0 & \ldots & 0 & 0\\
0 & 1 & \ldots & 0 & 0\\
\vdots & \vdots & \ldots &\vdots & \vdots\\
0 & 0 & \ldots & 1 & 0
\end{bmatrix},
\normalsize
\end{equation*}
and  $ f(i,j) $ is any function mapping the (row-index, column-index) pairs to the integers $\{0, \dots, p-1\}$. 

A column (or row) group is a column (or row) of circulant matrices. Each variable node has a label $(j,k)$ with $j\in \{0,..., c-1\}$ being the index of the corresponding column group and with $k\in \{0,..., p-1\}$ identifying the specific column within the group. Similarly, each check node has a label $(i,l)$ where $i\in \{0,..., r-1\}$ and $l\in \{0,..., p-1\}$.

This paper focuses on separable, circulant-based (SCB) codes, which are defined as follows:

\begin{definition}[Separable, Circulant-Based (SCB) Code]
An {SCB} code is a circulant-based LDPC code with the parity-check matrix $H_{p,f}^{r,c}$ in which  $ f(i,j) $ is separable, i.e., $f(i,j)=g_r(i) \cdot g_c(j) \mod p$.\hfill$\blacksquare$
\end{definition}

Parity check matrices of SCB codes with a specified circulant matrix {(i.e., a specified $p$)} can be viewed as originating from  a common SCB mother matrix $H_{p,f_m}^{p,p}$ with $ f_m(i,j)=i \cdot j \mod p$.  The functions $g_r(i)$ and $g_c(j)$ effectively specify which rows and columns of the mother matrix are selected for the resultant SCB matrix.  The ranges of $g_r(i)$ and $g_c(j)$ are both $\{0, \ldots, p-1\}$.

{SCB codes include, for example, the constructions in{~\cite{tanner04}, \cite{FossorierIT04}, and \cite{Fan00}.}}The girth of all SCB codes is guaranteed to be at least 6 by the SCB constraint on the submatrix exponent value $f(i,j)=g_r(i) \cdot g_c(j)$ (since all entries in each of $g_r$ and $g_c$ are distinct, \cite{dolecekIT10}).

The SCB structure imposes certain conditions \cite{dolecekIT10} on the variable and check nodes:

\textit{Bit Consistency:} The neighboring check nodes of a variable node must have distinct row-group ($i$) labels.

\textit{Check Consistency:} The neighboring variable nodes of a check node must have distinct column-group ($j$) labels.

\textit{Pattern Consistency:} (As shown in \cite{dolecekIT10}.) Since every entry in a row $(i,l)$ and a column $(j,k)$ in  the SCB mother matrix with the value 1 satisfies $k+ij=l \mod p$, if two variable nodes corresponding to columns $ (j_1,k_1) $ and $ (j_2,k_2) $ share a check node in row group $ i $, they must satisfy:
\begin{equation} \label{equpattern}
k_1+i j_1 = k_2+ i j_2 \mod p.
\end{equation}
\normalsize

{\textit{Converse of Pattern Consistency:} If \eqref{equpattern} is satisfied, then the two variable nodes $ (j_1,k_1) $ and $ (j_2,k_2) $ share a check node in row group $ i $ of the SCB mother matrix.}

\textit{Cycle Consistency:} As shown in \cite{dolecekIT10}, the equations of the form \eqref{equpattern} for any length-$2t$ cycle in an SCB mother matrix, which involve $t$ variable nodes with column-group labels $j_1$ through $j_t$  and $t$ check nodes with row-group labels $i_1$ through $i_t$, show that the cycle must satisfy:
\begin{equation} \label{equcycle}
\sum\limits_{m = 1}^t {i_m (j_{(m + 1)\;\bmod \;t}  - j_m )}  = 0\bmod p  .
\end{equation}
\normalsize

{\textit{Converse of Cycle Consistency:}  If \eqref{equcycle} is satisfied, a series of equations of the form \eqref{equpattern}  must also hold for some set $\{k_1, \ldots, k_t\}$ with $0\le k_m < p$.  For example, selecting any value of $0\le k_1 < p$ forces the values of $\{k_2, \ldots, k_t\}$.  These equations of the form  \eqref{equpattern} force the presence of the cycle by the converse of the pattern consistency condition above.}

{One particular type of SCB code considered in this paper is formed by selecting a set of $r$ rows from the SCB mother matrix.  We call such codes selected-row (SR) SCB codes.  Using the converse of cycle consistency, we can establish that all SR-SCB codes with $r\ge3$ have a girth of 6 (i.e., the smallest cycles are always length-6 cycles.)}
{\begin{lemma} \label{lemma:girth=6}
All SR-SCB codes with $r\ge3$ have girth 6.
\end{lemma}
\textit{Proof:}
It is known that the girth of such codes must be at least 6, see \cite{Fan00,dolecekIT10}.  That the girth is exactly 6 follows from the fact that if $r\ge3$, there is always a solution to
\begin{equation}
i_1(j_2-j_1) + i_2(j_3-j_2) + i_3(j_1 - j_3) = 0 \bmod p \, , 
\end{equation}
 which is $j_1=i_3$, $j_2=i_1$, and $j_3=i_2$. \hfill $\blacksquare$}
 
{Note that it may be possible to remove these length-6 cycles by removing columns from the SR-SCB code (i.e., by shortening the code).  Although already established in \cite{Fan00,dolecekIT10}, cycle consistency shows that length-4 cycles are not possible in any SCB codes because 
\begin{equation*}
i_1 (j_2-j_1) + i_2 (j_1-j_2) = (i_1-i_2)(j_2-j_1) \ne 0 \bmod p. 
\end{equation*}

After reviewing absorbing sets below, (\ref{equcycle}) is used to construct a matrix equation based on the cycles contained in an absorbing set that must be satisfied if that absorbing set is present in the SCB code.}


\subsection{Applying the cycle consistency matrix to absorbing sets} \label{mtrxrep}
{This section introduces the cycle consistency matrix as a way to capture the cycle behavior of a graphical structure.    While the cycle consistency matrix approach is generally applicable to graphical structures that contain cycles in SCB codes, we provide the specific example of absorbing sets to illustrate the technique since the absorbing sets are the focus of the analysis in the rest of the paper.}

An LDPC code with the parity-check matrix $ H $ is often viewed as {a {bipartite ({Tanner}) graph} $G_H = (V, F,E)$}, where the set $V$ represents the variable nodes, the set $F$ represents the check nodes, and the set  $E$ corresponds to the edges between variable and check nodes.

For a variable node subset $V_{\text{as}} \subset V$, analogous to $G_H$, let $G_{\text{as}} = (V_{\text{as}}, F_{\text{as}},E_{\text{as}})$ be the bipartite graph of the edges $E_{\text{as}}$ between the variable nodes $V_{\text{as}}$ and their neighboring check nodes $F_{\text{as}}$.  Let $o(V_{\text{as}})\subset F_{\text{as}}$ be the neighbors of $V_{\text{as}}$ with odd degree in $G_{\text{as}}$  and $e(V_{\text{as}})\subset F_{\text{as}}$ be the neighbors of $V_{\text{as}}$ with even degree in $G_{\text{as}}$.  We refer to the nodes in  $e(V_{\text{as}})$ as ``satisfied check nodes'' because they will satisfy their parity-check equations when all the nodes in $V_{\text{as}}$ are in error.

\begin{definition}[Absorbing Set {(cf.~\cite{dolecekIT10})}]
An $(n_v,n_o)$ absorbing set is a set $V_{\text{as}} \subset V$ with $|V_{\text{as}}|=n_v$ and $|o(V_{\text{as}})|=n_o$, where each node in $V_{\text{as}}$ has strictly fewer neighbors in $o(V_{\text{as}})$ than in $e(V_{\text{as}})$. \hfill$\blacksquare$
\end{definition}

Moreover, if each variable node in $ V \backslash V_{as} $  has strictly fewer neighbors in $o(V_{as})$ than in $F {\backslash} o(V_{as})$, an $(n_v,n_o)$ absorbing set is called an $(n_v,n_o)$ {\em fully} absorbing set {\cite{dolecekIT10}.} An important property of fully absorbing sets is that they are stable under bit-flipping decoding in which the bit values of variable nodes are flipped if a majority of their neighboring check nodes are not satisfied.

\begin{figure}
\centering
\includegraphics[width=0.31\textwidth]{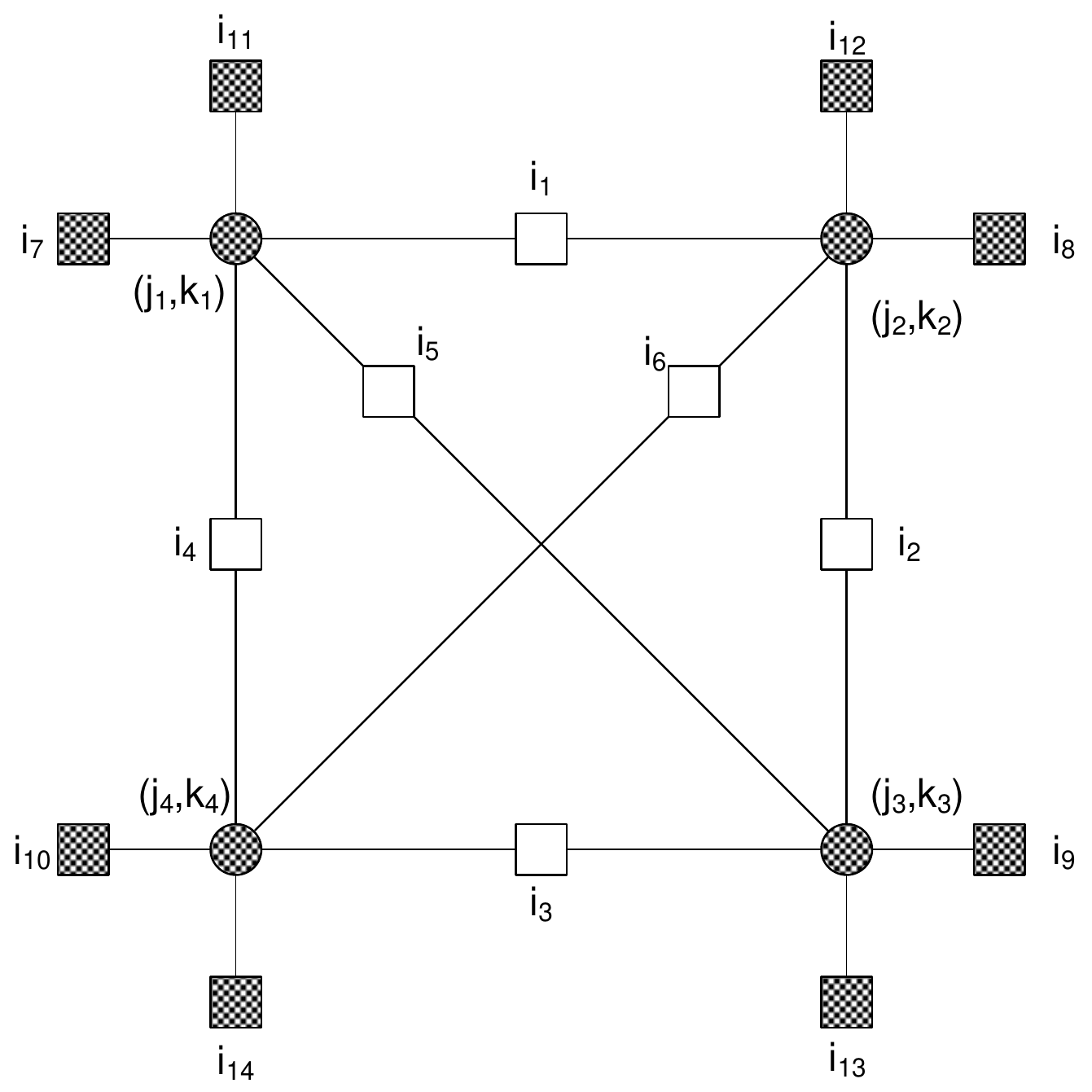}
\caption{Depiction of a $(4,8)$ absorbing set. Black circles are the four variable nodes (bit nodes) of the absorbing set. The white squares are the satisfied neighboring check nodes, and the black squares are the eight unsatisfied neighboring check nodes.\label{fig48}}
\vspace{-0.2in}
\end{figure}

Fig.~\ref{fig48} shows an example of {a} $ (4,8) $ absorbing set, which has 4 variable nodes and 8 unsatisfied check nodes.

Suppose there are $n_v$ variable nodes in an absorbing set. Let $j_1, \ldots, j_{n_v}$ be the column-group labels of these $n_v$ nodes in the SCB mother matrix. Define 
\begin{equation}\label{eq:u}
u_m=j_{m}-j_1 \text{ for  } m \in \{2,\cdots,n_v\}
\end{equation}
and define $ \mathbf{u}=[u_2,\cdots,u_{n_v}] $. For each cycle in the absorbing set, {by replacing the difference of $ j $'s with the difference of $ u $'s}, \eqref{equcycle} may be written as
\begin{equation} \label{equcycleu}
\sum\limits_{m = 2}^{t} (i_{m-1}-i_{m})u_{m} = 0 \mod p  ,
\end{equation}
\normalsize
where $2t$ is the length of that cycle.  Note that {the sequence of check-node row groups $\{i_1, i_2, \cdots, i_t\}$} will be different for different cycles reflecting the particular cycle trajectories.

Every cycle in the absorbing set satisfies an equation of the form \eqref{equcycleu}.  Taken together, these equations produce a matrix equation: $ \mathbf{M}\mathbf{u} = 0 \mod p $, where $ \mathbf{M}_{\ell m} $ is the coefficient of $ u_m $ in \eqref{equcycleu} for the $ \ell $th cycle.
{\begin{definition}[Cycle Consistency Matrix]
{A} Cycle Consistency Matrix (CCM) $\mathbf{M}$ of a graphical structure containing cycles in a bipartite graph of an SCB LDPC code, for example an absorbing-set graph $G_\text{as}$,  has a number of columns one less than the number of variable nodes in the structure.  Each row in $\mathbf{M}$ corresponds to  a cycle in the structure, specifying the coefficients of $\mathbf{u}$ in \eqref{equcycleu} for the corresponding cycle.  Furthermore, $ \mathbf{M}\mathbf{u} = 0 \mod p $ completely characterizes the requirement that every cycle in the structure satisfies \eqref{equcycleu}.
\hfill$\blacksquare$\end{definition}
\subsection{Efficient cycle consistency matrices} \label{sec:ECCM}
A key property of a CCM is that $ \mathbf{M}\mathbf{u} = 0 \mod p $ completely characterizes the requirement that every cycle in the structure satisfies \eqref{equcycleu}.  Even so, it is not necessary for $\mathbf{M}$ to include a row for every cycle in the structure. A cycle need not be included in $\mathbf{M}$ if it is a linear combination of cycles already included in $\mathbf{M}$. Thus, the number of rows needed in $\mathbf{M}$ is the number of linearly independent cycles in the structure.}

The notion from graph theory \cite{Diestelgraphtheory} of an incidence matrix is useful for establishing the number of linearly independent cycles in a graphical structure.
\begin{definition}[Incidence Matrix]
 For a graph with $n$ vertices and $q$ edges, the (unoriented) incidence matrix  is an $n \times q$  matrix $B$ with $ B_{ij}=1 $ if vertex $ v_i $ and edge $ x_j $ are incident and 0 otherwise.\hfill$\blacksquare$
\end{definition}

Note that since each edge is incident to exactly two vertices, each column of $B$ has exactly two ones.

The incidence matrix of a graph is useful for identifying the cycles in the graph because every cycle has the property that the indicator vector $\mathbf{x}_c$ of the edges in the cycle satisfies $B\mathbf{x}_c=0 \mod 2$.  In fact, the edges identified by the vector $\mathbf{x}_c$ form a cycle (or a union of cycles) if and only if $B\mathbf{x}_c=0 \mod 2$.  This is formalized in the definition below.
{\begin{definition}[Binary Cycle Space]
The binary cycle space (bcs) of a graph is the set of binary indicator vectors of the edges in a cycle or a union of cycles.  The bcs is the null space over $ GF(2) $ of the graph's incidence matrix.  \hfill$\blacksquare$
\end{definition}}

{An efficient CCM is the one with the least possible number of rows, which is the dimension of the binary cycle space $D_{\text{bcs}}$.
\begin{definition}[Efficient Cycle Consistency Matrix]
An Efficient Cycle Consistency Matrix $\mathbf{M}$ of a graph $G$ has $D_{\text{bcs}}$  rows. The rows of $\mathbf{M}$ correspond to $D_{\text{bcs}}$ linearly independent cycles in $G$.  Each row has the coefficients of $\mathbf{u}$ in \eqref{equcycleu} for the corresponding cycle. \hfill$\blacksquare$\end{definition}}

{ For any bipartite graph, the check nodes and variable nodes together can be the nodes used to construct an incidence matrix whose nullspace is  the binary cycle space.  As an example, for the absorbing set of Fig.~\ref{fig48}, an $18 \times 20$ incidence matrix $B_{(4,8)}$ can be formed from the 18 nodes (14 check nodes and 4 variable nodes) and 20 edges in that graphical structure.  The rank of $B_{(4,8)}$ is 17 which indicates that $D_{\text{bcs}} = 20-17=3$.  Thus, there are three linearly independent cycles in the $(4,8)$ absorbing set of  Fig.~\ref{fig48} and an efficient CCM for this absorbing set will have three rows.}

\subsection{The VN graph approach to finding $D_{bcs}$}\label{sec:VNgraph}
{The smallest absorbing sets for the SCB codes we study below all share the property that within the absorbing set, the check nodes all have degrees of two or less.    This is consistent with work such as \cite{Dekhordi12} in which elementary trapping sets, defined to have check node degrees of two or less \cite{Milenkovic07}, are shown to dominate the code performance.} 

{The constraint that the check node degree be two or less permits a simpler technique for identifying the smallest number of rows needed for a CCM than what was presented above.   Our examples of this technique are all absorbing sets, but the technique applies to elementary trapping sets and more generally to any graphical structure in which the check nodes all have degrees of two or less. }

{For any bipartite graph in which check nodes all have degree two or less a ``variable-node'' (VN) graph can be constructed whose only vertices are the variable nodes of the original graph and for which two vertices in the VN graph are connected iff there is a check node that connects them in the original graph. Under this constraint, for codes with girth greater than 4 {(including all SCB codes)}, multiple edges are not allowed between vertices of the VN graph since they would correspond to a length-4 cycle.} 

As long as the check node degree is constrained to be two or less, the null space of the incidence matrix for the VN graph identifies the same binary cycle space for the absorbing set as the null space of the incidence matrix for the original bipartite graph.  However, the VN graph greatly simplifies the process.

\begin{figure}
\centering
\includegraphics[width=0.2\textwidth]{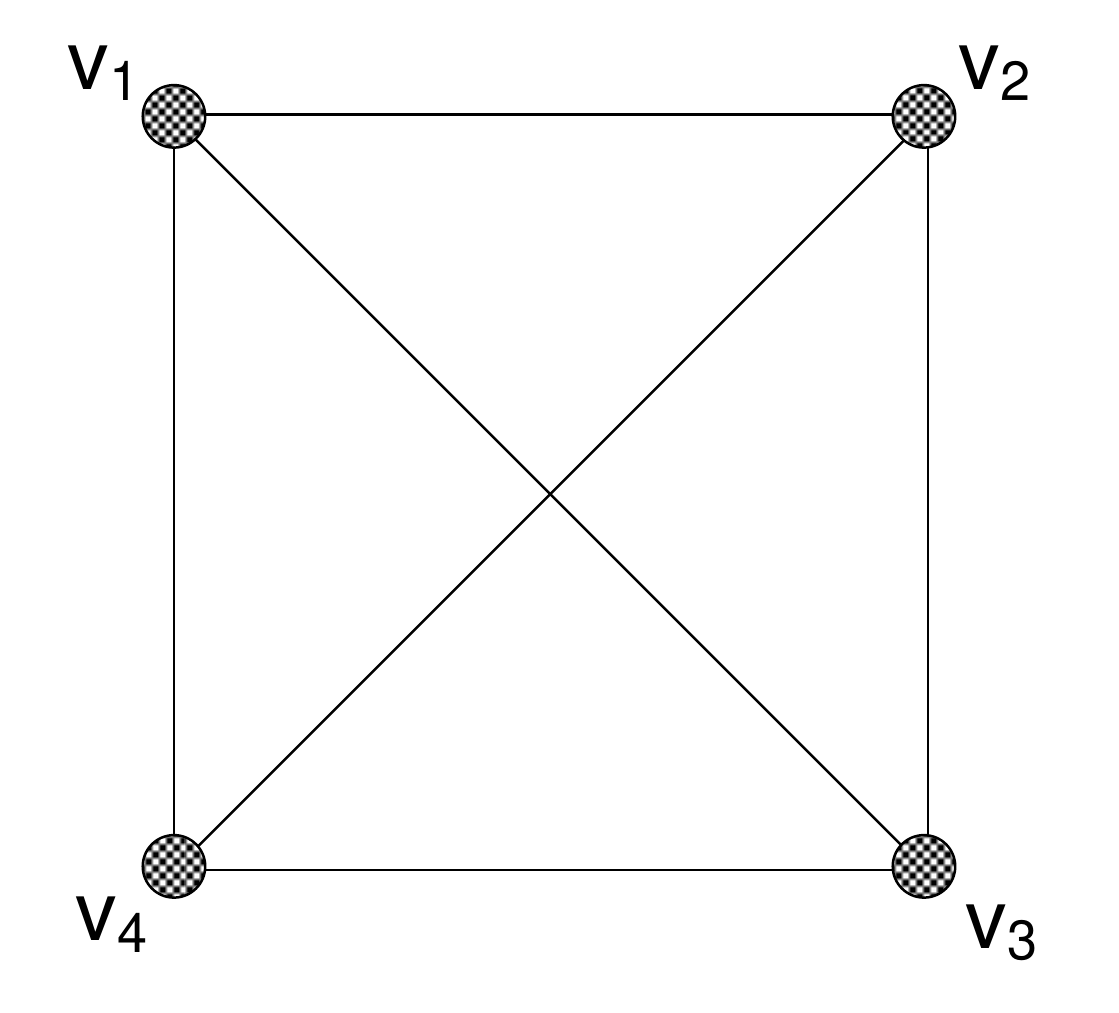}
\caption{Variable-node (VN) graph of the $(4,8)$ absorbing set of Fig. \ref{fig48}.}\label{fig:vn48}
\vspace{-0.2in}
\end{figure}

The $ (4,8) $ absorbing set provides an example of how to use the VN-graph approach to construct the CCM for a given absorbing set.
The absorbing set graph in Fig.~\ref{fig48} induces the variable-node (VN) graph {shown in Fig.~\ref{fig:vn48}}.  There are five cycles in the variable-node graph, but not all of these cycles need to be explicitly represented in the CCM.  The incidence matrix $B_{\text{as}}$ of the VN graph shown in Fig.~\ref{fig:vn48} with $q=6$ edges and $V_{\text{as}}=4$ nodes is
\begin{equation}\label{eq:Bas48}
B_{\text{as}}=\left[\begin{array}{cccccc}
1 & 1 & 1 & 0 & 0 & 0\\
1 & 0 & 0 & 1 & 1 & 0\\
0 & 1 & 0 & 1 & 0 & 1\\
0 & 0 & 1 & 0 & 1 & 1\\
\end{array}
\right].
\end{equation}\normalsize

The VN-graph incidence matrix $B_{\text{as}}$ is simply the transpose of the submatrix $\hat{H}_{\text{as}}$ of the parity-check matrix whose rows correspond to satisfied check nodes of $G_{\text{as}}$ (a subset of $F_{\text{as}}$) and whose columns correspond to the variable nodes in $V_{\text{as}}$:

\begin{equation}
\hat{H}_{\text{as}}=\left[\begin{array}{cccc}
1 & 1 & 0 & 0\\
1 & 0 & 1 & 0\\
1 & 0 & 0 & 1\\
0 & 1 & 1 & 0\\
0 & 1 & 0 & 1\\
0 & 0 & 1 & 1\\
\end{array}
\right].
\end{equation}

The rank of $B_{\text{as}}$ in \eqref{eq:Bas48} is $3$ under $GF(2)$. Thus the dimension of the binary cycle space is $D_{\text{bcs}}=q$$-$$\text{rank}(B_\text{as})=6$$-$$3 = 3$, which means {that three linearly independent cycles span the binary cycle space}.  This is the same answer we found above, but the VN graph approach uses a $4 \times 6$ incidence matrix rather than an $18 \times 20$ incidence matrix.

Applying \eqref{equcycleu} to the following three linearly independent cycles in Fig.~\ref{fig:vn48}: $v_1$$-$$v_2$$-$$v_3$$-$$v_1$, $v_1$$-$$v_2$$-$$v_4$$-$$v_1$, $v_1$$-$$v_3$$-v_4$$-$$v_1$ (here and in the remainder $v_l$ corresponds to the variable node labeled {$(j_l,k_l)$} {in Fig.~\ref{fig48}}) produces the three rows of the CCM in \eqref{eq48} and its determinant in \eqref{eq:detM48} as follows:
\begin{equation}\label{eq48}
\mathbf{M} = \begin{bmatrix}
i_1-i_2 & i_2-i_5 & 0\\
i_1-i_6 & 0 & i_6-i_4 \\
0 & i_5-i_3 & i_3-i_4
\end{bmatrix} \, ,
\end{equation}
and
\begin{equation}\label{eq:detM48}
\hspace{-.025in}
\small
\det\mathbf{M}=-\text{$(i_1$$-$$i_2)(i_6$$-$$i_4)(i_5$$-$$i_3)-(i_2$$-$$i_5)(i_1$$-$$i_6)(i_3$$-$$i_4)$}.
\normalsize
\end{equation}

\subsection{Necessary and sufficient conditions for absorbing sets}\label{sec:NASC}

Recall that the elements of $\mathbf{u}$ as defined in \eqref{eq:u} contain difference information about the column groups involved in the absorbing set: the value of the first column group is subtracted from each of the others. The vector $\mathbf{u}$ cannot be an all-zero vector because the Check Consistency condition requires that variable nodes sharing a check node have distinct column groups, and a zero entry indicates that the variable node is in the first ($j=j_1$) column group. Thus, a necessary condition for the existence of a given absorbing set in an SCB code is that its CCM matrix $\mathbf{M}$ does not have full {column-}rank in $GF(p)$.

\begin{definition}[Extensible VN Graph] If the VN graph of an absorbing set $G_\text{as}^1$ is a sub-graph of the VN graph of at least one other absorbing set $G_\text{as}^2$ with the same number of variable nodes, then the VN graph of $G_\text{as}^1$ is extensible. \hfill$\blacksquare$ \end{definition}
{Note that the incidence matrix $B_2$ of absorbing set $G_\text{as}^2$ is then the incidence matrix $B_1$ of absorbing set $G_\text{as}^1$ with some additional columns.}
\comment{
{When variable node degree is at least 2, each variable node in the absorbing set is incident to at least two satisfied check nodes. Each of these two check nodes has to connect to another variable node in the absorbing set. If there is no cycle in the absorbing set, the size of this absorbing set cannot be finite. Therefore each absorbing set necessarily contains a  cycle when the variable node degree is strictly larger than 1.}
}

Equipped with these definitions, the next theorem gives necessary and sufficient conditions for the existence of absorbing sets in SCB codes.

\begin{theorem} \label{thm:ccm}
Given a proposed $(n_v,n_o)$ absorbing set graph $G_{\text{as}} = (V_{\text{as}}, F_{\text{as}},E_{\text{as}})$, {where every variable node is involved in at least one cycle,}\footnote{{If the variable node degree is at least 2, then each variable node in a given absorbing set must be a part of at least one cycle.}} specified column group labels of the $n_v$ variable nodes in $V_{\text{as}}$ in the SCB mother matrix, and specified row-group labels of the check nodes in $F_{\text{as}}$ in the SCB mother matrix, the following are necessary conditions for the proposed absorbing set to exist in each daughter SCB LDPC code (with the parity check matrix $H$ that includes the specified row and column groups of that SCB mother matrix):
(1) The CCM for $G_{\text{as}}$ {does not have full column-rank};
(2) Variable nodes in  $V_{\text{as}}$ satisfy the Check Consistency condition and can form a difference vector ${\bf u}$ in the null space of the CCM; and
(3) Each check node in  $F_{\text{as}}$ satisfies the Bit Consistency condition.
Taken together, these conditions are also sufficient if the VN graph of this absorbing set is not extensible.
\end{theorem}

\textit{Proof:}
Each of the three conditions has already been shown above to be a necessary condition for the existence of $G_\text{as}$ in an SCB. Consider the sufficiency of the three conditions.  

If all three conditions are satisfied, all the cycles corresponding to the rows of the CCM exist in $G_H$ because they can be constructed as follows: Conditions (1) and (2) ensure a sequence of variable node column groups $ [j_1,j_2,\cdots,j_{n_v}] $ that form a vector $ [u_2,...,u_{n_v}] $ in the null space of the CCM.  For any fixed $ k_1 $, we can compute $ k_2,\cdots,k_{n_v} $ using the {converse of Cycle Consistency}.  Any linear combination of these cycles exists in $G_H$ as well.

These cycles cover every edge except edges that connect a variable node to a degree-1 check node in $G_{\text{as}}$.  If (1) a variable node's unsatisfied check node is actually the same as another variable node's unsatisfied check node, or (2) a variable node's unsatisfied check node is actually the same as a satisfied check node in the graph, or (3) two of the satisfied check nodes are actually the same, then there exist other independent cycles in the VN graph which extend the VN graph.

In these cases, the original structure is extensible and the three conditions are not sufficient to establish  the existence of the absorbing set $G_{\text{as}}$ originally considered.  The three conditions might instead be caused by the presence of an absorbing set whose VN graph contains the VN graph of the originally considered absorbing set as a  subgraph.  However, if the original VN graph is not extensible, the above constructed solution must establish the existence of the proposed absorbing set $G_{\text{as}}$ in $G_H$.  In this case the three conditions are sufficient. \hfill$\blacksquare$

\begin{corollary} \label{cor:ext}
{Suppose a non-extensible $ (n_{v_1},n_{o_1}) $ absorbing-set graph $G^{1}_{\text{as}} = (V^{1}_{\text{as}}, F^{1}_{\text{as}},E^{1}_{\text{as}})$ has the VN graph that is a sub-graph of the VN graph of another $ (n_{v_2},n_{o_2}) $ absorbing-set graph $G^{2}_{\text{as}} = (V^{2}_{\text{as}}, F^{2}_{\text{as}},E^{2}_{\text{as}})$. {(Note that since the VN graph of $G^{1}_{\text{as}}$  is non-extensible, $n_{v_1} \ne n_{v_2}$.)} Then, the existence of $G^{1}_{\text{as}}$ is a necessary condition {for} the existence of $G^{2}_{\text{as}}$}.
\end{corollary}

\textit{Proof:}
Suppose the CCMs of $G^{1}_{\text{as}}$ and $G^{2}_{\text{as}}$ are $ \mathbf{M}_1 $ and $ \mathbf{M}_2 $, respectively. If the VN graph of $G^{1}_{\text{as}}$  is a sub-graph of the VN graph of $G^{2}_{\text{as}}$, the independent cycles of $G^{1}_{\text{as}}$ will also be independent in $G^{2}_{\text{as}}$ and thus at least one valid $ \mathbf{M}_2 $ has $ \mathbf{M}_1 $ as a sub-matrix:
\begin{equation}
\mathbf{M}_2 = \left[
\begin{array}{cc}
\mathbf{M}_1 & 0\\
\mathbf{A} & \mathbf{B}
\end{array}
\right],
\end{equation}
{where the sub-matrix $ \left[ \mathbf{A} \quad \mathbf{B} \right] $ represents the other linearly independent cycles in $G^{2}_{\text{as}}$, which are not included in $G^{1}_{\text{as}}$.}
Therefore, if there exists a valid $ \mathbf{u}_2 $ such that $ \mathbf{M}_2 \mathbf{u}_2 = 0 \mod p$, the first 
elements would also be a valid $ \mathbf{u}_1 $ such that $ \mathbf{M}_1 \mathbf{u}_1 = 0 \mod p$.  {Because $G^{1}_{\text{as}}$ is non-extensible, a valid $ \mathbf{u}_1 $ such that $ \mathbf{M}_1 \mathbf{u}_1 = 0 \mod p$ is sufficient to force the existence of $G^{1}_{\text{as}}$.}  This shows that the existence of $G^{1}_{\text{as}}$ is a necessary condition of the existence of $G^{2}_{\text{as}}$.
\hfill$\blacksquare$


\section{Analytical results for $r=5$ SCB codes}\label{analysis5}

{
{We consider $r=3$, $r=4$ and $r=5$ SCB codes, but we begin with the $r=5$ case because simple row selection eliminates the smallest absorbing sets in this case.}  Thus, this section considers $r=5$ (five row groups) and shows how to design an SCB code with a specified circulant matrix that eliminates the smallest absorbing sets by selecting rows from the SCB mother matrix to force the CCMs associated with the dominant absorbing sets to be full column-rank.}  

This example of SCB code design involves two classes of SCB codes:
\begin{itemize}
\item Array-based codes \cite{Fan00} are the most elementary SCB codes in which the first $r$ rows of the SCB mother matrix $H_{p,f}^{p,p}, f(i,j)=i \cdot j $ comprise the parity-check matrix.  We will refer to this class as the elementary array-based (EAB) codes.
\item As shown in \cite{DOLECEKISTC10}, a careful selection of the $r$ row-groups from the overall {SCB} mother matrix can improve performance over the EAB codes.  Thus, selected-row (SR) SCB codes are our second class of SCB codes.  The parity-check matrix for these codes is $H_{p,f}^{r,p}, f(i,j)=g_r(i) \cdot j $ where $g_r(i)$ is called the row-selection function (RSF).  
{We will often represent an RSF as the vector $\begin{bmatrix} g_r(0)&g_r(1)&g_r(2)&g_r(3)&g_r(4)\end{bmatrix}$.}
\end{itemize}

Theorem~\ref{thm:ccm} shows that an absorbing set may be avoided {either} by forcing the associated CCM to {have a zero-dimension nullspace} or {by} precluding $\mathbf{u}$ from {being in} the null space of $\mathbf M$. Corollary~\ref{cor:ext} shows that if a non-extensible absorbing set does not exist, then all absorbing sets whose VN graphs contain the VN graph of this absorbing set also do not exist. The CCM approach carefully selects the RSF (and possibly also an analogous column selection function (CSF)) to preclude small absorbing sets, in the order of the size {(the value of $n_v$)} of the VN graph of the absorbing sets.

{Section  \ref{subsection48} establishes that $(4,8)$ absorbing sets are the smallest possible for a general $r=5$ code family.  Then Corollary~\ref{lemma48} shows that $(4,8)$ absorbing sets indeed exist for this $r=5$ array-based code family for each set of row group labels that satisfy the Bit Consistency conditions and $\det M = 0 \mod p$.  This theoretical result is also consistent with previous experimental results of a sum-product decoding algorithm implemented in software and on a hardware emulator~\cite{zhangTCOM} for which it was shown that decoding errors due to $(4, 8)$ absorbing sets dominate the low BER region of certain $r=5$ array-based codes.  
}

Sections \ref{subsection48}, \ref{subsection59} and \ref{subsection68}, respectively, show that $(4,8)$, $(5,9)$ and $(6,8)$ absorbing sets exist in the EAB code with $r=5$ and that carefully selecting row groups from the SCB mother matrix can preclude all these absorbing sets. Section~\ref{nonexistence3sets} provides several good RSFs that preclude these absorbing sets for $r=5$ SCB codes.  Section~\ref{tannerfor5} explores the absorbing set spectrum of the existing quasi-cyclic LDPC codes with the Tanner  \textit{et al.} construction \cite{tanner04}. Section~\ref{sec:equivmap5} identifies equivalence classes among SCB codes and uses the equivalence-class approach to enumerate all of the RSFs for $r=5$ and $p=67$ that preclude all $(4,8)$, $(5,9)$ and $(6,8)$ absorbing sets.


\subsection{$ (4,8) $ absorbing sets} \label{subsection48}
In this section we analyze $ (4,8) $ absorbing sets. The main result states that $ (4,8) $ absorbing sets exist in SCB codes 
{and specifically in} 
EAB codes (Corollary \ref{cor:eab48}), but that 
{$ (4,8) $ absorbing sets } 
can be provably eliminated from SR-SCB codes 
{using a suitable RSF,  as shown in Corollary \ref{cor:sr48}.}
These theoretical results will be substantiated by experimental results in Section~\ref{results}.

{Since Lemma \ref{lemma:girth=6} establishes that all SR-SCB codes have girth 6, $(4,8)$ absorbing sets are the smallest possible for SR-SCB codes with variable node degree $r =5$.  (Also see \cite{dolecekITA10}.) To see why this is true, note that the cases with $n_v<4$ are not possible because each of $n_v$ variable nodes needs to have at least three satisfied checks but this would require a girth of four. When $n_v=4$,
{ the girth constraint and the need for three satisfied checks prevent check node degrees larger than two and prevent any variable node from sharing  more than three checks with other variable nodes in the absorbing set. Hence,}  
$n_o=8$ and each pair of variable nodes in the absorbing set shares a distinct satisfied check.
Therefore, the $(4,8)$ absorbing set in Fig.~\ref{fig48} is the smallest possible absorbing set in the code family described by the parity check matrix $H_{p,f(i,j)}^{5,p}$. More general results regarding the minimality of absorbing sets are provided in~\cite{dolecekITA10}.}

This section shows that the $(4,8)$ absorbing set always exists in $r=5$ EAB codes.  However, by carefully selecting row groups from the SCB mother matrix, $r=5$ SR-SCB codes can preclude all $(4,8)$ absorbing sets.

Using Fig.~\ref{fig48} and \eqref{eq:detM48}, Theorem~\ref{thm:ccm} leads to the following corollary regarding $(4,8)$ absorbing sets:

\begin{corollary}\label{lemma48}
Consider an SCB mother matrix with a specified $p$.   The existence of a selection of integers for row group labels $i_1, \dots, i_6$ satisfying the Bit Consistency conditions associated with the absorbing set shown in Fig.~\ref{fig48} and satisfying $\det \mathbf{M}=0 \mod p$ (with $\det \mathbf{M}$ given in \eqref{eq:detM48}) is necessary and sufficient for the existence of the $(4,8)$ absorbing set of Fig.~\ref{fig48} in the bipartite graph of the SCB mother matrix.
\end{corollary}

\textit{Proof}:
Since the VN graph in Fig.~\ref{fig:vn48} of the $ (4,8) $ absorbing sets is a fully connected graph, it is not extensible without introducing a parallel edge, which would in turn imply a length-4 cycle in the corresponding bipartite graph. {This property would violate Lemma \ref{lemma:girth=6}}.  Because the VN graph is non-extensible, Theorem~\ref{thm:ccm} applies.  Hence $ \det \mathbf{M}$$=$$0 \mod p$ and Bit Consistency are both necessary conditions.  If the Check Consistency is also satisfied, the three conditions together are sufficient by Theorem 1.  The rest of the proof shows that identifying a selection of integers for row group labels $i_1, \dots, i_6$ satisfying the Bit Consistency conditions and satisfying $\det \mathbf{M}$$=$$0 \mod p$ implies the existence of $j_1, j_2,j_3, j_4$ satisfying the Check Consistency conditions.

Consider the Bit Consistency conditions associated with each of the four variable nodes in Fig.~\ref{fig48}.  Our concern is only regarding Bit Consistency conditions that involve satisfied check nodes.  Hence the six row-group labels of interest are $\{i_1, \dots, i_6\}$.  For example, the Bit Consistency conditions applied to $(j_1,k_1)$ require that $i_1$$\ne$$i_5$, $i_1$$\ne$$i_4$, and $i_4$$\ne$$i_5$.

There are six different row-group labels of interest, but only five possible row groups since $r=5$.  At least one pair of row-group labels must share the same row group. The inequalities implied by the Bit Consistency conditions allow only $i_1=i_3$, $i_2=i_4$, and $i_5 = i_6$.

{As shown in \cite{dolecekIT10} for $(4,4)$ absorbing sets, if all three of these equalities are satisfied, the Cycle Consistency conditions applied to the three length-8 cycles involving two pairs of same-row-group check nodes (i.e.,  $v_1$$-$$v_2$$-$$v_3$$-$$v_4$$-$$v_1$, $v_1$$-$$v_2$$-$$v_4$$-$$v_3$$-$$v_1$, and $v_1$$-$$v_4$$-$$v_2$$-$$v_3$$-$$v_1$),  yield three equations that require all variable nodes to have the same column group. This violates the Check Consistency conditions.}

Visualizing the $(4,8)$ absorbing set as a triangular-base pyramid with the four variable nodes as vertices, every point is symmetric to every other point and the three possible equalities allowed by the Bit Consistency conditions are all isomorphisms.  Thus, we only need to consider one case of one equality being satisfied and one case of two equalities being satisfied.

Thus, there are only two possible non-isomorphic row-group labelings  for the $(4,8)$ absorbing set as follows:  for  any bijective assignment of 
{five distinct row group labels}  to $\{ t, x, y, z, w \}$, $(i_1,i_2,i_3,i_4,i_5,i_6)$ can be either $(x,y,x,y,z,w)$   (Assignment 1, where two equalities are satisfied) or $(x,t,w,y,z,z)$  (Assignment 2, where only one equality is satisfied). {A detailed proof of the possible assignments can be found in \cite{DOLECEKISTC10}.} 
Applying  \eqref{eq:detM48} to these two assignments yields the following:
\begin{equation}\label{eq11}
\det \mathbf{M}={(y-x)}\left( (z - x)(w - y) + (z - y)(w - x)\right)
\end{equation}
for Assignment 1, and
 \begin{equation}\label{eq2}
\det \mathbf{M}=(z-w)(x-t)(y-z)-(y-w)(x-z)(z-t)
\end{equation}
for Assignment 2.

{Suppose} $ \det{\mathbf{M}}=0 \mod p $.  Then, there exists a non-zero solution to $\mathbf{M} \cdot \mathbf{u} = 0 \mod p$, where $\mathbf{u}$$=$$[u_2,u_3,u_4]^T $.   With either Assignment 1 or Assignment 2, the resulting $ \mathbf{M}$ has six nonzero entries such that if one of $\{ u_2,u_3,u_4 \}$ is nonzero, all three must be nonzero.  Furthermore, the structure of $\mathbf{M}$ ensures that any nontrivial solution to $\mathbf{M} \cdot \mathbf{u} = 0 \mod p$ will have $u_2$, $u_3$, and $u_4$ all distinct. Thus, for a fixed $ j_1 $, we can find $ j_2 $, $ j_3 $, and $ j_4 $ without contradiction to Check Consistency (since all are distinct).  Fixing one specific $k$ value will determine all other $k$'s according to Cycle Consistency, yielding the variable nodes of $ (4,8) $ absorbing sets in the code.  Therefore, a row-group labeling that satisfies Bit Consistency and $ \det{\mathbf{M}}$$=$$0 \mod p $ is a necessary and sufficient condition for the existence of $ (4,8) $ absorbing sets. \hfill $\blacksquare$

We will examine how Corollary \ref{lemma48} applies to all SR-SCB codes. To begin, consider the special case of EAB codes, which select the first five rows so that $\{x, y, z, w,t\} = \{0, 1, 2, 3, 4\}$.  For Assignment 1 there are no solutions that achieve $ \det{\mathbf{M}}$$=$$0 \mod p $ for prime $p$ large enough ($p$$>$$17$).  However, for Assignment 2 there are $8$ solution sets for $(x, y, z, w, t)$: $\left\{(4,3,2,0,1),(4,1,2,0,3),(3,4,2,1,0),(3,0,2,1,4), \right.$
$\left. (1,4,2,3,0),(1,0,2,3,4),(0,3,2,4,1),(0,1,2,4,3)\right\}$. Note that in these solutions $z$, the value of $i_5=i_6$,  is always $2$.

\begin{remark}\label{48structure}
{The eight Assignment 2 solutions above imply a relatively simple structure.  The six edges of the VN graph in Fig. \ref{fig:vn48} include three pairs of edges such that the edges in each pair do not share any variable nodes.  Each pair of edges corresponds to a pair of check nodes in  Fig.~\ref{fig:vn48}.  The check-node pairs are $(i_1, i_3)$, $(i_2,i_4)$, and $(i_5,i_6)$.  Including isomorphisms, the eight solutions above imply that one of the pairs of check nodes has both check nodes in row group 2, a second pair of labels must have check nodes in row groups 0 and 4, and the third pair must have check nodes in row groups 1 and 3.  The eight solutions above assume that the $(i_5,i_6)$ pair has both check nodes in row group 2, but choosing one of the other pairs to have  both check nodes in row group 2 merely produces an isomorphism.} 
\end{remark}

As an example, \deleted{let's} \added{let us} find one set of $ j $'s and $ k $'s so that $ \det \mathbf{M}$$=$$0 \mod p $. Among the EAB Assignment 2 solutions above, consider the last one: $(x $$=$$0, ~y$$=$$1, ~z$$=$$2,~w$$=$$4,~t$$=$$3)$ so that $(i_1,i_2,i_3,i_4,i_5,i_6) = (x,t,w,y,z,z)=(0,3,4,1,2,2)$. For this solution set,

\begin{equation}\label{eq48b}
\mathbf{M} = \begin{bmatrix}
i_1-i_2 & i_2-i_5 & 0\\
i_1-i_6 & 0 & i_6-i_4 \\
0 & i_5-i_3 & i_3-i_4
\end{bmatrix}
= \begin{bmatrix}
-3& 1 & 0\\
-2 & 0 & 1\\
0 & -2 & 3
\end{bmatrix}
\, ,
\end{equation}
and
\begin{equation}\label{eq:detM48example}
\det \mathbf{M}=-(-3)(1)(-2)-(1)(-2)(3) = 0\, .
\end{equation}
\normalsize

As shown in \eqref{eq:detM48example}, $ \det \mathbf{M}$$=$$0 $ for any $ p $. Suppose $ p=47 $ and $ u_2=1 $. Solving $ \mathbf{M} \mathbf{u}=0 $ gives us $ u_3=3$ and  $u_4=2 $. Selecting $ j_1=0 $ gives $ j_2=1,j_3=3,j_4=2 $. From Pattern Consistency, taking $ k_1=0 $ gives $ k_2=0, k_3=41, k_4=45 $. Thus, once a nonzero solution to $ \mathbf{M} \mathbf{u}=0 \mod p $ is specified (for example by selecting $u_2$ and solving for $\mathbf{u}$ as above), specifying any $j_1$ and $k_1$ leads directly to a solution that identifies a specific absorbing set.

Continuing with this particular choice of $(i_1,i_2,i_3,i_4,i_5,i_6)$, there are $p-1$ ways to select $u_2$, each of which yields a distinct nonzero solution to $ \det \mathbf{M}$$=$$0 \mod p $.  (Selecting $u_2$$=$$0$ gives the all-zeros solution.) For each of these solutions there are $p$ ways to select each of $j_1$, and $k_1$, yielding a total of $p^2(p-1)$ solutions.  

Thus, there are there $\mathbf{\Theta}(p^3)$ such solutions. {Here and in the remainder of the paper we use the standard asymptotic notation for $\mathbf{\Theta}$: a positive function $f(p)$ is $\mathbf{\Theta}(p^k)$ if there exist constants $c_1$ and $c_2$ such that $0 <c_1 \leq c_2 < \infty$ for which $c_1p^k \leq f(p) \leq c_2p^k$ for all $p \ge 0$.}  For each of the eight Assignment 2 solutions above there are $\mathbf{\Theta}(p^3)$ absorbing sets.  The overall number of absorbing sets is thus $\mathbf{\Theta}(p^3)$. 


To avoid the $(4,8)$ absorbing set we need to consider alternatives to the EAB selection of row-group labels in  $\{0, 1, 2, 3, 4\}$.  In general, SR-SCB codes can select the row-group labels to force the  CCM  for the $(4,8)$ absorbing sets to have a zero-dimension nullspace (equivalent to forcing the determinant to be nonzero for this square CCM). The goal is to identify a set of five (since $r=5$) row-group labels such that any valid assignment of those labels to $i_1$ through $i_6$ satisfies $ \det{\mathbf{M}}$$\ne$$0 \mod p $  {in both \eqref{eq11} and \eqref{eq2}}.  One such example is the RSF given by $[0, 1, 2, 4, 6 ]$.  For a valid assignment of $\{0, 1, 2, 4, 6 \}$ to $i_1$ through $i_6$, recall that the Bit Consistency conditions allow only $i_1=i_3$, $i_2=i_4$, and/or $i_5 = i_6$.  Any such valid assignment of these labels has $ \det{\mathbf{M}}$$\ne$$0 \mod p $  {in both \eqref{eq11} and \eqref{eq2}} for any prime $p$ greater than $23$.   Therefore in SR-SCB codes with a well chosen RSF there are no $(4,8)$ absorbing sets for sufficiently large $p$. Another example is the RSF  $[0, 1, 3, 8, 19]$ which avoids all $(4,8)$ absorbing sets for $p=47$ and certain other $p$'s as described in \cite{DOLECEKISTC10}. 

The above analysis proves the following corollaries.
\begin{corollary} \label{cor:eab48}
The $(4, 8)$ absorbing sets exist in {all $r=5$} SCB mother matrices and in particular in all EAB codes described by the parity check matrix $H_{p,i \cdot j}^{5,p}$, and their multiplicity scales as $\mathbf{\Theta}(p^3)$ in the EAB codes.
\end{corollary}
\begin{corollary} \label{cor:sr48}
There are no $(4,8)$ absorbing sets in the SR-SCB codes described by the parity check matrix $H_{p, g_r(i)\cdot j}^{5,p}$, for prime $p$ large enough and with {a} proper choice of  $g_r(i)$.
\end{corollary}

\subsection{$ (5,9) $ absorbing sets} \label{subsection59}
Assuming an RSF that avoids the $(4,8)$ absorbing sets, this section proves that the $(5,9)$ absorbing sets are the smallest remaining. {The CCM approach shows that the $(5,9)$ absorbing sets always exist in the EAB codes (Corollary \ref{cor:eab59}), but are avoided for SR-SCB codes by some of the RSFs that precluded the $(4,8)$ absorbing sets (Corollary \ref{cor:sr59}).}
We start by establishing a series of intermediate results, proven in Corollaries~\ref{lemma5b},~\ref{cor5b} and~\ref{lemma59}.

\begin{corollary}\label{lemma5b}
{For an $r=5$ SR-SCB code in which} $ (4,8) $ absorbing sets do not exist, $(5,b)$ absorbing sets do not exist for $ b<9 $.
\end{corollary}
{\textit{Proof:}  For $r=5$, the total number of edges in a  $(5,b)$ absorbing set is 25.   The number of edges that connect to satisfied check nodes is even, with two such edges associated with a pair of connected variable nodes.  Since the girth is six for SR-SCB codes, each pair of variable nodes can be connected at most once.  There are ten such pairs.  
Any group of four variable nodes forms a (4,8) absorbing set iff all six of its variable-node pairs are connected.  Thus, if $ (4,8) $ absorbing sets do not exist, each such group of four variable nodes must have one unconnected pair.  This implies that at least two of the ten possible pairs must be unconnected.  Thus, there are at most eight connected variable-node pairs, which implies at most sixteen edges connected to satisfied check nodes, which means at least nine edges connected to unsatisfied check nodes, and therefore $b\ge9$.}   
\hfill$\blacksquare$

\begin{corollary}\label{cor5b}
{For an $r=5$ SR-SCB code in which} $ (4,8) $ absorbing sets do not exist, the $(5,9)$ absorbing set is the smallest possible, and there are no other $ (5,b) $ absorbing sets.
\end{corollary}

\textit{Proof:}
Because each of the five variable nodes must have three satisfied checks, fifteen of the twenty five edges must go to satisfied checks.  Thus, the number of edges that go to unsatisfied check nodes is at most ten. From Corollary~\ref{lemma5b}, $b$$\ge$$9$ in the absence of $ (4,8) $ absorbing sets.   Since $ b $ is odd, the smallest possible $ (5,b) $ absorbing set is the $ (5,9) $ absorbing set.  We {now} prove that the $ (5,9) $ absorbing set does not contain the $(4,8)$ absorbing-set as a sub-graph.

Suppose a $(5,9)$ absorbing set has a $ (4,8) $ absorbing set as a subgraph. We label the five variable nodes as $ v_1, v_2, v_3, v_4 $ and $v_5  $. Without loss of generality, let $ v_1, v_2, v_3 $ and $ v_4 $ form a $(4,8)$ absorbing set {as in Fig.~\ref{fig48}}. If $ v_5 $ connects to a satisfied check node in the $ (4,8) $ absorbing set, for example check node {$ i_1 $} that connects $ v_1 $ and $ v_2 $, $ v_1 $ and $ v_2 $ will have more unsatisfied check nodes than satisfied check nodes. Then $ v_5 $ has {also} to connect to one of $ v_1 $'s unsatisfied check nodes ({$i_7 $ or $i_{11}$}) to make $ v_1 $ satisfy the absorbing set property. However, $ v_1, i_1 , v_5$ and $i_7 $ or $ v_1, i_1 , v_5$ and  $i_{11}$ would then form a length-4 cycle, which leads to a contradiction {of  Lemma \ref{lemma:girth=6}}. 

Thus $ v_5 $ cannot connect to the satisfied check nodes of $ v_1, v_2, v_3 $ and $ v_4 $. Since $ v_5 $ has at least three satisfied check nodes, it has to connect to at least three of the formerly unsatisfied check nodes of $ v_1, v_2, v_3 $ and $ v_4 $. This configuration makes the total number of unsatisfied check nodes seven or less and certainly not nine. This property again leads to a contradiction.  Therefore, $(5,9)$ absorbing sets cannot have $ (4,8) $ absorbing sets as a subset. \hfill$\blacksquare$

\begin{figure}
\centering
\includegraphics[width=0.31\textwidth]{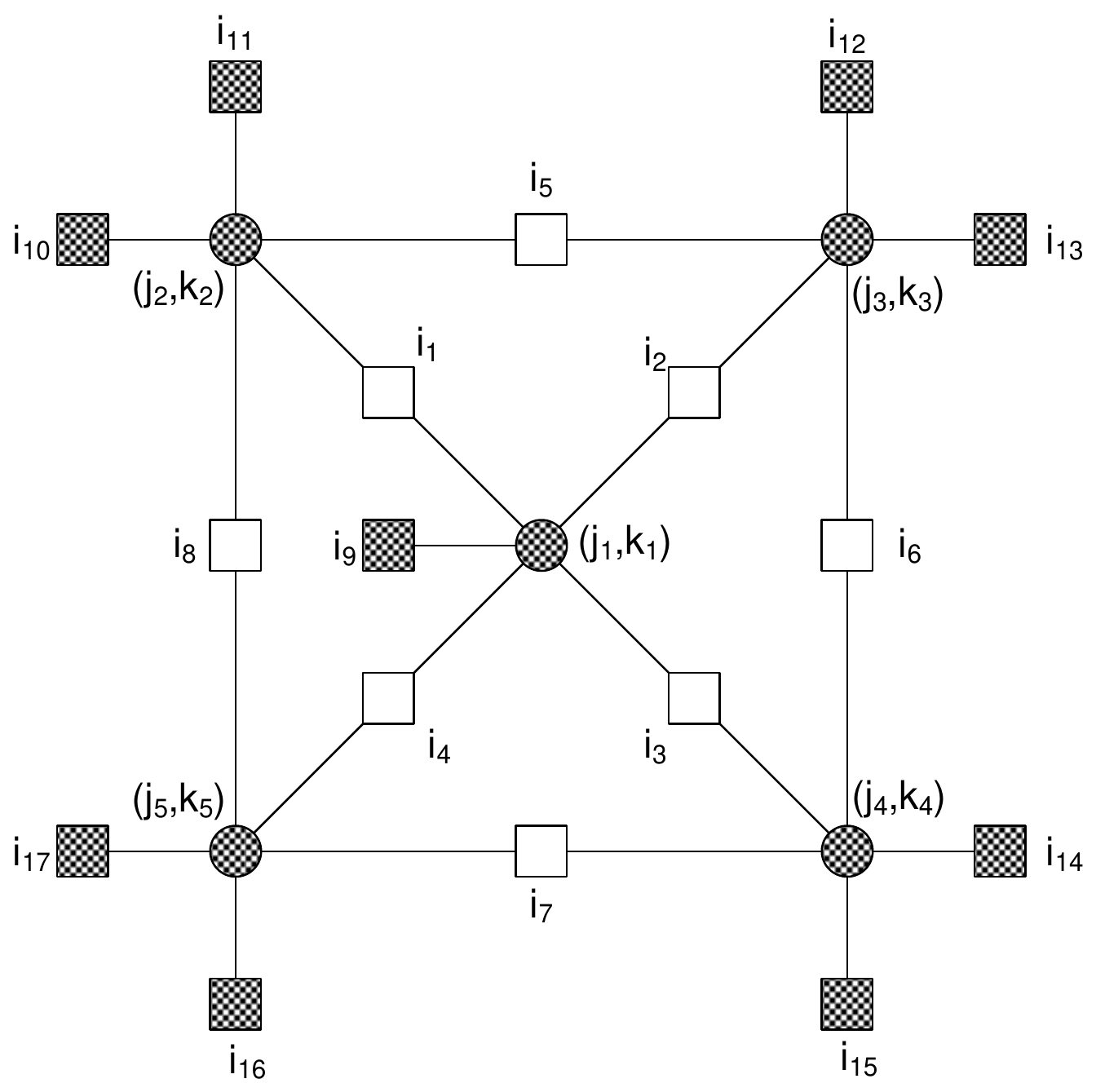}
\caption{Depiction of a $(5,9)$ absorbing set. Black circles are bit nodes in the absorbing set, white squares are their satisfied checks, and black squares are their unsatisfied checks.\label{fig59}}
\vspace{-0.2in}
\end{figure}

The only remaining possible configuration of a $(5,9)$ absorbing set is {where} one variable node  has four satisfied check nodes and four variable nodes have three satisfied check nodes. Fig.~\ref{fig59} depicts this configuration, which has a VN graph that does not contain the VN graph of a $ (4,8) $ absorbing as a sub-graph. {Corollary~\ref{lemma59}} below establishes  the necessary and sufficient conditions of this absorbing set.

\begin{corollary}\label{lemma59}
For matrix $\mathbf{M}$ in \eqref{eq59} below, a row-group label assignment that satisfies the Bit Consistency conditions and  satisfies $ \det \mathbf{M}=0 \mod p$ is necessary and sufficient for the existence in an SR-SCM code of the $ (5,9) $ absorbing set shown in Fig.~\ref{fig59}.
\end{corollary}

\textit{Proof:}
With an analysis similar to that of $ (4,8) $ absorbing sets, the binary cycle space for Fig.~\ref{fig59} has dimension $4$. We construct the following CCM by selecting the following linearly independent cycles: {$v_1$$-$$v_2$$-$$v_3$$-$$v_1$, $v_1$$-$$v_2$$-$$v_5$$-$$v_1$, $v_1$$-$$v_3$$-$$v_4$$-$$v_1$, and $v_1$$-$$v_4$$-$$v_5$$-$$v_1$}:
\begin{equation}\label{eq59}
\mathbf{M} =\left[
\begin{array}{cccc}
i_1-i_5 & i_5-i_2 & 0 & 0\\
i_1-i_8 & 0 & 0 & i_8-i_4 \\
0 & i_2-i_6 & i_6-i_3&0\\
0 & 0 & i_3-i_7 & i_7-i_4
\end{array}
\right].
\end{equation}
\normalsize

As in the proof of Corollary~\ref{lemma48}, we can show that a row-group label assignment that satisfies the Bit Consistency conditions and  $ \det \mathbf{M} = 0 \mod p $ is necessary and sufficient for the existence of a $(5,9)$ absorbing set, where
\begin{equation} \label{eq59m}
\begin{split}
\det \mathbf{M} =& (i_1-i_5)(i_8-i_4)(i_2-i_6)(i_3-i_7)\\
&-(i_1-i_8)(i_5-i_2)(i_6-i_3)(i_7-i_4) \mod p.
\end{split}
\end{equation}
\normalsize
\hfill$\blacksquare$

Under the Bit Consistency conditions, there are five possible non-isomorphic check labeling patterns of $(i_1,i_2,i_3,i_4,i_5,i_6,i_7,i_8)$ as follows: $(x,y,z,w,z,w,x,y)$, $(x,y,z,w,t,w,x,y)$,$(x,y,z,w,t,w,x,z)$,$(x,y,z,w,t,x,t,y)$, and $(x,y,z,w,t,w,t,y)$, where different letters correspond to distinct row-group labels.

The EAB codes have the set of row-group labels $\{ x, y, z, w,t \}$ drawn from the set $ \{0, 1, 2, 3, 4\}$. For the $ 4 $th pattern  $(x,y,z,w,t,x,t,y)$, there are eight solution sets $(x, y, z, w, t)$ $ \in$ $\left\{(4,0,1,3,2),(4,0,3,1,2),(3,1,4,0,2),(3,1,0,4,2), \right.$ $\left. (1,3,4,0,2),(1,3,0,4,2),(0,4,1,3,2),(0,4,3,1,2)\right\}$ that always have $\det \mathbf{M}$$=$$0$ in \eqref{eq59m}. 
{The other four patterns have a nonzero determinant for $ p $ large enough.} Once the labels of the check nodes are selected so that  $\det \mathbf{M}$$=$$0$ in \eqref{eq59m} (cf. Fig.~\ref{fig59}), the variable node labels (pairs $(j_1,k_1)$ through $(j_5,k_5)$) can be selected in $\mathbf{\Theta}(p^3)$ ways as in the $(4,8)$ case.

For SR-SCB codes with a proper choice of RSF, $ \det \mathbf{M}$$\neq$$0 \mod p $ for $ p $ large enough. One such example is using an RSF of $[0,1,2,4,7] $ where $ \det \mathbf{M}$$\neq$$0 \mod p $ for the prime $p$ greater than $89$ except for the set $p \in\{ 101,103,131,179\}$. Therefore we can conclude with {the} following two corollaries.

\begin{corollary} \label{cor:eab59}
$(5,9)$ absorbing sets exist in EAB codes described by the parity check matrix $H_{p,i \cdot j}^{5,p}$, and their number scales as $\mathbf{\Theta}(p^3)$.
\end{corollary}
\begin{corollary} \label{cor:sr59}
There are no $(5,9)$ absorbing sets in the SR-SCB codes described by the parity check matrix $H_{p, g_r(i)\cdot j}^{5,p}$, for prime $p$ large enough and with {a}  proper choice of RSF $g_r(i)$.
\end{corollary}

\subsection{$ (6,8) $ absorbing sets} \label{subsection68}
This section considers the $(6,8)$ absorbing sets, which are the smallest remaining after the $(4,8)$ and $(5,9)$ absorbing sets.  We will investigate the $(6,8)$ absorbing sets both for EAB codes and for SR-SCB codes that preclude the $(4,8)$ and $(5,9)$ absorbing sets. The six subsections that follow examine respectively the six candidate configurations of $(6,8)$ absorbing sets to be studied. 

\textit{Section summary.} {We first prove in subsection~\ref{app:cand1} that  {in (6,8) absorbing sets in $r=5$ SR-SCB codes  that preclude $(4,8)$ absorbing sets the largest possible check node degree is $2$.}
Combinatorial and consistency arguments show (subsections~\ref{app:cand3} and ~\ref{app:cand4}) that {three} of the remaining five configurations are not present for $p$ sufficiently large in either the EAB code or in SR-SCB codes that preclude the $(4,8)$ and $(5,9)$ absorbing sets. 

The other two configurations have the cardinality $\mathbf{\Theta}(p^3)$ in the EAB code. However, both of these configurations contain a $(4,8)$ absorbing set as a subset and thus cannot be present in SR-SCB codes that preclude the $(4,8)$  absorbing sets. These two configurations are analyzed in subsections~\ref{app:cand2} and~\ref{app:cand5}.

With this comprehensive analysis of $(6,8)$ absorbing sets we can conclude that $r=5$ SR-SCB codes can avoid all $(4,8)$, $(5,9)$, and  $(6,8)$ absorbing sets for $p$ large enough if the five row groups are properly chosen.


\subsubsection{Configuration 1 -  Check nodes with degree$>$2}\label{app:cand1} 
This case is precluded by the following lemma:

\begin{lemma}\label{lemma68deg2}
If a $(6,8)$ absorbing set has a check node that connects to more than two variable nodes in the absorbing set graph,  {it must contain} a $(4,8)$ absorbing set as a sub-graph. 
\end{lemma}

\textit{Proof:} {Consider a check node that connects to more than two variable nodes in such an absorbing set. It is either a satisfied check connected to either 4 or 6 variable nodes or an unsatisfied check connected to either 3 or 5 variable nodes.}

{If this check is satisfied, it cannot be connected to all 6 variable nodes because any additional satisfied check would complete a length-4 cycle and violate the girth constraint.  Suppose now that the check in question is satisfied and connected to 4 (out of 6) variable nodes.  As with the degree-6 case discussed above, these 4 variable nodes (forming a 4-clique) cannot have any other checks in common because of the girth constraint. }

{The nodes in this 4-clique each have at least 3 satisfied checks, by the absorbing set constraint. For each of these 4 variable nodes, two satisfied (degree-2) checks must necessarily be shared with the remaining 2 variable nodes. By the girth constraint, these 4 variable nodes then have exactly 3 satisfied and 2 unsatisfied checks each. The two remaining variable must share an additional satisfied check (so that all their checks are satisfied) to make the total number of unsatisfied checks be equal to 8. Consider two variable nodes from the 4-clique and the two remaining variable nodes outside of the clique. Each of these 4 variable nodes pairwise shares a satisfied check  with the other 3 variable nodes. These checks are distinct and the configuration induced by these 4 variable nodes is indeed a $(4,8)$ absorbing set.}

 {Now, we suppose that the check connected to more than 2 variable nodes is unsatisfied. It cannot be connected to 5 variable nodes because to have three satisfied checks, pairs of these five variable nodes would need to share at least one additional check, violating the girth constraint.  Suppose that this check is connected to 3 variable nodes. These 3 variable nodes (creating a 3-clique) must each share a distinct satisfied check with each of the remaining 3 variable nodes, by the absorbing set property. By the girth constraint, each variable node in the clique has an additional unsatisfied check not shared with any other variable node. The total number of unsatisfied checks incident to the 3-clique is thus 4.  Since the total number of unsatisfied checks is 8, the remaining 3 variable nodes likewise share one unsatisfied check (thus creating another 3-clique) and each have an additional unsatisfied check not shared with any other variable node. Consider two variable nodes from each of the two 3-cliques. In this group of 4 variable nodes, every pair of variable nodes is connected via a distinct check, forming a $(4,8)$ absorbing set that is a subgraph of the $(6,8)$ absorbing set.}

Hence, all cases have been considered, and any $(6,8)$ absorbing set with a check node having degree larger than two must contain a $(4,8)$ absorbing set as a sub-graph. 
\hfill$\blacksquare$

Based on Lemma~\ref{lemma68deg2},  attention is now restricted to cases where all check nodes in the absorbing set graph have degree at most 2.



In a candidate $(6,8)$ absorbing set,  variable nodes can have 3, 4 or 5 satisfied checks.  
{To maintain 8 unsatisfied checks}
, there can be at most 2 variable nodes with 5 satisfied checks. Suppose first that there are two such variable nodes. Since there are a total of 8 unsatisfied checks, the other 4 variable nodes must each have 3 satisfied and 2 unsatisfied checks. This necessarily implies the configuration shown in Fig.~\ref{cand2} which we discuss next.

\subsubsection{Configuration 2 - Fig.~\ref{cand2}}\label{app:cand2}

\begin{figure}
\centering
\includegraphics[width=0.27\textwidth]{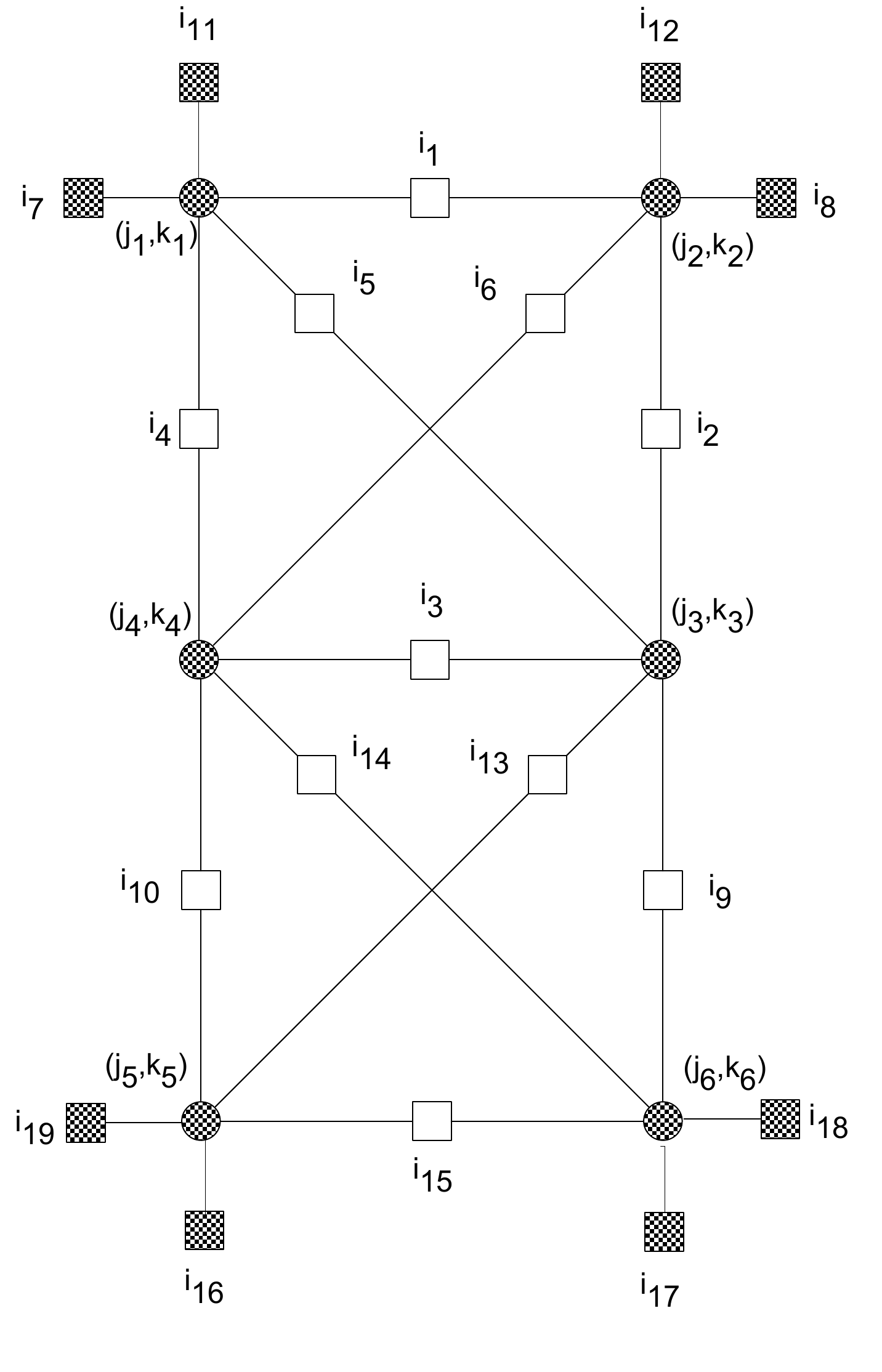}
\caption{Configuration 2 of a $(6,8)$ absorbing set.  In this configuration two variable nodes have 5 satisfied check nodes which forces the remaining four variable nodes each to have exactly two satisfied check nodes.  Note the two $(4,8)$ absorbing sets $(v_1, v_2, v_3, v_4)$ and  $(v_3, v_4, v_5, v_6)$.}\label{cand2}
\vspace{-0.2in}
\end{figure}

The configuration in Fig.~\ref{cand2} has two overlapping  $ (4,8) $ absorbing sets. Thus the existence of this $ (6,8) $ absorbing set {requires zero determinants for the CCMs of these two $ (4,8) $ absorbing sets}.   Thus, SR-SCB codes that preclude $(4,8)$ absorbing sets also preclude this $(6,8)$ absorbing set configuration.  This is formalized by the following corollary.

\begin{corollary} \label{cor:sr68cand2}
There are no $(6,8)$ absorbing sets with the structure of Fig.~\ref{cand2} in SR-SCB codes that preclude $(4,8)$ absorbing sets.
\end{corollary}

We can also show the following result:
\begin{corollary} \label{lemmacand2}
The existence of a selection of integers for row group labels satisfying the Bit Consistency conditions associated with the absorbing set shown in Fig.~\ref{cand2} and satisfying $ \det \mathbf{M}_1=0 \mod p$ and $ \det \mathbf{M}_2=0 \mod p$, where $ \mathbf{M}_1 $ and $ \mathbf{M}_2 $ are CCMs of the two internal $ (4,8) $ absorbing sets constitute a necessary and sufficient condition for the existence of the $ (6,8) $ absorbing set in Fig.~\ref{cand2}.
\end{corollary}
The proof of Corollary \ref{lemmacand2} follows that of Corollary~\ref{lemma48}.  

For EAB codes, we can show the following striking result:
\begin{corollary}
For EAB codes, there are exactly $p^2(p-1)$ $(6,8)$ absorbing sets of the type shown in Fig.~\ref{cand2} for $p>17$.
\end{corollary}

\textit{Proof:}  The $(6,8)$ absorbing set in Fig.~\ref{cand2} includes two $(4,8)$ absorbing sets, one spanning the 4 variable nodes $(j_1,k_1)$, $(j_2,k_2)$, $(j_3,k_3)$ and $(j_4,k_4)$ and the other spanning  $(j_3,k_3)$, $(j_4,k_4)$, $(j_5,k_5)$ and $(j_6,k_6)$.

Recall from Remark \ref{48structure} in Section~\ref{subsection48} that any $(4,8)$ absorbing set in an EAB code requires two check nodes connecting two disjoint pairs of variable nodes to be both in row group 2.  Another two check nodes connecting two disjoint pairs of variable nodes need to be in row groups 1 and 3.  The third and final two check nodes connecting two disjoint pairs of variable nodes need to be in row groups 0 and 4.

Applying this remark simultaneously to the two $(4,8)$ absorbing sets in Fig.~\ref{cand2}  reveals that the only possibility that does not violate the Bit Consistency condition is for  $i_1$, $i_{3}$, and $i_{15}$ in Fig.~\ref{cand2} to be in row group 2.  Neglecting isomorphisms, with  $i_1=i_{3}=i_{15}=2$, the required structure permits only two distinct row group labelings for the remaining satisfied check nodes.   Both have $i_4=i_9=1$ and  $i_2=i_{10}=3$.  However, one labeling has 
$i_5=i_{14}=4$ and  $i_6=i_{13}=0$ while the other has $i_5=i_{14}=0$ and  $i_6=i_{13}=4$.

For each of these two isomorphisms, there are $p^2(1-p)/2$ ways to distinctly select $u_2$, $j_1$, and $k_1$ (Symmetry causes half of the $p^2(1-p)$ choices for $u_2$, $j_1$, and $k_1$ to produce a labeling that is a 180 degree rotation of the labeling resulting from a different choice of  $u_2$, $j_1$, and $k_1$.)  Thus the total number of absorbing sets of the type in Fig.~\ref{cand2} is exactly $p^2(p-1)$. 
\hfill$\blacksquare$

 Suppose now that there is exactly one variable node in the absorbing set having all five checks satisfied. The variable nodes in the absorbing set must necessarily be arranged either as in Fig.~\ref{figa3} or Fig.~\ref{figa4}.
 
\subsubsection{$ (6,8) $ configuration 3 - Fig.~\ref{figa3}}\label{app:cand3} For this configuration, we have the following result.

\begin{figure}
\center
\includegraphics[width=0.31\textwidth]{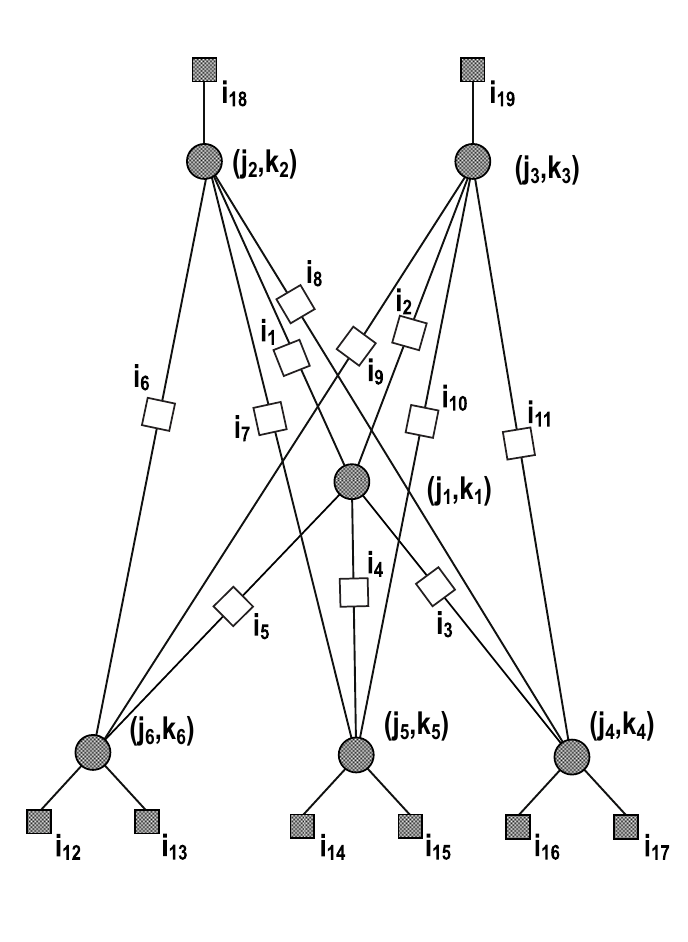}
\caption{Configuration 3 of a $(6,8)$ absorbing set.  This is the first of two configurations with exactly one variable node that has 5 satisfied check nodes.}
\label{figa3}
\vspace{-0.2in}
\end{figure}

\begin{lemma}\label{lemma2a}
In the Tanner graph corresponding to $H_{p, f(i,j)}^{5,p}$ for the EAB codes and SR-SCB codes, there are no $(6,8)$ absorbing sets for $p$ large enough of the type shown in Fig.~\ref{figa3}.
\end{lemma}

\textit{Proof:}
Without loss of generality we may assign check node labels for the checks emanating from the variable node {$(j_1,k_1)$} as follows:
$i_1=x$,$i_2=y$, $i_{3}=z$, $i_{4}=w$, and $i_5=t$, where $x,y,z,w,t$ are the five distinct check labels.

The binary cycle space for Fig.~\ref{figa3} has dimension $6$. We construct the CCM of \eqref{eq2:CCM_figa3} by selecting the linearly independent cycles  $v_1$$-$$v_2$$-$$v_6$$-$$v_1$, $v_1$-$v_3$-$v_6$-$v_1$, $v_1$-$v_2$-$v_5$-$v_1$, $v_1$-$v_3$-$v_4$-$v_1$, $v_1$-$v_3$-$v_5$-$v_1$, and $v_1$-$v_2$-$v_4$-$v_1$ as follows:

\begin{equation}\label{eq2:CCM_figa3}
\mathbf{M} =
\begin{bmatrix}
x-i_6 & 0 & 0 & 0 & i_6-t\\
0 & y-i_9 & 0 & 0 & i_9-t \\
x-i_7 & 0 & 0 & i_7-w & 0 \\
0 & y-i_{11} & i_{11}-z & 0 & 0 \\
0 & y-i_{10} & 0 & i_{10}-w & 0 \\
x-i_8 & 0 & i_8-z & 0 & 0
\end{bmatrix}.
\end{equation}
\normalsize

The rank of the matrix is at most 5 so in fact we may consider the $5 \times 5$ submatrix formed by the first five rows (call it $B$). If the matrix $B$ is full rank, then $\mathbf{M}$ is full rank and has a zero-dimension nullspace. Hence $\det(B)=0$ is necessary for the existence of the absorbing sets of this type.  This condition can be expressed as
\begin{equation}\label{eq2a111}
\begin{split}
-(i_{11}-z) [ -(x-i_6)(i_9-t)(i_7-w)(y-i_{10})&\\
+(x-i_7)(i_6-t)(y-i_9)(i_{10}-w)]= 0 \mod p.&
\end{split}
\end{equation}

Also consider the bottom-left $4 \times 4$ submatrix (call it $ A $). Note that the Bit Consistency conditions ensure that every element of $\mathbf{M}$ that is not explicitly zero in \eqref{eq2:CCM_figa3} must be nonzero.  Thus, if the matrix $ A $ is full rank, then $\mathbf{M}$ is full rank (rank-5) because either of the top two rows provides a row linearly independent from the bottom four rows.  Hence, $\det(A)=0$ is necessary for the existence of absorbing sets of the type shown in Fig.~\ref{figa3}.  This condition can be expressed as
\begin{equation}\begin{split}\label{eq2a112}
 -(x-i_7)(i_{10}-w)(y-i_{11})(i_{8}-z)&\\
 +(x-i_8)(i_7-w)(i_{11}-z)(y-i_{10})&= 0 \mod p.
\end{split}
\end{equation}
For the values of $i_6,i_7,i_9,i_{10}$ and $i_{11}$ in the {available label set $\{x,y,z,w,t\}$}
both for  the EAB and for the SR-SCB codes, and such that the $i$ labels of check nodes sharing a variable node are distinct (see Fig.~\ref{figa3}), \eqref{eq2a111} and \eqref{eq2a112} evaluate to zero for only a finite number of values of the parameter $p$.

For $g_r(i)=i$ (the EAB code) $\det(B) \neq 0$ for $p>23$. We can also find many SR-SCB codes where \eqref{eq2a111} and \eqref{eq2a112} evaluate to zero only for a finite number of values of the parameter $p$. 
{For example, with the SR-SCB code defined by the RSF $ [0,1,2,4,7] $, $\det(A) \neq 0$ for $p>89$.}
Thus for $p$ sufficiently large, neither the EAB code nor well-designed SR-SCB codes contain $(6,8)$ absorbing sets of the type shown in Fig.~\ref{figa3}.
\hfill$\blacksquare$

Incidentally, 
{there do exist values of $p$ for certain RSFs for which both  $\det(A) = 0$ and $\det(B) = 0$.  For example, for $p=11,17,19,23$ with the RSF $[0,1,2,3,4]$ both conditions hold. As we show in Corollary \ref{corollary2a} below,  $\det(A) = 0$ and $\det(B) = 0$ together are a sufficient condition for the existence of absorbing sets of the type shown in Fig.~\ref{figa3} if $(4,8)$ absorbing sets have been excluded by the RSF.
{We also identified RSFs such as $[0,1,2,4,8]$ where both determinants evaluate to zero (even before applying the$\mod p$) for some choice of check node labels, and thus the associated codes have these $(6,8)$ absorbing sets regardless of the choice of $p$.}}

\begin{corollary}\label{corollary2a}
In the Tanner graph corresponding to $H_{p, f(i,j)}^{5,p}$,  {if $(4,8)$ absorbing sets have been excluded by the RSF,} $(6,8)$ absorbing sets of the type shown in Fig.~\ref{figa3} exist if and only if row labels $i_1, \dots, i_{11}$ can be specified to satisfy Bit and Check Consistency conditions and to satisfy $\det(A)=0$ and $\det(B)=0$, {where $A$ and $B$ are as defined in the proof of Lemma~\ref{lemma2a}}.
\end{corollary}

\textit{Proof:}
We already established above that both $\det(A) = 0$ and $\det(B) = 0$ are necessary for the existence of $(6,8)$ absorbing sets of the type shown in Fig.~\ref{figa3}.  We only need to establish that these two conditions are also sufficient.

The rank of any matrix is lower bounded by the rank of any of its submatrices.  Recalling that every element of $\mathbf{M}$ that is not explicitly zero in \eqref{eq2:CCM_figa3} must be nonzero, $C$ is a full-rank 4 by 4 matrix. Thus the rank of $ \mathbf{M} $ is no less than 4.

The first three rows of $A$ are always linearly independent.  (Recall that every element of $\mathbf{M}$ that is not explicitly zero in \eqref{eq2:CCM_figa3} must be nonzero.) Thus if $ \det(A)=0 $ the bottom row of $ A $ is a linear combination of the other 3 rows.  Thus, the bottom row of $ \mathbf{M} $ is a linear combination of the three rows of $ \mathbf{M} $ just above it, and $ rank(B)=rank(\mathbf{M})$.

Thus, if  $ \det(A)$ and $ \det(B)$ are zero, $ \mathbf{M} $ is not full rank, and there is a non-zero $ \mathbf{u} $ in the nullspace of the $\mathbf{M}$.  If the selected RSF precludes $(4,8)$ absorbing sets, then the $(6,8)$ absorbing set in Fig.~\ref{figa3} is not extensible.  Thus by Theorem \ref{thm:ccm}  $ \det(A)=0 $ and $ \det(B)=0 $ is a sufficient condition for the existence of  the $(6,8)$ absorbing set in Fig.~\ref{figa3}. \hfill$\blacksquare$


\subsubsection{$ (6,8) $ configuration candidate 4 - Fig.~\ref{figa4}}\label{app:cand4}

Note that the configuration shown in Fig.~\ref{figa4} contains a subgraph that is the $(4,8)$ absorbing set of Fig. \ref{fig48}.  This implies the following:
 \begin{corollary}
For SR-SCB codes that do not contain $ (4,8) $ absorbing sets,  the configuration of Fig.~\ref{figa4} is not possible.
\end{corollary}

\begin{figure}
\center
\includegraphics[width=0.35\textwidth]{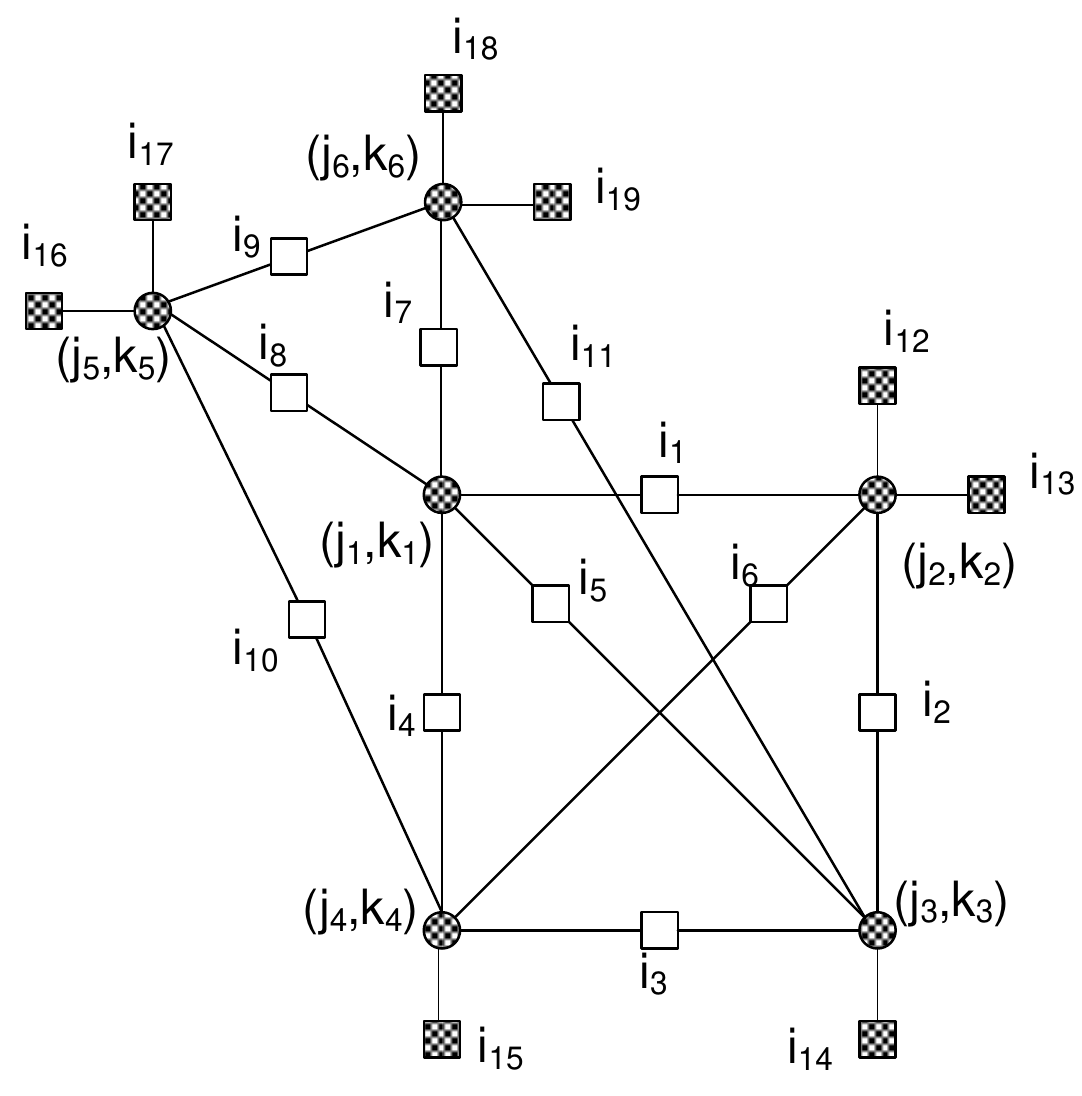}
\caption{Configuration 4 of a $(6,8)$ absorbing set.  This is the second of the two configurations with exactly one variable node that has 5 satisfied check nodes.}
\label{figa4}
\vspace{-0.2in}
\end{figure}

Even though EAB codes contain $(4,8)$ absorbing sets, the following lemma states that EAB codes avoid the $(6,8)$ absorbing set in Fig.~\ref{figa4}.
\begin{lemma}\label{lemmacand4}
In the EAB codes corresponding to $H_{p,i \cdot j}^{5,p}$, there are no $(6,8)$ absorbing sets with the topology shown in Fig.~\ref{figa4} for $p$ large enough.
\end{lemma}

\textit{Proof:}
The bcs for Fig.~\ref{figa4} has dimension $6$. Using linearly independent cycles {$v_1$$-$$v_2$$-$$v_3$$-$$v_1$, $v_1$$-$$v_2$$-$$v_4$$-$$v_1$, $v_1$$-$$v_3$$-$$v_4$$-$$v_1$, $v_1$$-$$v_3$$-$$v_6$$-$$v_1$,  $v_1$$-$$v_4$$-$$v_5$$-$$v_1$, $v_1$$-$$v_5$$-$$v_6$$-$$v_1$,} we construct the following CCM: 

\small
\begin{equation}\label{eq:CCM_figa4}
\mathbf{M} =
\begin{bmatrix}
i_1-i_2 & i_2-i_5 & 0 & 0 & 0\\
i_1-i_6 & 0 & i_6-i_4 & 0 & 0 \\
0 & i_5-i_3 & i_3-i_4 & 0 & 0 \\
0 & i_{11}-i_5 & 0 & 0 & i_7-i_{11} \\
0 & 0 & i_{10}-i_4 & i_{8}-i_{10} & 0 \\
0 & 0 & 0 & i_9-i_8 & i_7-i_9
\end{bmatrix}.
\end{equation}
\normalsize

Let the top-left $3 \times 3$ submatrix of $ \mathbf{M} $ be $A$.  Note that $A$ is exactly the CCM in \eqref{eq48} for the $(4,8)$ absorbing set of Fig. \ref{fig48}.
Corollary \ref{cor:ext} requires that $ \det A = 0 \mod p $ for the $(6,8)$ absorbing set of Fig.~\ref{figa4} to exist.

As shown in Corollary~\ref{cor:eab48}, EAB codes always contain $(4,8)$ absorbing sets.  Thus there is an assignment of EAB row group labels to $i_1,i_2,i_3,i_4,i_5,i_6$  such that $ \det A = 0 \mod p $. 

When $ \det A = 0 \mod p $, the second row of $ \mathbf{M} $ is a linear combination of the first and third rows. Therefore the rank of $ \mathbf{M} $ only depends on the submatrix $ \mathbf{\hat{M}} $ that contains all the rows of $ \mathbf{M} $ except the second row. 

{It suffices to consider the case when the labels $i_1,i_2,i_3,i_4,i_5,i_6$ adopt one of the  following two assignments: $(i_1,i_2,i_3,i_4,i_5,i_6)= (x,t,w,y,z,z)$ or
$(z,t,z,y,x,w)$.
For the first assignment,
\small
\begin{equation}
\begin{split}
\det{\mathbf{\hat{M}}}=&(x-t)[ (z-w)(i_{10}-y)(i_7-i_{11})(i_9-i_8)\\
&-(i_{11}-z)(w-y)(i_8-i_{10})(i_7-i_9)],
\end{split}
\end{equation}
\normalsize
and for the second assignment, 
\small
\begin{equation}
\begin{split}
\det{\mathbf{\hat{M}}}=&(z-t)[ (x-z)(i_{10}-y)(i_7-i_{11})(i_9-i_8)\\
&-(i_{11}-x)(z-y)(i_8-i_{10})(i_7-i_9)].
\end{split}
\end{equation}
\normalsize
For both assignments and $p>41$, $\det{\mathbf{\hat{M}}} \ne 0 \mod p$ for $i_1$ to $i_{11}$ taking values in the set $\{ 0,1,2,3,4\}$ such that the Bit Consistency constraints are satisfied. Thus, this $(6,8)$ absorbing set is not present in EAB codes for $p>41$.}  \hfill$\blacksquare$

\subsubsection{$(6,8)$ configurations 5 and 6 -- Figs. \ref{figa5} and \ref{figa6}}\label{app:cand5}

Finally, we consider the case when no variable node in the absorbing set has all five satisfied checks. This implies one of the two configurations shown in Figs.~\ref{figa5} and \ref{figa6}. 

Fig.~\ref{figa5} contains a $(4,8)$ absorbing set.  Thus any SCM code avoiding the $(4,8)$ absorbing set will also avoid this (6,8) absorbing set.  The EAB code, which does not avoid the $(4,8)$ absorbing set of Fig. \ref{fig48}, has $\mathbf{\Theta}(p^3)$  $(6,8)$ absorbing sets of the type shown in Fig.~\ref{figa5} for $p$ large enough to guarantee that the needed $(4,8)$ absorbing set is present.

{Using arguments similar to those in Lemma \ref{lemma2a}, we can establish that for $p$ large enough  there are no $(6,8)$ absorbing sets of the type shown in Fig.~\ref{figa6} in either the EAB codes or in the SR-SCB codes.}
We omit detailed analysis of these cases for brevity.  
\begin{figure}
\center
\includegraphics[width=0.31\textwidth]{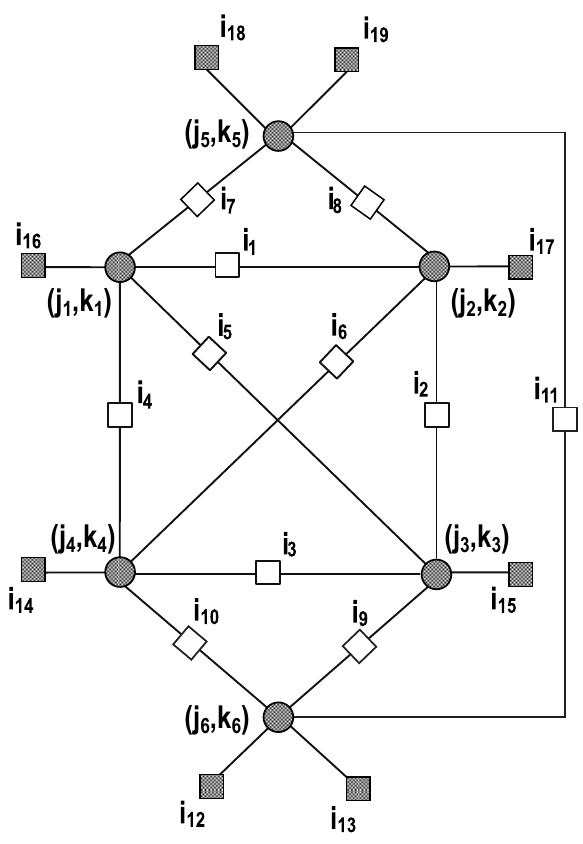}
\caption{Configurations 5 of a $(6,8)$ absorbing set.  This is the first of two configurations with no variable nodes that have 5 satisfied check nodes.}
\label{figa5}
\vspace{-0.2in}
\end{figure}

\begin{figure}
\center
\includegraphics[width=0.31\textwidth]{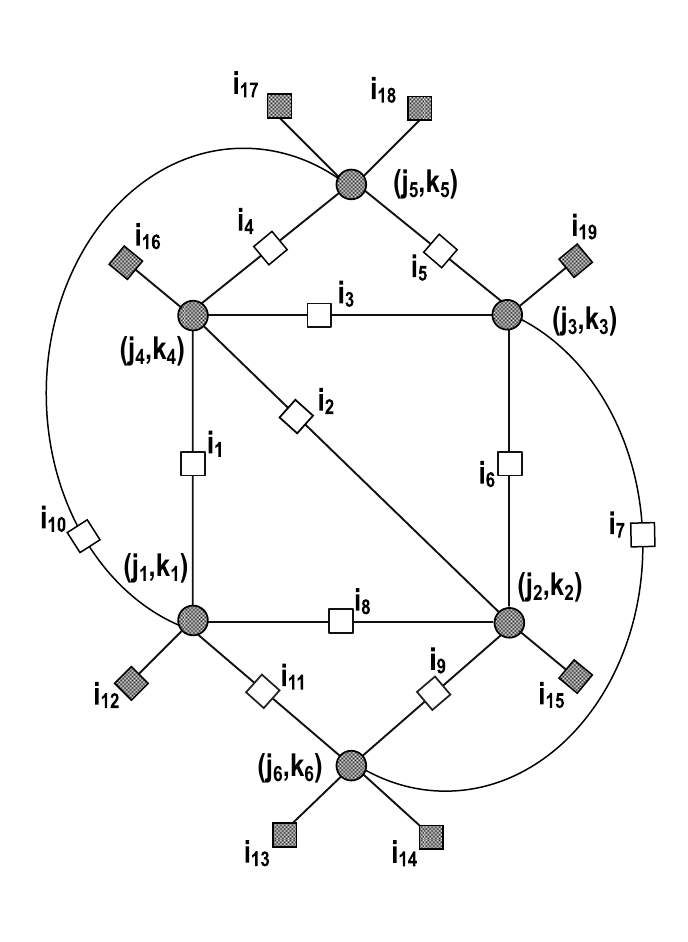}
\caption{Candidate 6 configuration of a $(6,8)$ absorbing set.  This is the second of two configurations with no variable nodes that have 5 satisfied check nodes.}
\label{figa6}
\end{figure}
{
\subsection{Summary for $ (4,8) $, $ (5,9) $ or $ (6,8) $ absorbing sets} \label{nonexistence3sets}

We have now considered all possible $ (4,8) $, $ (5,9) $, and $(6,8)$ configurations.  The following theorem is a consequence of the results established above:
\begin{theorem}\label{thm_r5}
In $r=5$ EAB codes the number of $(4,8)$, $(5,9)$ and $(6,8)$ absorbing sets scales as $\mathbf{\Theta}(p^3)$.  However, for SR-SCB codes with a properly selected RSF (such as the examples given in Table \ref{tbl:r=5codes}) the relevant determinants for the CCMs (or submatrices of CCMs) for all $(4,8)$, $(5,9)$ or $(6,8)$ absorbing sets  are nonzero  allowing a $p$ to be selected so that these determinants are all nonzero $\mod p$, thus precluding all $(4,8)$, $(5,9)$ and $(6,8)$ absorbing sets. \hfill$\blacksquare$
\end{theorem}

Table~\ref{tbl:r=5codes} presents some examples of RSFs that produce $r=5$ SR-SCB codes for which  $ (4,8) $, $ (5,9) $ and $ (6,8) $ absorbing sets do not exist.  These examples provide the smallest $p$ that precludes the absorbing sets.  The precluded absorbing sets can still appear again for larger values of $p$.  However, in each case there is a $p$ large enough that the absorbing sets are precluded for all larger values of $p$, as stated in the following corollary:

\begin{corollary} 
When all of the relevant determinants are nonzero, there is some $p$ for which the associated absorbing sets are precluded for all larger values of $p$.  \hfill$\blacksquare$
\end{corollary}
As examples, with RSF { $ [0,1,2,4,7] $ and  $p>179$ and with RSF $ [0,1,3,4,5] $ and $p>271$}  all $ (4,8) $, $ (5,9) $ and $ (6,8) $ absorbing sets are precluded.

\begin{table}
\caption{Row selection functions for SR-SCB codes with $r=5$ and the smallest $p$ precluding all $(4,8)$, $(5,9)$, and $(6,8)$ absorbing sets. \label{tbl:r=5codes}}
\begin{center}
\begin{tabular}{|l|c|l|}
\hline
$p$&Rate $\frac{k}{n}=1-\frac{rp-r+1}{p^2}$&RSF\\
\hline \hline
67&0.9263&$[0,1,2,4,17]$\\
\hline
73&0.9323&$[0,1,2,3,11]$\\
\hline
79&0.9373&$[0,1,2,6,7]$\\
\hline
83&0.9403&$[0,1,2,3,7]$\\
\hline
89&0.9443&$[0,1,2,4,11]$\\
\hline
\end{tabular}
\end{center}
\end{table}



\begin{remark}
For sufficiently small $ p $'s, $ (4,8) $, $ (5,9) $ and $ (6,8) $ absorbing sets cannot be eliminated simultaneously for the SR-SCB codes. \comment {For smaller $ p $'s, $ (4,8) $, $ (5,9) $ and $ (6,8) $ absorbing sets cannot be simultaneously eliminated in the SCB codes $H_{p, g_r(i) \cdot j}^{5,p}$ with any choice of RSF. }Here are a few examples that remove most of these absorbing sets for $ p<50 $: (1) for $ p=47 $ with RSF= $ [0,1,2,3,5] $, only the absorbing sets in Fig.~\ref{fig59} and Fig.~\ref{figa3} exist; (2) for  $ p=43 $ with RSF=$ [0,1,2,4,6] $, only the absorbing set in Fig.~\ref{fig59} exists.
 \hfill$\blacksquare$
\end{remark}

\subsection{Absorbing set spectrum in the Tanner  \textit{et al.} construction \cite{tanner04}}\label{tannerfor5}
We now apply our analysis to the Tanner \textit{et al.} construction in \cite{tanner04}, whose codes are in the family of SCB codes. 
{The Tanner \textit{et al.} codes  have the parity-check matrix $H_{p, a^i \cdot b^j}^{r,c}$ with $a,b \in GF(p)$ and $ o(a)=r $ and $ o(b)=c $, where $ o(a) $ indicates the multiplicative order of $ a $ in $GF(p)$. 
}

\begin{lemma}\label{lemmaqc1}
In the Tanner graph corresponding to quasi-cyclic LDPC codes with the parity check matrix given by $H_{p, {a^i \cdot b^j}}^{5,o(b)}$ in \cite{tanner04}, no $ (4,8) $ or $(6,8)$ absorbing sets exist with parameters selected in Table I of \cite{tanner04}.
\end{lemma}

\textit{Proof:} For the codes in  Table I of \cite{tanner04} with girth greater than 6, neither $ (4,8) $ nor $ (6,8) $ absorbing sets exist since they contain length-6 cycles.   For the codes in  Table I of \cite{tanner04} with girth $ =6 $, we have analyzed each code using the CCM approach and confirmed that in each case neither $ (4,8) $ nor $ (6,8) $ absorbing sets exist.  {We note that the girth = 6 codes of Table I in \cite{tanner04} span rates from 0.17 to 0.55.} The full analysis is too long to include, but we provide the following example.

Consider the girth $=6$ code with {$ p=31, c=6,r=5 $} as an example, with $ a=2, b=6 $. The resulting matrix $H_{31, f(i,j)}^{5,6}$ is a sub-matrix of the SR-SCB  code with RSF $[1,2,4,8,16]$.  Thus $f(i,j)=a^ib^j$ is achieved because the RSF enforces $a^i$ and the submatrix of the SR-SCB code selects the columns to enforce $b^j$.
We set up a system of equations as before, and conclude that for  $ (4,8) $ absorbing sets, the only possible labeling for $ p=31 $ is $ (i_1, i_2, i_3, i_4, i_5, i_6) = (x,t,w,y,z,z) $. Five non-isomorphic solutions, $ (x,y,z,w,t) = (2,4,1,16,8), (1,8,2,4,16), (1,2,4,16,8), (1,4,8,2,16)$, and  $(1,2,16,8,4) $
 set $ \det{\mathbf{M}}=0 \mod 31 $ for $ \det{\mathbf{M}}$ in \eqref{eq2}. 

{Each of the solutions corresponds to the CCM equation:
\small 
\begin{equation}\label{eqqcccm}
\mathbf{M} \mathbf{u} =
\begin{bmatrix}
x-t & t-z & 0\\
x-z & 0 & z-y \\
0 & z-w & w-y
\end{bmatrix}
\begin{bmatrix}
u_2\\
u_3\\
u_4
\end{bmatrix}
=0 \mod 31.
\end{equation}
\normalsize
Since $ \det{\mathbf{M}}=0 \mod 31$ and $ \text{rank}(\mathbf{M})=2 $, equation~\eqref{eqqcccm} is equivalent to 
\small
\begin{equation}\label{eqqcccm2}
\begin{bmatrix}
x-t & t-z & 0\\
x-z & 0 & z-y 
\end{bmatrix}
\begin{bmatrix}
u_1\\
u_2\\
u_3
\end{bmatrix}
=0 \mod 31.
\end{equation}
\normalsize
If we expand the $ u $'s with $ u_i= j_i-j_1 $, equation~\eqref{eqqcccm2} is equivalent to
\small
\begin{equation}\label{eqqcccm3}
\begin{bmatrix}
z-x &  x-t & t-z & 0 \\
y-x  & x-z & 0 & z-y
\end{bmatrix}
\begin{bmatrix}
j_1\\
j_2\\
j_3\\
j_4
\end{bmatrix}
=0 \mod 31.
\end{equation}
\normalsize

We denote the matrix in equation~\eqref{eqqcccm3} as $ \mathbf{Q_n} $ where the index $1 \le \mathbf{n} \le 5$ indexes the five realizations of $\mathbf{Q}$ corresponding to the five possible choices of $(x,y,z,w,t)$ values.  The nullspace of $ \mathbf{Q_n}$ is $ \mathbf{Q^{\perp}_n}$. For any  $ (4,8) $ absorbing set, $ (j_1,j_2,j_3,j_4) $ should be in $ \mathop  \cup \limits_{1 \leq \mathbf{n} \leq 5} \mathbf{Q^{\perp}_n} $.  Based on the code construction (cf. \cite{tanner04}), the column groups for this code are the powers of 6 mod 31.  Denote this set by $ Y=\{1,6,5,30,25,26\} $.   {The 4-tuple } $(j_1,j_2,j_3,j_4) $ must be a selection of four distinct elements of $Y$.  However, since there is no vector in $ \mathop  \cup \limits_{1 \leq \mathbf{n} \leq 5} \mathbf{Q^{\perp}_n} $ that consists of four distinct elements of $Y$, $ (4,8) $ absorbing sets do not exist and consequently the  $ (6,8) $ absorbing sets in Figs. \ref{cand2}, \ref{figa4} and \ref{figa5} also do not exist for this Tanner \textit{et al.} code.

%
{Similarly,  for the CCMs constructed in \eqref{eq2:CCM_figa3} for configuration 3 in Fig. \ref{figa3} and for configuration 6 in Fig. \ref{figa6} the null spaces do not include vectors that contain four distinct elements of $Y$, which precludes the absorbing sets in Fig. \ref{figa3} and \ref{figa6}.} \hfill$\blacksquare$}

Codes listed in Table I of \cite{tanner04} are mostly moderate-rate codes (the rate being around $0.5$). The construction of codes with higher rates  from this code family 
{ requires more columns to be selected from the SCB mother matrix. However, when more columns are selected (by choosing a higher-order $b$), the Tanner \textit{et al.} construction cannot guarantee the absence of certain absorbing sets.} 

\begin{lemma}\label{lemmaqc2}
In the Tanner graph corresponding to quasi-cyclic LDPC codes with the parity check matrix $H_{p, {a^i \cdot b^j}}^{5,o(b)}$ in \cite{tanner04}, selecting a higher-order $b$ to achieve higher rates can introduce $ (4,8) $, $ (5,9) $ and $ (6,8) $ absorbing sets.	
\end{lemma}
\textit{Proof:} {Continuing with the example above with $p=31$ and $a=2$ (which implies $r=5$),  we increase the rate by replacing $b=6$ (which implies $c=6$) with $b=15$  (which implies $c=10$.)} This new choice of $b$ yields the column selection set $Y= \{1,15,8,27,2,30,16,23,4,29\}$.  With this $Y$, $ (4,8) $ absorbing sets exist because there are vectors in $ \mathop  \cup \limits_{1 \leq \mathbf{n} \leq 5} \mathbf{Q^{\perp}_n} $ that consist of four distinct elements of $Y$.  Similarly, (6,8) absorbing sets exist  because corresponding solutions for Figs. \ref{figa3} and \ref{figa6} can be constructed from distinct elements of $Y$.\hfill$\blacksquare$


With the same row selection, the parity check matrix of the SR-SCB code has the parity check matrix of the Tanner \textit{et al.} construction as a submatrix. According to the conclusions of Theorem~\ref{thm_r5}, we can modify Tanner \textit{et al.} construction by using an optimized row selection 
{that precludes all the $ (4,8) $, $ (5,9) $ and $ (6,8) $ absorbing sets for appropriate values of $ p $.} 

\begin{lemma}\label{lemmaqc3}
In the Tanner graph corresponding to quasi-cyclic LDPC codes (in \cite{tanner04})  described by $H_{p, {a^i \cdot b^j}}^{5,o(b)}$ with higher rates and large enough $ p $, all of $ (4,8) $, $ (5,9) $ and $ (6,8) $ absorbing sets can be eliminated by only modifying the mapping sequence of the rows.
\end{lemma}

\textit{Proof:} By replacing the row mapping sequence $ g_r(i) = \alpha^{i-1} $ by $ g_r(i)=  [ 0,1,2,4,17 ] $ and choosing $p=67$, the quasi-cyclic LDPC $H_{p, \hat{f}(i,j)}^{5,o(b)}$ becomes a new class of SCB LDPC codes with $ \hat{f}(i,j)= g_r(i)\cdot b^j $. Since $H_{p, \hat{f}(i,j)}^{5,o(b)}$ is a submatrix of the mother {matrix} $H_{p,i \cdot j}^{p,p}$,  it follows that $ (4,8) $,$ (5,9) $ and $ (6,8) $ absorbing sets can be structurally eliminated for $ p $ large enough using the exactly the same analysis as in Sections~\ref{subsection48} through~\ref{subsection68}. \hfill$\blacksquare$

\begin{remark}
 {We already showed that one can easily check the existence of certain absorbing sets in the Tanner \textit{et al.} construction \cite{tanner04} (Lemmas~\ref{lemmaqc2} and \ref{lemmaqc3}). Other popular quasi-cyclic LDPC codes, such as codes in \cite{FossorierIT04} and  \cite{Fan00} can also be viewed as being constructed as a selection of certain rows and columns of the SCB matrix.
{Codes in Section III-B2/B3 of \cite{FossorierIT04} use the SCB structure with the parity-check matrix $ H_{p, f(i,j)}^{r,c} $, where $f(i,j)=i \cdot j$, $0\leq i \leq r-1$, $0\leq j \leq c-1 $ or $ f(i,j)=(a^i-1)(b^j-1)$, $0 \leq i \leq r-1$, $0\leq j \leq c-1 $.}
 Array codes~\cite{Fan00} use the first $ r $ row groups of $H_{p, i \cdot j}^{p,p}$. The approach 
 {developed for the SCB construction} 
 can therefore be easily applied to codes in \cite{FossorierIT04} and \cite{Fan00}.}



\end{remark}


\subsection{Equivalence Classes for SR codes} \label{sec:equivmap5}
We now present three code equivalence conditions and use them to fully characterize the $p=7$, $r=5$ case.

Since the order of the elements in the RSF only permutes the rows of the matrix $H_{p, g_r(i) \cdot j}^{r,c}$ and does not change the code properties, we can assume that the RSF vector $ [a_1,a_2,...,a_r] $ is ordered in ascending order. 

Consider a \textit{difference matrix} $ D $ of the RSF, where $ D_{ij} \equiv a_j-a_i \mod p , 1 \leq i,j \leq r$. If $ \tilde{D}=D $ or $ \tilde{D} $ is  the reflection of $ D $  on its antidiagonal, we say that $ \tilde{D} $ and $ D $ are equivalent difference matrices.   The following lemma establishes some absorbing-set equivalence classes.
\begin{lemma}\label{lemmarowmap}
The following pairs of RSFs are equivalent in the sense that they have the same absorbing sets:
\begin{enumerate}
\item $ [a_1,a_2,...,a_r] \equiv  [a_1,a_2,...,a_r]+ \text{constant}  \mod   p  $
\item $ [a_1,a_2,...,a_r] \equiv  [a_1,a_2,...,a_r] \times \text{constant$\ne$$0$}  \mod  p  $
\item $ [a_1,a_2,...,a_r]  \equiv   [\tilde{a}_1,\tilde{a}_2,...,\tilde{a}_r] $ if they have equivalent difference matrices.
\end{enumerate}
\end{lemma}

\textit{Proof:}
the existence of any particular absorbing set hinges on whether the determinant of the associated CCM (or certain submatrices of the CCM) is zero. 
Since the determinant is only a function of the differences between the elements of the RSF, if two RSFs $ [a_1,a_2,...,a_r] $ and $  [\tilde{a}_1,\tilde{a}_2,...,\tilde{a}_r]$ share any of the three equivalent conditions, $ [a_1,a_2,...,a_r] $ leads to a zero determinant if and only if $  [\tilde{a}_1,\tilde{a}_2,...,\tilde{a}_r]$ leads to a zero determinant.
\hfill$\blacksquare$

\begin{remark}
{Since the null space of CCM also only depends on the difference of the column-group indices, analogous equivalence conditions can be established column-wise.}\hfill$\blacksquare$
\end{remark}

The following is a consequence of Lemma~\ref{lemmarowmap}.
\begin{corollary}\label{cor:equivRSF}
Any RSF is equivalent to an RSF $ [0,1,x,y,z] $, where $ x,y,z $ are positive integers.
\end{corollary}

\textit{Proof:}
With condition (1) in Lemma~\ref{lemmarowmap}, a mapping vector is equivalent to any mapping vector that differs by a constant in $ GF(p) $. Thus we can subtract the smallest value in the mapping vector and obtain a $ 0 $ in the mapping vector. Then, with condition (2) in Lemma~\ref{lemmarowmap} and the multiplicative property of $ GF(p) $, we can multiply the mapping vector by some value in $ GF(p) $ to make one of the non-zero elements equal to $ 1 $. Therefore any mapping vector is equivalent to a mapping vector that contains $ 0 $ and $ 1 $. 
\hfill$\blacksquare$

This result enables a reduced search of structured matrices with good error-floor properties. For example, a row-mapping vector $ [0,1,2,4,6] $ is equivalent to $ [1,2,3,5,7] $, and to $ [0,2,4,8,12] $.

{Here is an example for $ p=67 $. There are  $ 67 \choose 5  $ $ = 9657648 $ RSFs, but there are only 2192 equivalence classes. These equivalence classes can be represented by their class leaders which start with 0 and 1, according to Corollary\ref{cor:equivRSF}. Only 63 out of 2192 classes preclude (4,8), (5,9) and (6,8) absorbing sets. Table~\ref{tab_r5p67} lists all the 63 class leaders, where the common 0 and 1 are omitted to save space. To generate all the equivalent RSFs, we can couple the 0 and 1 to each 3-element set and form a 5-element RSF vector. Then by multiplying $ i $ and/or adding $ j $ in $ GF(p) $, where $ i,j = 1,2,...,p-1 $, we can reconstitute all the equivalent RSF vectors that preclude (4,8), (5,9) and (6,8) absorbing sets.}

\begin{table}
\caption{The 63 class leaders (excluding 0 and 1) for RSF equivalence classes precluding (4,8), (5,9) \& (6,8) absorbing sets for $ p=67 $.}
\label{tab_r5p67}
\centering
\begin{tabular}{|c|c|c|c|c|c|c|}
\hline
2,4,17 & 2,12,17 & 2,16,48 & 3,8,42 & 3,16,39 & 3,27,49 & 3,54,63\\
\hline
2,4,48 & 2,12,24 & 2,17,45 & 3,8,54 & 3,17,30 & 3,27,52 & 4,13,47\\
\hline
2,4,53 & 2,12,38 & 2,24,27 & 3,8,57 & 3,17,47 & 3,29,50 & 4,13,61\\
\hline
2,4,64 & 2,12,41 & 2,25,41 & 3,10,50 & 3,17,49 & 3,29,51 & 4,16,19\\
\hline
2,5,17 & 2,13,42 & 2,28,38 & 3,10,54 & 3,18,39 & 3,30,38 & 4,19,42\\
\hline
2,5,44 & 2,13,44 & 2,31,32 & 3,15,18 & 3,20,21 & 3,30,40 & 4,20,26\\
\hline
2,5,60 & 2,13,48 & 3,4,17 & 3,15,46 & 3,21,40 & 3,42,49 & 4,44,54\\
\hline
2,9,32 & 2,16,32 & 3,4,20 & 3,15,63 & 3,21,59 & 3,46,50 & 4,47,48\\
\hline
2,9,44 & 2,16,38 & 3,8,37 & 3,16,20 & 3,27,29 & 3,52,59 & 4,53,61\\
\hline
\end{tabular}
\end{table}

Because code design is an offline process, complexity is not the primary consideration.  However, after establishing the CCM equations for the absorbing sets of interest, identifying all good SR-SCB codes for r=5 and $p=67$  shown in Table \ref{tab_r5p67} is completed in about five minutes using a simple Matlab implementation.


\section{Theoretical results for $ r =4 $}\label{analysis4}

This section provides an example with $r=4$ (four row groups) that shows how to design an SCB code with a specified circulant matrix that eliminates the dominant absorbing sets by selecting rows and columns from the SCB mother matrix to force the CCMs associated with the dominant absorbing sets to have zero-dimension nullspaces.

In addition to the EAB codes and SR-SCB codes introduced in Section~\ref{analysis5}, removing a few column groups from an SR-SCB code provides further improvement.  Hence, shortened SR (SSR) SCB codes form our third class of SCB codes.  The parity-check matrix for these codes is $H_{p,f}^{r,c}$, with $f(i,j)=g_r(i) \cdot g_c(j) $ where $g_c(j)$ is called the column-selection function (CSF).  Note that for a $ p \times p $ circulant matrix, EAB and SR-SCB codes have $p$ column groups ($p^2$ binary columns), but SSR-SCB codes have fewer {column} groups since $g_c(j)$ selects a subset of the possible column groups.

Section~\ref{sec:nosmaller} identifies the $(6,4)$ absorbing sets as dominant for EAB codes with $r = 4$.  Section~\ref{sec:abs64} analyzes the three possible $(6,4)$ absorbing set configurations and shows how carefully selecting four row groups from the SCB mother matrix can eliminate two of the three possible configurations. Section~\ref{sec:abs64col} provides an efficient provable algorithm to eliminate all $(6,4)$ absorbing sets by combining the row selection of Section~\ref{sec:abs64} with column selection in which some column groups of the SCB mother matrix are removed.

\subsection{Identifying the dominant absorbing sets} \label{sec:nosmaller}
From the previous results in~\cite{dolecekIT10}, $(6,4)$ absorbing sets are the smallest possible structure for EAB codes with $r=4$ for $p>19$.  Hardware simulations{~\cite{zhangTCOM}} also demonstrate that $(6,4)$ absorbing sets are the dominant cause of the error floor for example $r=4$ EAB codes.

Based on these results, a key goal will be to design an $r=4$ SCB code that avoids all $(6,4)$ absorbing sets.  The lemma below establishes that the new code design approach does not introduce other smaller absorbing sets that were avoided by the codes discussed above.

\begin{lemma} \label{lemma_modnosmall}
In the Tanner graph corresponding to $H_{p,g_r(i) \cdot j}^{4,p}$, there is no absorbing set smaller than $ (6,4) $ for $p$ large enough with a careful choice of the row-selection function $g_r(i)$.
\end{lemma}

\textit{Proof:}
The smallest possible absorbing sets for an  SCB code specified by $H_{p,g_r(i) \cdot j}^{4,p}$ are $ (4,4)$, $(5,2) $, $(5,4) $ and $(6,2)$ absorbing sets (cf.~\cite{dolecekIT10}). The $ (4,4)$, $(5,4) $ and $(6,2)$ absorbing sets for $r=4$ are sub-graphs of the $(4,8)$, $(5,9)$ and $(6,8)$ absorbing sets, respectively, for $r=5$.  From the analysis of absorbing sets for $r=5$  {in the previous section (see also~\cite{DOLECEKISTC10})}, a careful choice of the $r=5$ row-selection function (RSF) $\tilde{a}(i)$ eliminates the $(4,8)$, $(5,9)$ and $(6,8)$ absorbing sets for $p$ large enough.  Taking any 4-element subset $g_r(i)$ of such an $\tilde{a}(i)$, for example the RSF $g_r(i)$ where $[g_r(0), g_r(1), g_r(2), g_r(3)]=[0,1,3,4]$, {as a subset of $ [0,1,3,4,5] $ that can eliminate the $(4,8)$, $(5,9)$ and $(6,8)$ absorbing sets for $ p>271 $,} 
avoids the $ (4,4)$, $(5,4) $ and $(6,2)$ absorbing sets for $r=4$. Any RSF that avoids the $(4,4)$ absorbing set also avoids the $(5,2)$ absorbing set~\cite{dolecekIT10}. Thus the resulting $H_{p, g_r(i) \cdot j}^{4,p}$  avoids $ (4,4) $, {$ (5,2) $, $ (5,4)$}  and $ (6,2) $ absorbing sets. \hfill$\blacksquare$
\begin{remark}
Since the SSR code is a shortened version of the SR code, obtained by removing certain variable nodes, no smaller absorbing set will be introduced in the SSR code.\hfill$\blacksquare$
\end{remark}

In Section~\ref{sec:abs64}, we show that SR codes always have $ (6,4) $ absorbing sets, irrespective of the choice of $g_r(i)$. Avoidance of all such configurations using shortening is the subject of Section~\ref{sec:abs64col}.
\subsection{$(6,4)$ absorbing sets in SR-SCB codes}\label{sec:abs64}
Three distinct configurations of $(6,4)$ absorbing sets are possible for $r=4$ SCB codes.  This section shows which of these configurations are possible in EAB and SR-SCB codes.  The first configuration exists in the EAB code and in every possible SR-SCB code.  The second configuration exists in the EAB code but can be avoided by a proper choice of the RSF for the SR-SCB code.  The third configuration does not exist in either the EAB code or the SR codes.
\subsubsection{The first (6,4) configuration}
Fig.~\ref{fig64a} shows the first configuration of a $(6,4)$ absorbing set in an $r=4$ SCB code.  The following lemma establishes that the EAB code and all SR codes have this configuration.
\begin{figure}
\center
\includegraphics[width=0.31\textwidth]{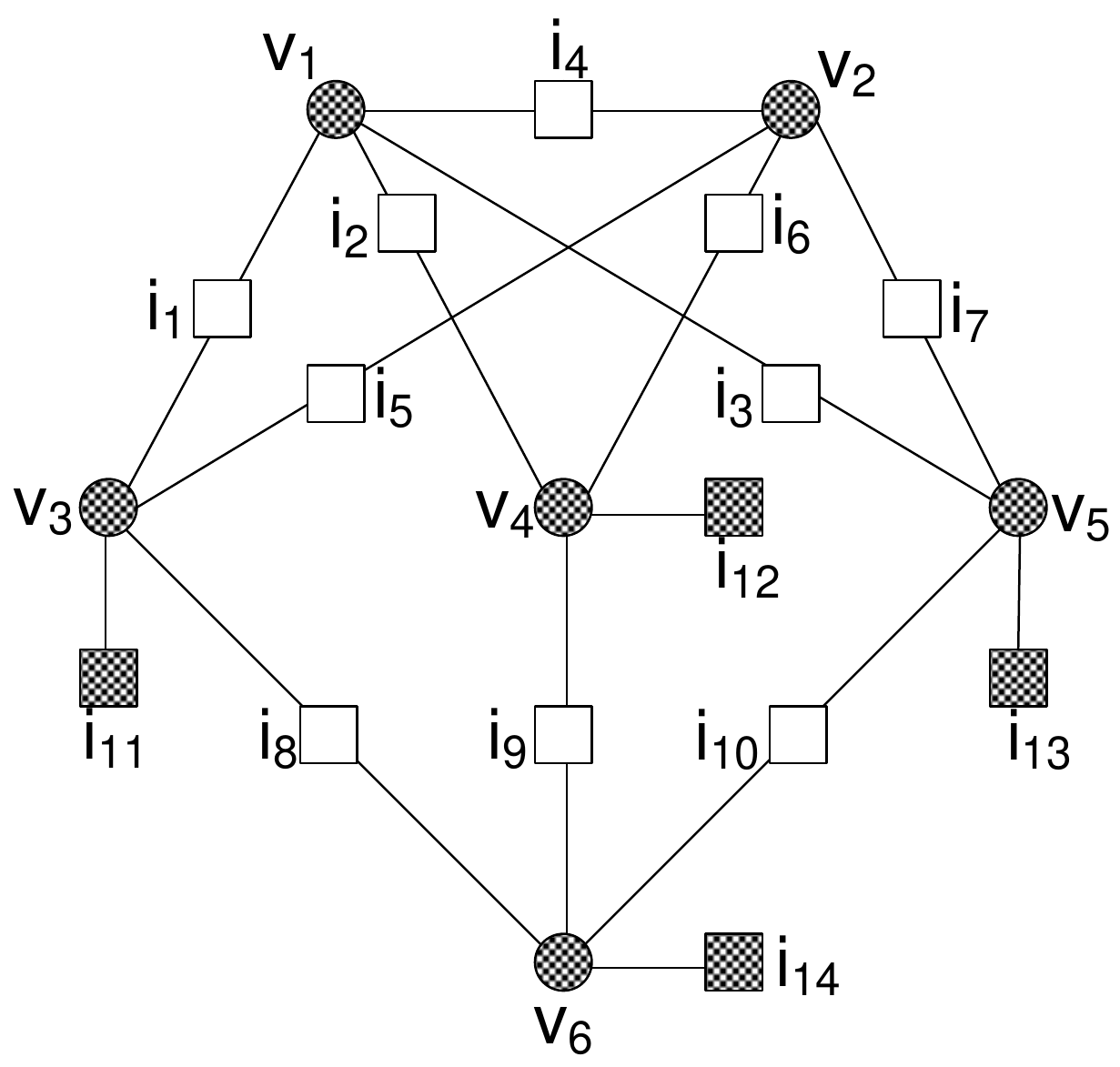}
\center
\caption{Depiction of the first (6, 4) absorbing set configuration.}
\label{fig64a}
\vspace{-0.2in}
\end{figure}

\begin{lemma}\label{lemma64a}
In the Tanner graph corresponding to the EAB code and all SR codes with $H_{p, f(i,j)}^{4,p}$ there are $(6,4)$ absorbing sets for any $p$ with the configuration shown in Fig.~\ref{fig64a}.
\end{lemma}

\textit{Proof:}
Using the technique of Section~\ref{mtrxrep} we construct the CCM for this configuration.  The bcs for Fig.~\ref{fig64a} has dimension $5$.  Using the five linearly independent cycles: $v_1$$-$$v_2$$-$$v_3$$-$$v_1$, $v_1$$-$$v_2$$-$$v_4$$-$$v_1$, $v_1$$-$$v_2$$-$$v_5$$-$$v_1$, $v_1$$-$$v_3$$-$$v_6$$-$$v_4$$-$$v_1$, $v_1$$-$$v_5$$-$$v_6$$-$$v_4$$-$$v_1$, we construct the following CCM:

\small
\begin{equation}\label{eq64am}
\mathbf{M} =\begin{bmatrix}
i_4-i_5 & i_5-i_1 & 0 & 0 & 0\\
i_4-i_6 & 0 & i_6-i_2 & 0 & 0 \\
i_4-i_7 & 0 & 0 & i_7-i_3 & 0 \\
0 & i_1-i_{8} & i_9-i_2 & 0 & i_8-i_9 \\
0 & 0 & i_2-i_9 & i_{10}-i_3 & i_9-i_{10}
\end{bmatrix}.
\end{equation}
\normalsize 

To avoid this absorbing set we need  to force the determinant of the CCM  to be nonzero.  Note that $\det(\mathbf{M})$ is computed as
\small
\begin{align*}
&\mathbf{M}_{11}\mathbf{M}_{23}\mathbf{M}_{34}\mathbf{M}_{42}\mathbf{M}_{55}\\
&-\mathbf{M}_{12}\bigl(\mathbf{M}_{23}\mathbf{M}_{31}\mathbf{M}_{45}\mathbf{M}_{54}-\mathbf{M}_{21}\mathbf{M}_{34}(\mathbf{M}_{43}\mathbf{M}_{55}-\mathbf{M}_{45}\mathbf{M}_{53})\bigr),
\end{align*}
\normalsize
where $\mathbf{M}_{ij}$ denotes the $(i,j)$ entry in $\mathbf{M}$.

From the proof of Lemma 8 in \cite{dolecekIT10}, there are only two non-isomorphic row-group labelings for the check nodes of Fig.~\ref{fig64a}. It is sufficient to consider only $(i_1, i_2, i_3, i_4, i_5, i_6, i_7, i_8, i_9, i_{10}) = (x, y, z, w, y, z, x, z, x, y)$ or $(x, y, z, w, y, z, x, z, w, y)$ 
with $x,y,z,$ and $w$ taking distinct values in the range of RSF. 
The first labeling yields
\begin{equation}\label{eq64am1}
\mathbf{M} =\left[
\begin{array}{ccccc}
w-y & y-x & 0 & 0 & 0\\
w-z & 0 & z-y & 0 & 0 \\
w-x & 0 & 0 & x-z & 0 \\
0 & x-z & x-y & 0 & z-x \\
0 & 0 & y-x & y-z & x-y
\end{array}
\right],
\end{equation}
\normalsize
and $ \det(\mathbf{M}) = 0 $ regardless of the specific values of $w,x,y,z$ taken as mutually distinct integers$\mod p$.

Thus, there exists a non-zero solution to $ \mathbf{M} \cdot \mathbf{u} \equiv 0 \mod p $. One solution to this equation is
\small
\begin{equation}\label{eq64asol}
\begin{bmatrix}
u_2\\
u_3\\
u_4\\
u_5\\
u_6
\end{bmatrix}
\equiv
\begin{bmatrix}
(x-y)(z-y)(x-z)\\
(w-y)(z-y)(x-z)\\
(w-z)(y-x)(x-z)\\
(w-x)(y-x)(z-y)\\
(w-y)(x-z)(z-y)+(y-x)(w-z)(x-y)
\end{bmatrix}
\end{equation}
\normalsize

For this absorbing set, the Check Consistency 
condition requires $ u_2 \neq 0 $, $ u_3 \neq 0 $, $ u_4 \neq 0 $, $ u_5 \neq 0 $, $ u_2 \neq u_3 $, $ u_2 \neq u_4 $, $ u_2 \neq u_5 $, $ u_3 \neq u_6 $, $ u_4 \neq u_6 $, and $ u_5 \neq u_6 $.  These requirements as well as the Bit Consistency inequalities are met since $x,y,z,w$ are mutually distinct.

The solution {in \eqref{eq64asol}} satisfies the Bit, Check and Cycle Consistency constraints.  This is a sufficient condition for the existence of a $(6,4)$ absorbing sets with the configuration of Fig.~\ref{fig64a}.  Any four distinct values between $0$ and $p-1$ for $\{x,y,z,w\} $ identify a labeling of this first type that identifies an absorbing set in the EAB code and every SR {SCB} code.

Consider $(i_1, i_2,\ldots, i_{10}) = (x, y, z,w, y, z, x, z,w, y)$, the second labeling.  In this case
$\det(\mathbf{M})\ne 0$ for {$ \{ x,y,z,w \} = \{ 0,1,2,3 \} $}. {Thus $\det(\mathbf{M})\not\equiv 0 \mod p$ for $ p $ large enough}, and there is no such $(6,4)$ configuration with this labeling in the EAB code for $ p $ large enough. 
The EAB code is one possible SR code.  Other careful choices of the RSF, produce other SR codes that likewise do not have this $(6,4) $ configuration. \hfill$\blacksquare$

\subsubsection{Two additional configurations}
Fig.~\ref{fig64b} shows the second {possible} configuration of a $(6,4)$ absorbing set in an $r=4$ SCB code.  Similar arguments to those above establish that the EAB code has this configuration but well-designed SR codes avoid it.  One such example is the RSF $ [0,1,3,4] $ which avoids this configuration for $ p>31 $.
\begin{figure}
\center
\includegraphics[width=0.32\textwidth]{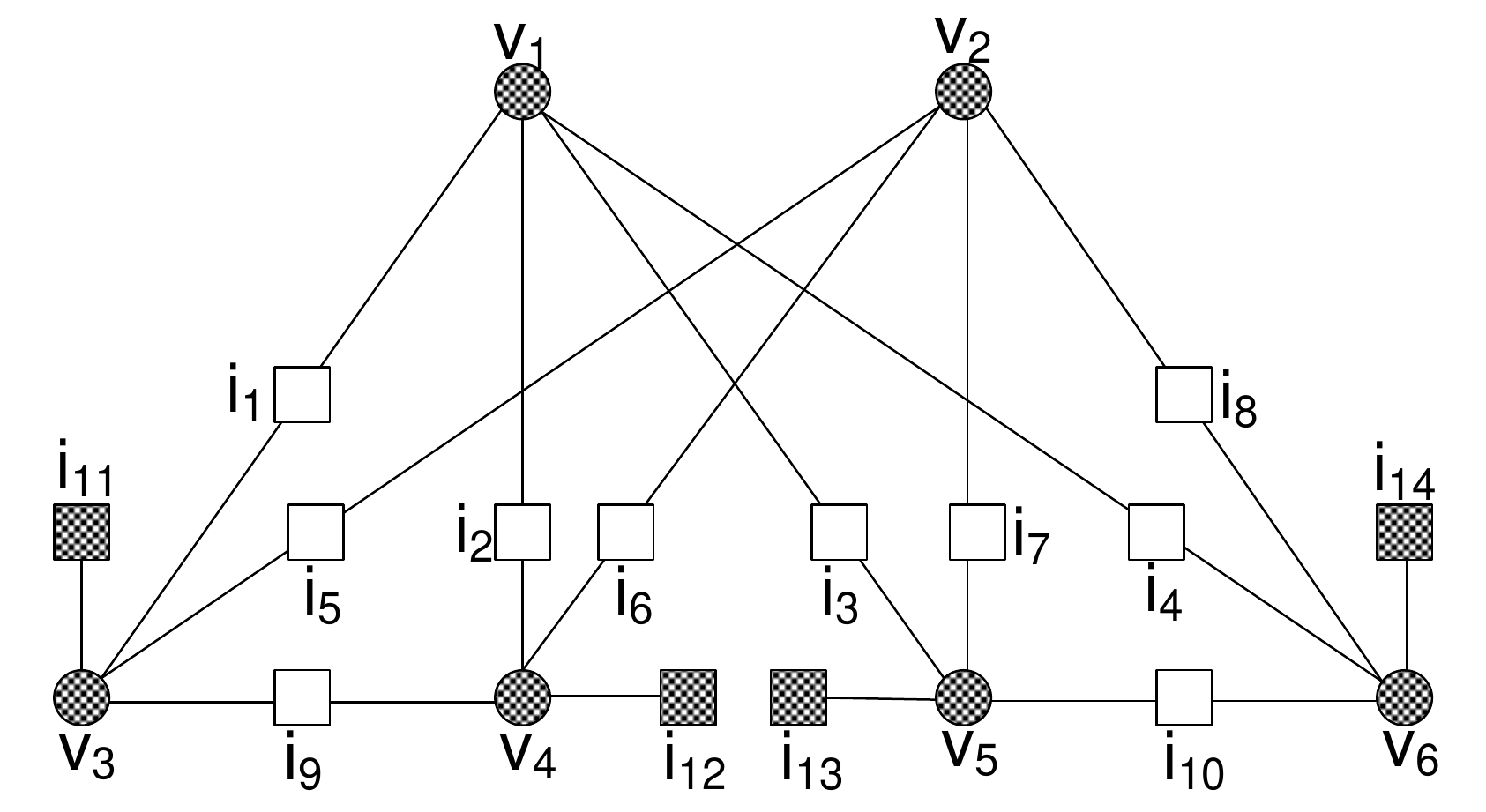}
\center
\caption{Depiction of the second candidate (6, 4) absorbing set.}
\label{fig64b}
\vspace{-0.2in}
\end{figure}

Fig.~\ref{fig64c} shows the third configuration of a $(6,4)$ absorbing set in an $r=4$ SCB code.  Similar arguments to those above establish that neither the EAB nor well-designed SR codes have this configuration for $p$ sufficiently large.  In fact, $ p>13 $ is sufficient for the absence of this configuration in the EAB codes.  SR-SCB codes can avoid the configuration in Fig.~\ref{fig64c} with a careful choice of the row mapping. One such example is the EAB RSF itself.   Another is to choose the RSF  $ [0,1,3,4] $, which can avoid this configuration for $ p>19 $.
\begin{figure}
\center
\includegraphics[width=0.35\textwidth]{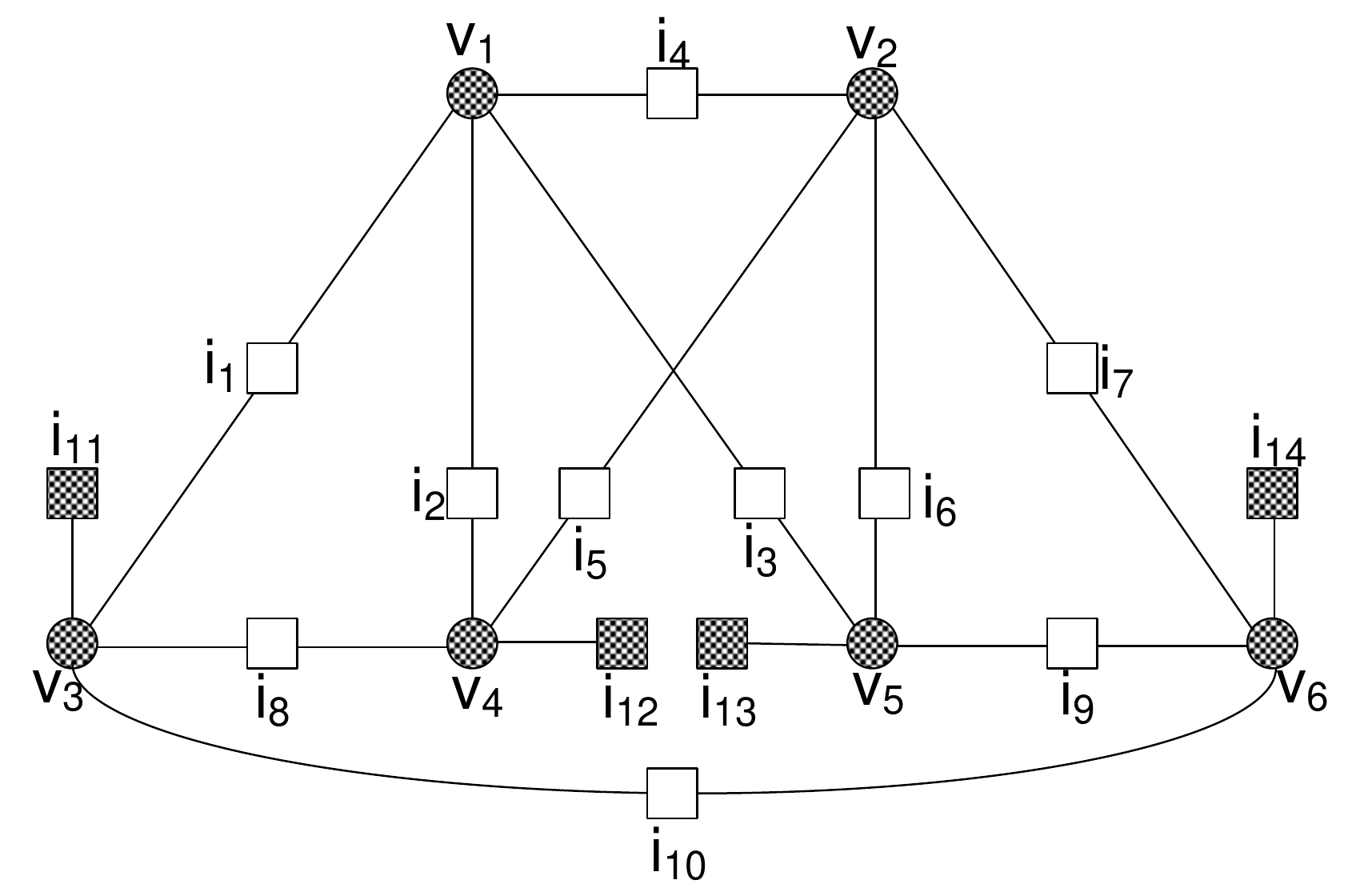}
\center
\caption{Depiction of the third candidate (6, 4) absorbing set.}
\label{fig64c}
\vspace{-0.2in}
\end{figure}

\subsection{Eliminating $ (6,4) $ absorbing sets with shortening}\label{sec:abs64col}
For a sufficiently large $p$, well-designed SR codes avoid the $ (6,4) $ absorbing set configurations in Figs.~\ref{fig64b} and \ref{fig64c}. However, {as shown in Lemma~\ref{lemma64a}}, SR codes cannot eliminate the $ (6,4) $ absorbing set configuration in Fig.~\ref{fig64a}. We now consider shortened SR (SSR) codes that retain only certain column groups from the SCB mother matrix (reducing the rate).  A well-chosen column selection $ g_c(j) $ allows the Tanner graph corresponding to $H_{p, g_r(i) \cdot g_c(j)}^{4,c}$ to avoid all $ (6,4) $ absorbing sets.

We begin with an SR code using well-selected RSF $ g_r(i) $, for instance $ [0,1,3,4] $, that already avoids the $ (6,4) $ absorbing set configurations in Figs.~\ref{fig64b} and \ref{fig64c} for $ p $ large enough. We then choose a CSF $ g_c(j) $ to also avoid the $ (6,4) $ absorbing set configurations in Fig.~\ref{fig64a}. Choosing a column selection $ g_c(j) $ {reduces} to choosing a submatrix of $H_{p, g_r(i) \cdot j}^{4,p}$ by eliminating certain variable nodes.  This operation {cannot} introduce smaller absorbing sets.

One solution to $ \mathbf{M} \cdot u \equiv 0 \mod p $ is equation~\eqref{eq64asol}. The rank of $\mathbf{M}$ in~\eqref{eq64am1} is 4, and therefore this single solution forms a basis of the null space. Multiplying $u$ by a constant $ c $, for $1 \leq c \leq p-1 $, also results in a solution. These $ p-1 $ solutions cover all of the feasible solutions described by the null space. For fixed $ u_1,\cdots,u_5 $, we can choose $ j_1 $ from $ 0,1,\cdots,p-1 $ and obtain $ j_2,\cdots,j_6 $. Thus, there are $ p(p-1) $ ways to find $ j_1 $ to $ j_6 $ for a fixed $ \{ x,y,z,w \} $. Since there are $ 4! $ ways to assign check node labels based on the set $ \{ x,y,z,w \} $ for a fixed row mapping, there are at most $ 24p(p-1) $ possible vectors $ [j_1,j_2,\cdots,j_6]  $ that can form the configuration in Fig.~\ref{fig64a}.
These vectors form the set $\tilde{V}$ of vectors, which {completely characterizes the instances of this} absorbing set configuration.

If a CSF is applied, each {variable} node  group label $ j $ is in a set $ J $ where $ J \subset \{ 0,1,...,p-1 \} $ and we can only choose $ [j_1,j_2...,j_6]  $ such that $ j_m \in J, m=1,...,6$. There are $ {|J|\choose 6} $ possible $ [j_1,j_2...,j_6]  $ {vectors} and they form a set of vectors $ {V} $. If $ V \cap \tilde{V} = \emptyset $, the new code does not have the $ (6,4) $ configuration in Fig.~\ref{fig64a}.
We can find such a CSF with the greedy column-cutting procedure described in Algorithm~\ref{columncut} or the column-adding procedure described in Algorithm~\ref{columnadd}.

\begin{algorithm}
\caption{ Greedy column-cutting algorithm.} \label{columncut}
\begin{algorithmic}[1]
\STATE \% Initialization: $C$ begins with all columns.
\STATE {$ C =\{ 0,1,...,p-1 \} $. }
\STATE {Collect all the vectors $ [j_1,j_2...,j_6]  $, that form the configuration in Fig.~\ref{fig64a} with $ j_n \in C, n=1,...,6 $  and they form a set $ W{=\tilde{V}}$. }
\STATE \% Proceed one column at a time removing columns that \% preclude vectors in $\tilde{V}$ until no vector in $\tilde{V}$ is possible.
\WHILE {$ |W|>0 $}
\STATE{Find the most frequent $ j $ in $ W $, say $ j_m $. }
\STATE{Replace $ C$ by $ C \setminus j_m $.}
\STATE{Remove every $ [j_1,j_2...,j_6]  $ that involves $ j_m $ from $ W $.}
\ENDWHILE
\STATE{ \% $C$ contains the column groups of the designed code.}
\end{algorithmic}
\end{algorithm}

\begin{algorithm}
\caption{ Column-adding algorithm.} \label{columnadd}
\begin{algorithmic}[1]
\STATE \% Initialization: Select an initial six columns for  $C$ that \% do not form a vector in $ \tilde{V}$.
\STATE {$ J =\{ 0,1,...,p-1 \} $. }
\STATE {Collect all the ordered vectors $ [j_1,j_2...,j_6]  $, that form the configuration in Fig.~\ref{fig64a} with $ j_n \in J, n=1,...,6 $, into the set $ \tilde{V}$. }
\STATE {Choose a distinct $6$-element set $ C $, $ C \subseteq J $. }
\WHILE{ for some ordering $[ \hat{j}_1,\hat{j}_2...,\hat{j}_6] $ of the elements of $C $, $ [\hat{j}_1,\hat{j}_2...,\hat{j}_6] \in \tilde{V} $ }
\STATE {Choose another distinct $6$-element set $ C $ randomly such that {$ C \subseteq J $}.}
\ENDWHILE
\STATE \% Proceed one column group at a time, adding columns \% to $C$ that do not introduce the absorbing set.
\STATE {$ J= \{ 0,1,...,p-1 \}  \setminus C $}
\WHILE {$ |J|>0   $}
\STATE{Select a $ j_m $ randomly from $ J $. }
\IF  {every $ [\hat{j}_1,\hat{j}_2...,\hat{j}_6] \not\in {\tilde{V}} $ for every $ \hat{j}_1,\hat{j}_2...,\hat{j}_6 \in \{ C \cup j_m \} $}
\STATE{$ C= C \cup j_m $}
\ENDIF
\STATE{$ J=J \setminus j_m $}
\ENDWHILE
\STATE{ \% $C$ contains the column groups of the designed code.}
\end{algorithmic}
\end{algorithm}

\begin{remark}
A similar technique could be applied to increase the girth~\cite{MilenkovicIT06} instead of eliminating the smallest absorbing set. However, simply increasing the girth would not guarantee a better performance, see e.g.,~\cite{NguyenIT11}.
\hfill$\blacksquare$
\end{remark}
\subsection{SSR codes with the Tanner \textit{et al.} construction} \label{sec:tannersr4}
This section shows that for $ r=4 $, the row-selection function $ g_r(i) $ of the Tanner \textit{et al.} construction \cite{tanner04} will always introduce $(4,4)$ absorbing sets for the case set forth in the following lemma. {For $a$ an element of $GF(p)$ let $o(a)$ denote its multiplicative order in 	$GF(p)$.}

\begin{lemma}\label{lemmatanner4}
In the Tanner graph corresponding to the quasi-cyclic LDPC code with the parity check matrix $H_{p, g_r(i) \cdot g_c(j)}^{4,p-1}$, where $ g_r(i)=a^i, o(a)=4, g_c(j)=b^j, o(b)=p-1 $, $ (4,4) $ absorbing sets as shown in Fig.~\ref{fig44} always exist.	
\end{lemma}

\begin{figure}
\center
\includegraphics[width=0.31\textwidth]{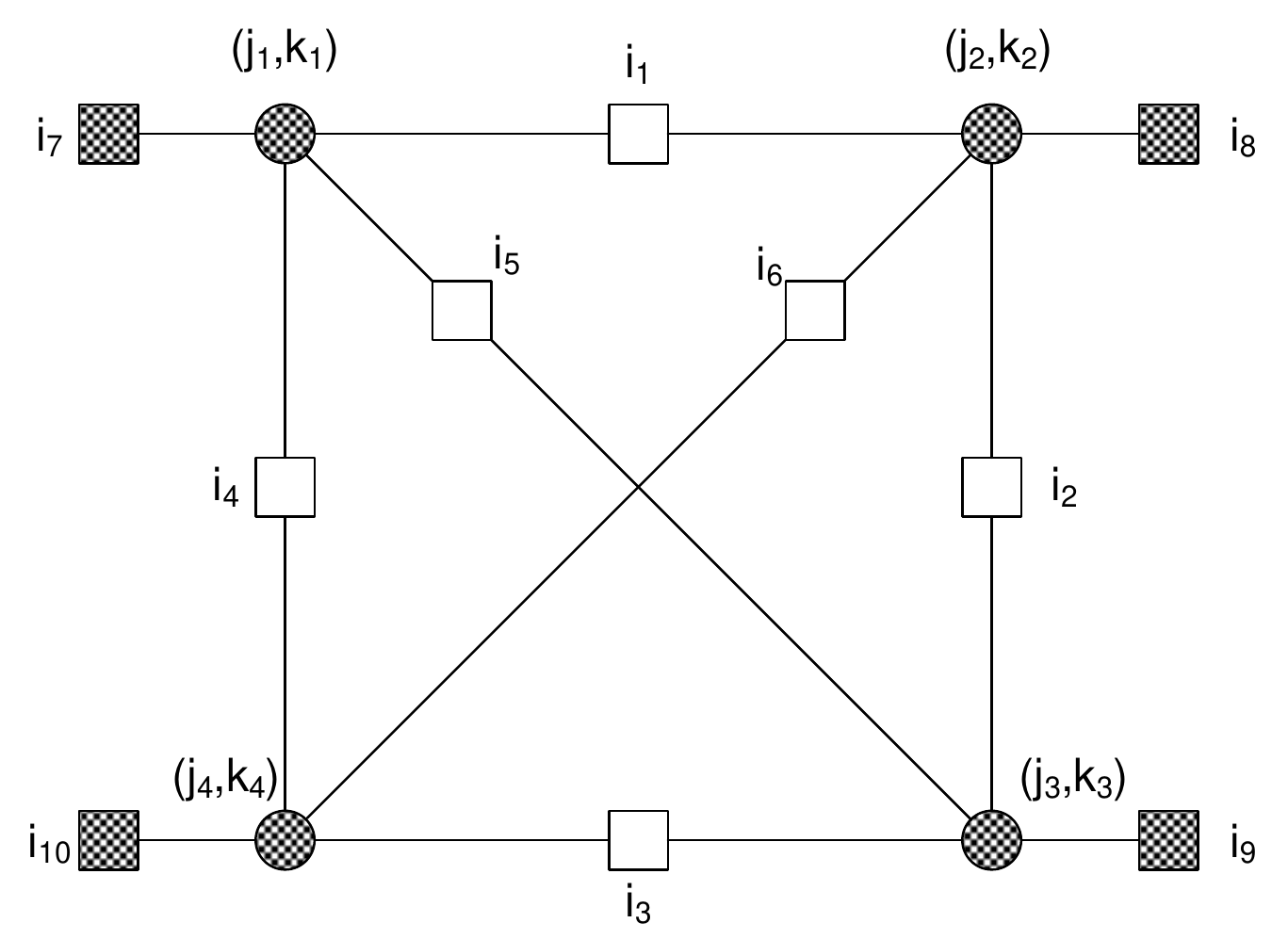}
\caption{Depiction of the $(4,4)$ absorbing set configuration.}
\label{fig44}
\vspace{-0.2in}
\end{figure}

\textit{Proof:}
The proof uses the same techniques as that of Corollary~\ref{lemma48}. Details are omitted for brevity.\hfill$\blacksquare$

\begin{remark}
Tanner \textit{et al.} constructions with $o(b)<p-1 $ and $ r=4 $ sometimes can avoid all $ (4,4) $ absorbing sets by either increasing the girth or by making the intersection between null space of $ \mathbf{M} $ and variable node space to be empty. However, $ (4,4) $ absorbing sets may still exist in some constructions with $o(b)<p-1 $ and $ r=4 $.  For example, with $ p=67, g_r(i)=11^i, g_c(j)=5^j, o(11)=4, o(5)=30 $, the resulting code has $ (4,4) $ absorbing sets.
\end{remark}

%

\section{Discussion for $ r=3 $} \label{section3}
}

{We briefly remark on the case where the column weight is $3$. First, we note that for the EAB codes with $ r=3 $, the $ (3,3) $ and $ (4,2) $ absorbing sets in Fig.~\ref{fig33} are the smallest ones, \cite{dolecekIT10}. { It is easy to show that the same absorbing sets are also the smallest possible for SR-SCB codes (of girth 6).
The CCM expression describing a $(3,3)$ absorbing set now takes on a particularly simple form:
\[ i_1 (j_3 - j_1) + i_3(j_2-j_3) + i_2(j_1-j_2) \equiv 0 \mod p,\]
which always has the solution $ j_1 = i_3 $, $ j_2 = i_1 $ and $ j_3 = i_2 $. %
{The number of $(3,3)$ absorbing sets scales as $\Theta(p^3)$.} This expression precisely describes a length-6 cycle and thus establishes a one-to-one relationship between $(3,3)$ absorbing sets and length-6 cycles. It is therefore necessary to increase the girth for the elimination of $(3,3)$ absorbing sets. The resulting shortened codes would have much lower rates and as such are not of interest in this paper. 
}
}

{Similarly, we can apply the CCM approach to the $ (4,2) $ absorbing sets and %
{easily prove their existence in any EAB and SR-SCB codes. 
The CCM equation now becomes
\begin{equation}\label{eq42}
\mathbf{M} = \begin{bmatrix}
i_1-i_2 & i_2-i_5 & 0\\
0 & i_5-i_3 & i_3-i_4
\end{bmatrix} \, ,
\end{equation}
and always has a nontrivial nullspace. Thus this absorbing set cannot be avoided in the SR-SCB codes. The number of $(4,2)$ absorbing sets scales as $\Theta(p^3)$. 
}}

\begin{figure}
\center
\includegraphics[width=0.18\textwidth]{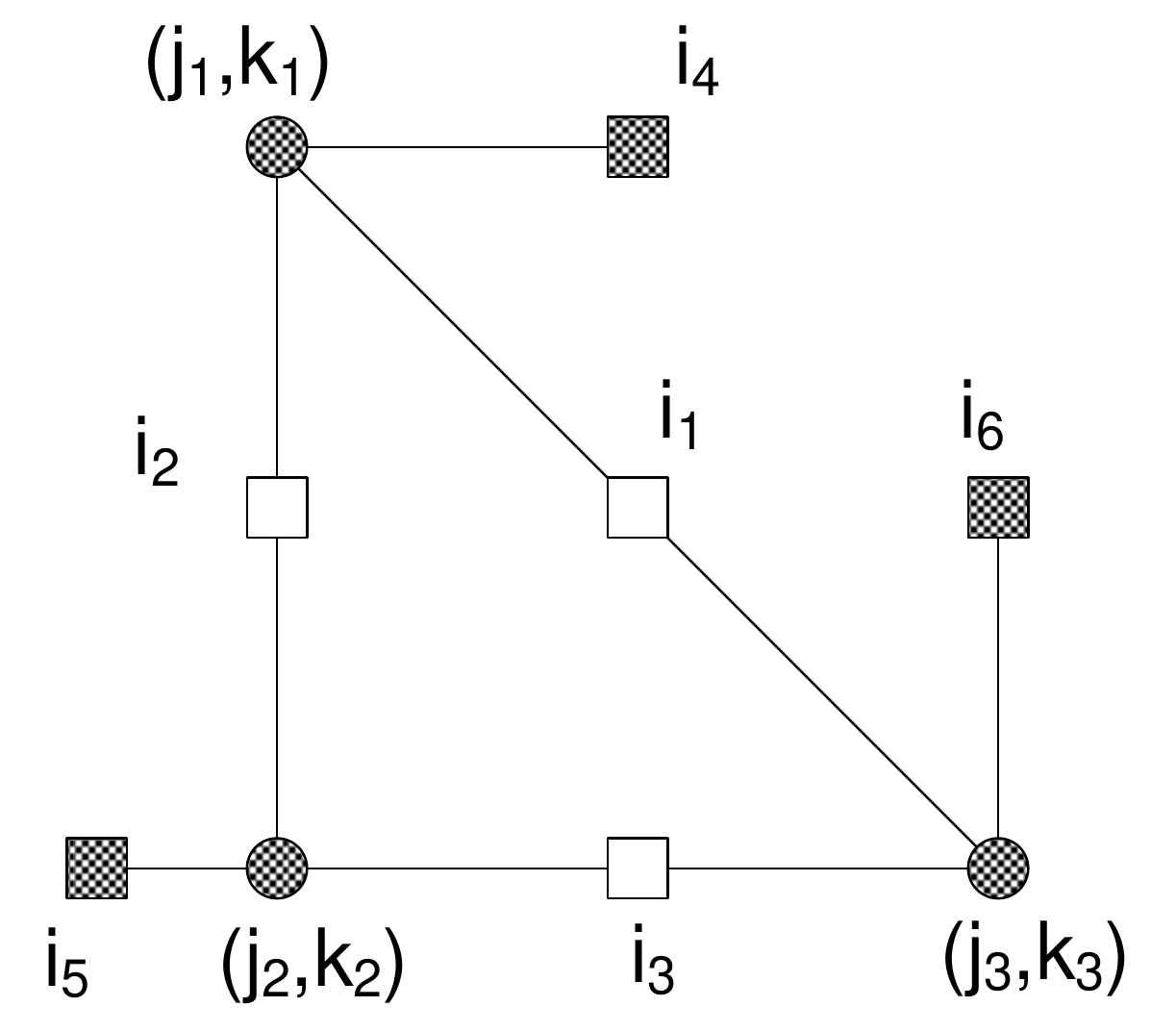}
\includegraphics[width=0.2\textwidth]{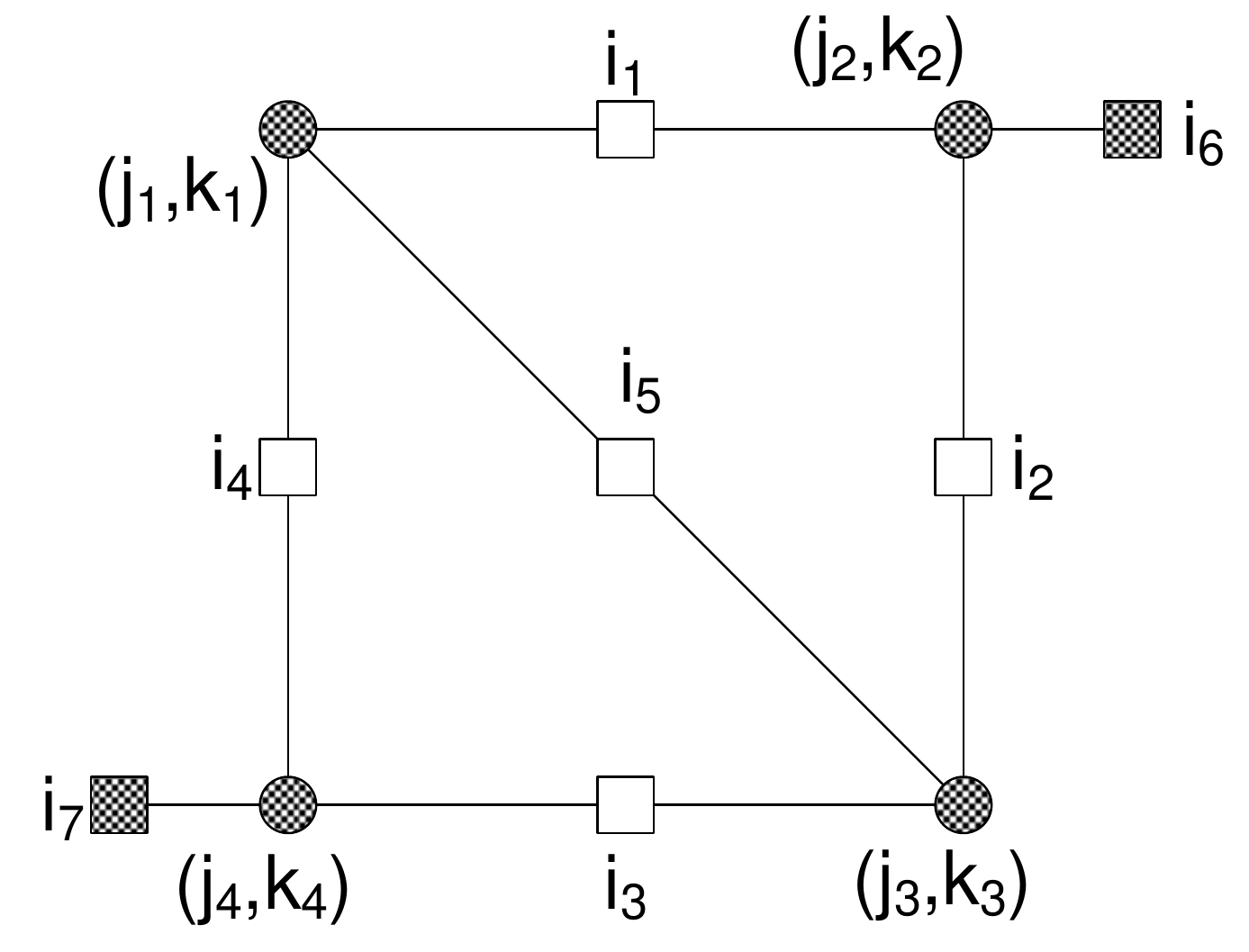}
\caption{Depiction of the $(3,3)$ and $(4,2)$ absorbing set configurations.}
\label{fig33}
\end{figure}



\section{Simulation Results}\label{results}

{In this section we experimentally demonstrate the performance improvement achieved by the CCM design approach for limited-precision decoders using both the sum-product algorithm (SPA) ~\cite{zhangTCOM} and the soft xor algorithm (SXOR)~\cite{Mansour}.  The  limited-precision decoders in the simulations use 200 iterations and a $ Q4.2 $  quantization, 4 bits to the left of the radix point to represent integer values and 2 bits to the right of the radix point to represent fractional values. We simulate SPA and SXOR for a variety ofdifferent codes.}  We also identify the absorbing sets responsible for each error in the error floor and report an error profile for each simulation.  As expected, when an SR-SCB code or SSR-SCB code is designed to preclude certain absorbing sets, they do not appear in the error profile obtained by simulation and the error floor improves.

\subsection{SCB codes for $r=5$}\label{resultsr5}


\begin{table}
\caption{{Software} error profiles for three $(2209,1978)$ codes: the EAB code, an SR-SCB code and the Huang \textit{et al.} construction. The total number of collected errors is denoted n.e. The number of runs is denoted n.r.}
\label{tab_sw_r5p47}
\centering
\begin{tabular}{|p{0.23in}|p{0.24in}|p{0.15in}|p{0.15in}|p{0.15in}|p{0.15in}|p{0.18in}|p{0.15in}|p{0.18in}|p{0.15in}|}
\hline
\multicolumn{10}{|c|}{$p=47$ $(2209,1978)$  EAB code} \\ \hline
SNR & n.r. & n.e. &(4,8)&(5,9)&(6,8)&(6,10)&(7,9)&(7,11)&(8,6)\\
\hline
5.2dB& 3.3E8 & 378 & 17 & 0 & 174 & 0 & 68 & 0 & 0 \\
5.4dB& 3.8E8 & 230 & 5 &  0 & 113 & 0 & 72 & 0 & 0 \\
\hline
\hline
\end{tabular}
\begin{tabular}{|p{0.23in}|p{0.24in}|p{0.15in}|p{0.15in}|p{0.15in}|p{0.15in}|p{0.18in}|p{0.15in}|p{0.18in}|p{0.15in}|}
\hline
\multicolumn{10}{|c|}{$p=47$ $(2209,1978)$  Huang \textit{et al.}  code} \\ \hline
SNR & n.r. & n.e. &(4,8)&(5,9)&(6,8)&(6,10)&(7,9)&(7,11)&(8,6)\\
\hline
5.2dB& 3.0E8 & 350 & 51 & 25 & 106 & 6 & 70 & 11 & 3 \\
5.4dB& 5.2E8 & 301 & 39 & 29 & 115 & 2 & 61 & 15 & 0 \\
\hline
\hline
\end{tabular}
\begin{tabular}{|p{0.23in}|p{0.24in}|p{0.15in}|p{0.15in}|p{0.15in}|p{0.15in}|p{0.18in}|p{0.15in}|p{0.18in}|p{0.15in}|}
\hline
\multicolumn{10}{|c|}{$p=47$ $(2209,1978)$ SR-SCB code RSF = [0,1,3,8,19]} \\ \hline
SNR & n.r. & n.e. &(4,8)&(5,9)&(6,8)&(6,10)&(7,9)&(7,11)&(8,6)\\
\hline
5.2dB& 8.1E8 & 241 & 0 & 0 & 4 & 69 & 9 & 13 & 29 \\
5.4dB& 2.1E9 & 165 & 0 & 0 & 0 & 56 & 2 & 15 & 8\\
\hline
\end{tabular}
\end{table}

\begin{figure}
\center
\includegraphics[width=0.37\textwidth]{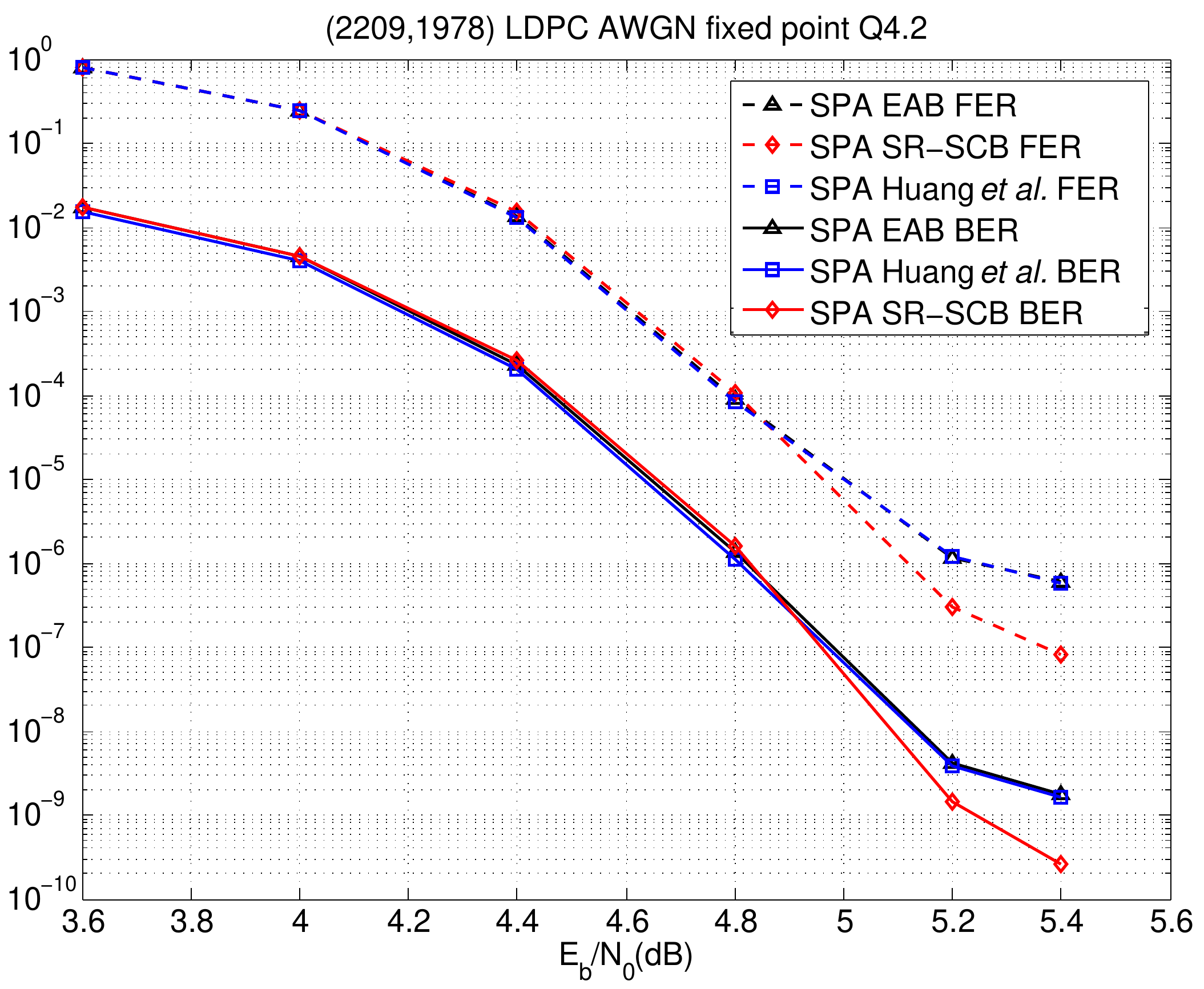}
\caption{Comparison of the $ (2209,1978), $ $r=5$, rate-0.895 EAB, SR-SCB LDPC codes and the Huang \textit{et al.}  construction \cite{HuangIT2012} using SPA decoder. }\label{fig:perf1}
\vspace{-0.15in}
\end{figure}

{Consistent with the results presented at the end of Section \ref{subsection48}, Table \ref{tab_sw_r5p47} and Fig.~\ref{fig:perf1} show the {software, limited-precision} SPA performance of two $(2209, 1978)$ SCB codes:  the EAB  code and an SR-SCB code. Both codes have  the following parameters: check node degree  = $47$, and bit node degree = $5$. The EAB code uses $g_r(i) = i$, and SR-SCB code uses the RSF $[ 0,1,3,8,19]$}. 
{Another $(2209, 1978)$ quasi-cyclic code using the Huang \textit{et al.}  construction \cite{HuangIT2012} with the same code parameters as our SCB codes is constructed and simulated with the same decoder.} 
{The improved error-floor slope under limited-precision SPA decoding of the SR-SCB code as compared to the EAB code and the Huang \textit{et al.} construction is consistent with the removal of all (4,8) absorbing sets.}

We also simulated all three codes using a {limited-precision} SXOR decoder in software. SXOR results are omitted in the interest of space.}  Irrespective of the decoding algorithm (SPA or SXOR), the SR-SCB code that provably eliminates all $(4,8)$ absorbing sets performs the best among the three codes at every SNR point measured. Both EAB and the Huang \textit{et al.}  code contain $(4,8)$ absorbing sets. The Huang \textit{et al.}  code offers visible performance advantage over the EAB code under 
 SXOR, in part due to the suppression of the effect that the $(4,8)$ absorbing sets of this code have under SXOR decoding.
 Under SPA (and as shown in Figure~\ref{fig:perf1}) this performance advantage vanishes.
 


In addition to software simulations, we also performed SPA FPGA simulations for $(2209,1978)$ EAB and SR-SCB codes (results omitted in the interest of space, see also \cite{DOLECEKISTC10}).  These results also demonstrate that the BER improvement is about one order of magnitude at 5.8 dB for SR-SCB codes relative to EAB codes.  This is the SNR point where SR-SCB codes reach BER of $10^{-11}$. The majority of collected errors for the EAB code in the low BER region is again due to $(4,8)$ absorbing sets. The FPGA simulations collected more errors overall so that $(5,9)$ absorbing set errors were observed with both the EAB and SR-SCB codes.


 

While Table \ref{tab_sw_r5p47} and Fig.~\ref{fig:perf1}  showed that performance improvement can be obtained by precluding only (4,8) absorbing sets, Table~\ref{tab_fpga_new4k} and Fig.~\ref{fig:perf4} show {hardware} simulations demonstrating the benefit of precluding  all $ (4,8) $, $ (5,9) $ and $ (6,8) $ absorbing sets.  Here we compare  longer block-length EAB and SR-SCB codes wither $r=5$ and $p=67$.  The SR-SCB code precludes all $ (4,8) $, $ (5,9) $ and $ (6,8) $ absorbing sets using the RSF $[0,1,2,4,17]$ with $p=67$ which was identified in Table \ref{tbl:r=5codes} of Section \ref{nonexistence3sets} as completely avoiding these three absorbing sets.  As expected, the error profile as shown in Table ~\ref{tab_fpga_new4k} shows that the SR-SCB code completely eliminates the $ (4,8) $, $ (5,9) $ and $ (6,8) $ absorbing sets.  Fig.~\ref{fig:perf4} shows the corresponding BER improvement. {The improved error-floor slope under limited-precision SPA decoding of the SR-SCB code is consistent with the removal of the smallest absorbing sets.}

\subsection{SCB codes for $ r =4 $}\label{resultsr4}

Fig.~\ref{figr4hard} and Table~\ref{tab:2k}  show the performance of $(2209, 2024)$, $p=47$ EAB code and SR-SCB codes both with check node degree  = $47$ and bit node degree = $4$.  The EAB code has $g_r(i) = i$ and the SR-SCB code uses RSF $[0,1,2,4]$.  Consistent with the theoretical analysis, the $ (6,4) $ absorbing sets dominate the error floor of the EAB code.  Precluding the (6,4) absorbing sets of Figs. \ref{fig64b} and \ref{fig64c} provides the SR performance improvement.   Precluding the (6,4) absorbing set of Fig. \ref{fig64a} as well requires shortening which lowers the code rate somewhat.   Such a code is discussed in the next paragraph.

\comment{
\begin{table}
\begin{center}
\caption{Hardware error profiles for the EAB $(1849,1638)$,code (top), and the SR-SCB code (bottom).}\label{tab_fpga_new3}
\begin{tabular}{|p{0.19in}|p{0.22in}|p{0.14in}|p{0.14in}|p{0.14in}|p{0.18in}|p{0.14in}|p{0.18in}|p{0.14in}|p{0.14in}|p{0.18in}|}
\hline
\multicolumn{11}{|c|}{$p=43$ $(1849,1638)$  EAB code} \\ \hline
SNR & n.e. &(4,8)&(5,9)&(6,8)&(6,10)&(7,9)&(7,11)&(8,6)&(8,8)&(8,10)\\
\hline
5.6dB& 236 & 98 & 24 & 37 & 11 & 15& 5 & 8 & 12 & 5\\
5.8dB& 182 & 103 & 19 & 29 & 7 & 7& 4 & 0 & 1 & 5\\
6.0dB& 160 & 98 &  22 & 15 & 3 & 4&  1 & 4 & 3 & 2\\
\hline
\hline
\end{tabular}
\begin{tabular}{|p{0.19in}|p{0.22in}|p{0.14in}|p{0.14in}|p{0.14in}|p{0.18in}|p{0.14in}|p{0.18in}|p{0.14in}|p{0.14in}|p{0.18in}|}
\hline
\multicolumn{11}{|c|}{$p=43$ $(1849,1638)$  SR-SCB code  RSF $=[0,1,2,4,6]$ } \\ \hline
SNR & n.e. &(4,8)&(5,9)&(6,8)&(6,10)&(7,9)&(7,11)&(8,6)&(8,8)&(8,10)\\
\hline
5.6dB& 126 & 0 & 23 & 0 & 23 & 17& 8 & 5 & 8 & 10\\
5.8dB& 92 & 0 & 27 & 0 & 15 & 13& 6 & 6 & 6 & 6\\
6.0dB& 48 & 0 &  19 & 0 & 14 & 4&  3 & 3 & 1 & 1\\
\hline
\end{tabular}
\end{center}
\end{table}
}

\comment{
\begin{figure}
\center
\includegraphics[width=0.37\textwidth]{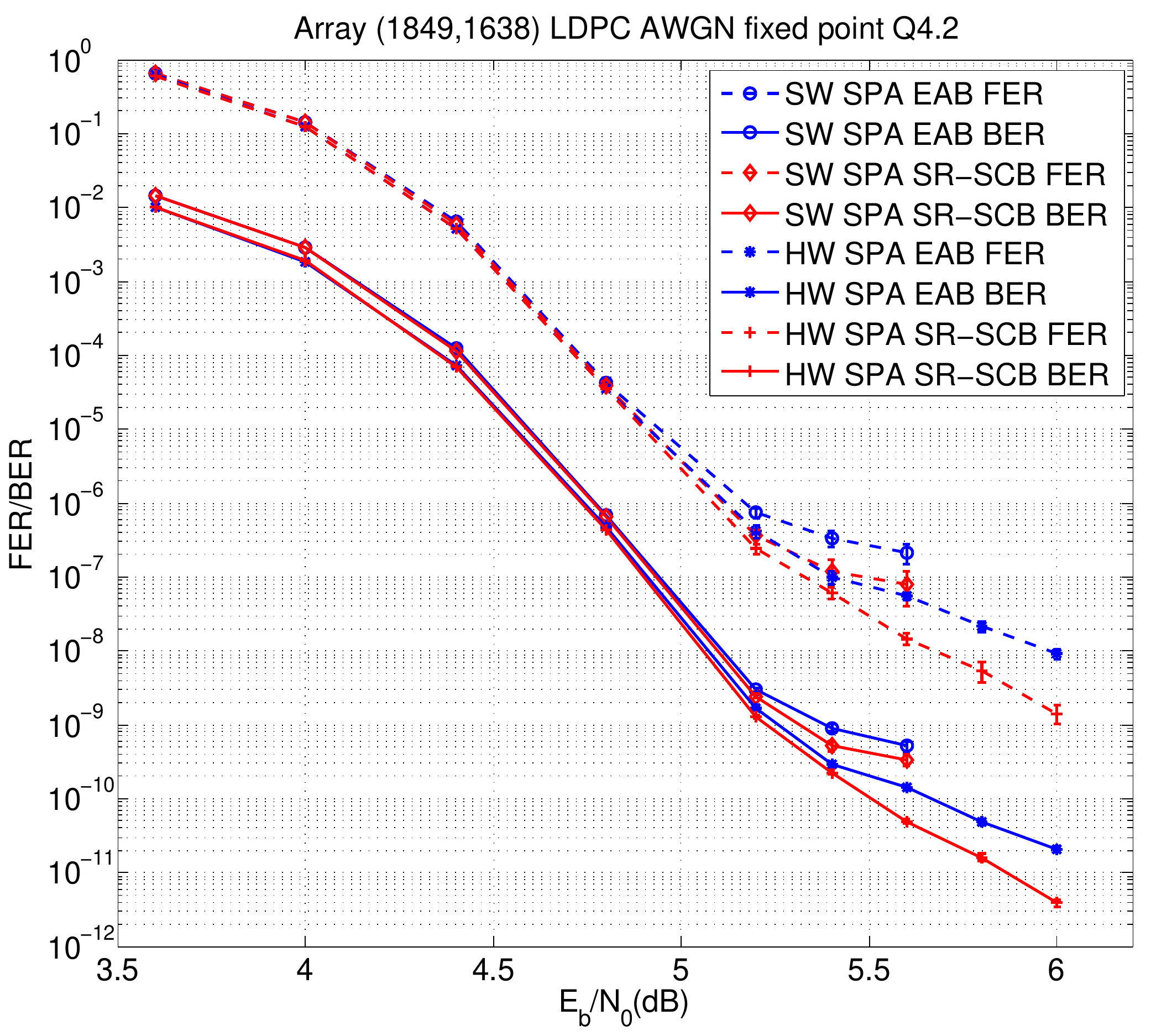}
\caption{Performance comparison of the $ (1849,1638) $ EAB and SR-SCB LDPC codes . }\label{fig:perf3}
\vspace{-0.15in}
\end{figure}
}

\begin{table}
\hspace{-0.5in}
\caption
{Error profiles for the EAB SCB $(4489,4158)$, code (top), and the SR-SCB code (bottom), both with $p=67$. n.e. is the number of collected errors. {n.r. is the number of runs.}}\label{tab_fpga_new4k}
\begin{tabular}{|p{0.18in}|p{0.24in}|p{0.15in}|p{0.13in}|p{0.14in}|p{0.14in}|p{0.18in}|p{0.14in}|p{0.17in}|p{0.13in}|p{0.13in}|}
\hline
\multicolumn{11}{|c|}{ $(4489,4158)$  EAB code} \\ \hline
SNR & n.r. & n.e. &(4,8)&(5,9)&(6,8)&(6,10)&(7,9)&(7,11)&(8,6)&(8,8)\\
\hline
5.6dB&1.1E9 & 150 & 67&17&22&7&6&5&6&6\\
5.8dB&2.1E9 & 139 & 83&18&16&6&5&1&3&3\\
6.0dB&4.3E9 & 131 & 77&18&22&5&1&1&2&1\\
6.2dB&8.6E9 & 107 & 85&10&5&4&2&0&0&0\\
\hline
\hline
\end{tabular}
\begin{tabular}{|p{0.18in}|p{0.24in}|p{0.15in}|p{0.13in}|p{0.14in}|p{0.14in}|p{0.18in}|p{0.14in}|p{0.17in}|p{0.13in}|p{0.13in}|}
\hline
\multicolumn{11}{|c|}{ $(4489,4158)$  SR-SCB code  RSF $=[0,1,2,4,17]$ } \\ \hline
SNR & n.r. & n.e. &(4,8)&(5,9)&(6,8)&(6,10)&(7,9)&(7,11)&(8,6)&(8,8)\\
\hline
5.6dB&4.3E9 & 106 & 0&0&0&25&15&6&15&13\\
5.8dB&2.6E10 & 140 & 0&0&0&35&29&14&16&6\\
6.0dB&3.4E10 & 60 & 0&0&0&25&7&5&9&5\\
\hline
\end{tabular}
\vspace{-0.1in}
\end{table}

\begin{figure}
\center
\includegraphics[width=0.35\textwidth]{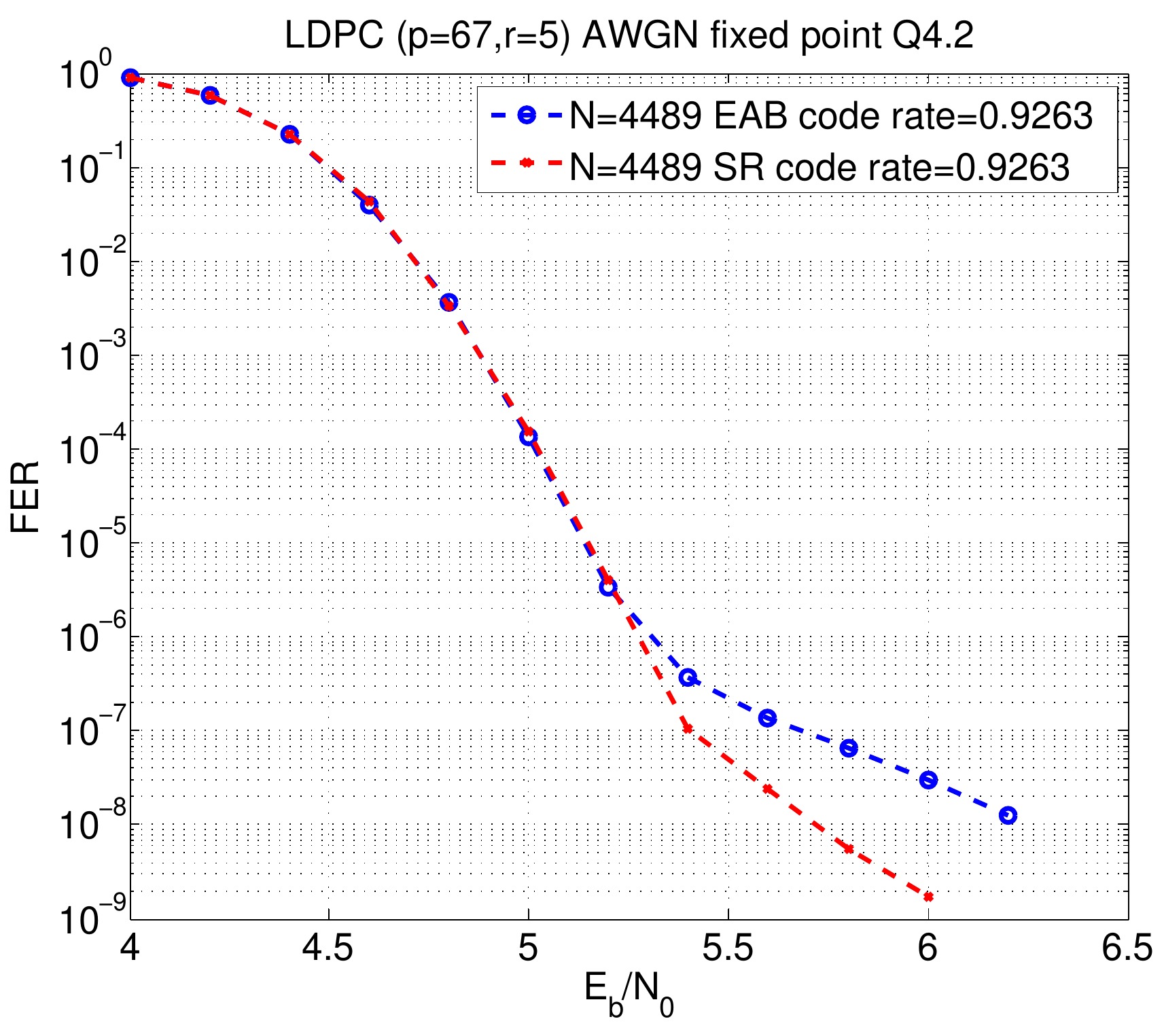}
\caption{Performance comparison of $ (4489,4158) $ EAB and SR-SCB codes {using limited-precision SPA decoder}. }\label{fig:perf4}
\vspace{-0.1in}
\end{figure}


Fig.~\ref{figr4hard}  also compares the performance of a high-rate quasi-cyclic (QC) code under the Tanner \textit{et al.} construction~\cite{tanner04} with a similar-rate shortened SR  (SSR) code that precludes all (6,4) absorbing sets.  The QC code has the following parameters: $ p=61 $, $f(i,j)=a^i \cdot b^j$, $ a=11 $, $ b=5 $, $ o(a)=4 $ and $ o(b)=30 ${, where $ o(a) $ indicates the multiplicative order of $ a $ in $GF(p)$}.  
Using the CCM based analysis, one can show that this code does not have $ (6,4) $ absorbing sets, although it does have $ (4,4) $ absorbing sets (due to an inappropriate row mapping).
The SSR-SCB code is obtained by shortening the SR-SCB  code with parameters $ p=79$ and  RSF = $[0,1,3,4]$ using Algorithm 1. This SSR-SCB code has CSF $ [2,6,7,\,14,17,18, \, 22, 26,27, \,30,36,37,38,46,47,\,49,55,56,$ $57,58,61,62,65,66,67,76,77,78 ] $.  We thus obtain a code with the same variable-node degree as the QC code, a similar block length ($ N=2212 $ vs. $N=1830$ for the QC code) and a similar rate ($0.8585$ vs. $0.8683$ for the QC code). The SSR-SCB code provably eliminates $ (6,4) $  absorbing sets without introducing smaller absorbing sets. The profiles in Table~\ref{tab:2k} confirm this claim. {Similar analysis can be applied to the codes in~\cite{FossorierIT04} as well.}

{The $ (1944,1620) $ quasi-cyclic code from 802.11n standard is also simulated as a reference. This code has larger block length and lower rate than our SSR-SCB codes. The error floor is dominated by the small absorbing sets such as $(3,3)$ and $(4,1)$ absorbing sets and the $ (2,2) $ trapping set which consists of a pair of degree-2 variable nodes. This code has a relatively early error floor compared to the SSR code.}

{Lastly, we plot the same SSR, Tanner  \textit{et al.} construction, and 802.11n codes~\footnote{\added{The 802.11n code was designed to primarily have good waterfall performance. This example is used here to further illustrate that such a code can have a high error floor if it is not optimized properly for a limited-precision decoder.}}  again in Fig. \ref{figr4soft}, now with a full precision decoder.} We notice that {(as expected) the performance of full precision-decoded codes improves for all codes considered. It is interesting to observe that the SSR code again eventually outperforms the 802.11n code, although the improvement is much less than in the limited precision case, as the importance of absorbing sets is not as significant under full precision as it is with a limited-precision decoder.}
\added{Construction of a rate-1/2 LDPC code (not a central focus of this paper) with a low error floor was recently presented in \cite{Bani12a}.}

\begin{table}
\caption{Error profiles for  $(2209,2024)$ EAB ($p=47$),  SR-SCB ($p=47$), and SSR-SCB ($p=79$) codes. 
} \label{tab:2k}
\centering
\hspace{-0.26in}
\begin{tabular}{|p{0.18in}|p{0.18in}|p{0.13in}|p{0.13in}|p{0.13in}|p{0.13in}|p{0.13in}|p{0.13in}|p{0.13in}|p{0.18in}|p{0.18in}|}
\hline
 SNR&n.r.&n.e.&(6,4)&(6,6)&(7,4)&(8,2)&(8,4)&(9,4)&(10,4)&(12,4)\\
\hline
\multicolumn{11}{|c|}{ $(2209,2024)$  EAB code} \\ \hline
5.6dB& 2.0E8& 322 & 236&2&2&27&3&1&37&1\\
6.0dB& 8.0E8 & 329 & 329&0&0&0&0&0&0&0\\
\hline
\hline
\multicolumn{11}{|c|}{ $(2209,2024)$  SR-SCB code  RSF $=[0,1,2,4]$ } \\ \hline
5.6dB& 2.0E8& 167 & 38&3&0&40&45&3&2&0\\
6.0dB& 8.0E8 & 88 & 4&0&0&2&48&3&0&0\\
\hline
\hline
\multicolumn{11}{|c|}{ $(2212,1899)$  SSR-SCB code  RSF $=[0,1,3,4]$ } \\ \hline
5.2dB& 8.0E8& 98  &  0 & 6 & 5 & 21 & 23 & 1 & 16 & 5 \\
5.6dB& 1.6E9 & 32  &  0 & 0 & 0 & 0 & 11 & 0 & 0 & 0 \\
\hline
\end{tabular}
\end{table}

\begin{figure}
\centering
\includegraphics[width=0.35\textwidth]{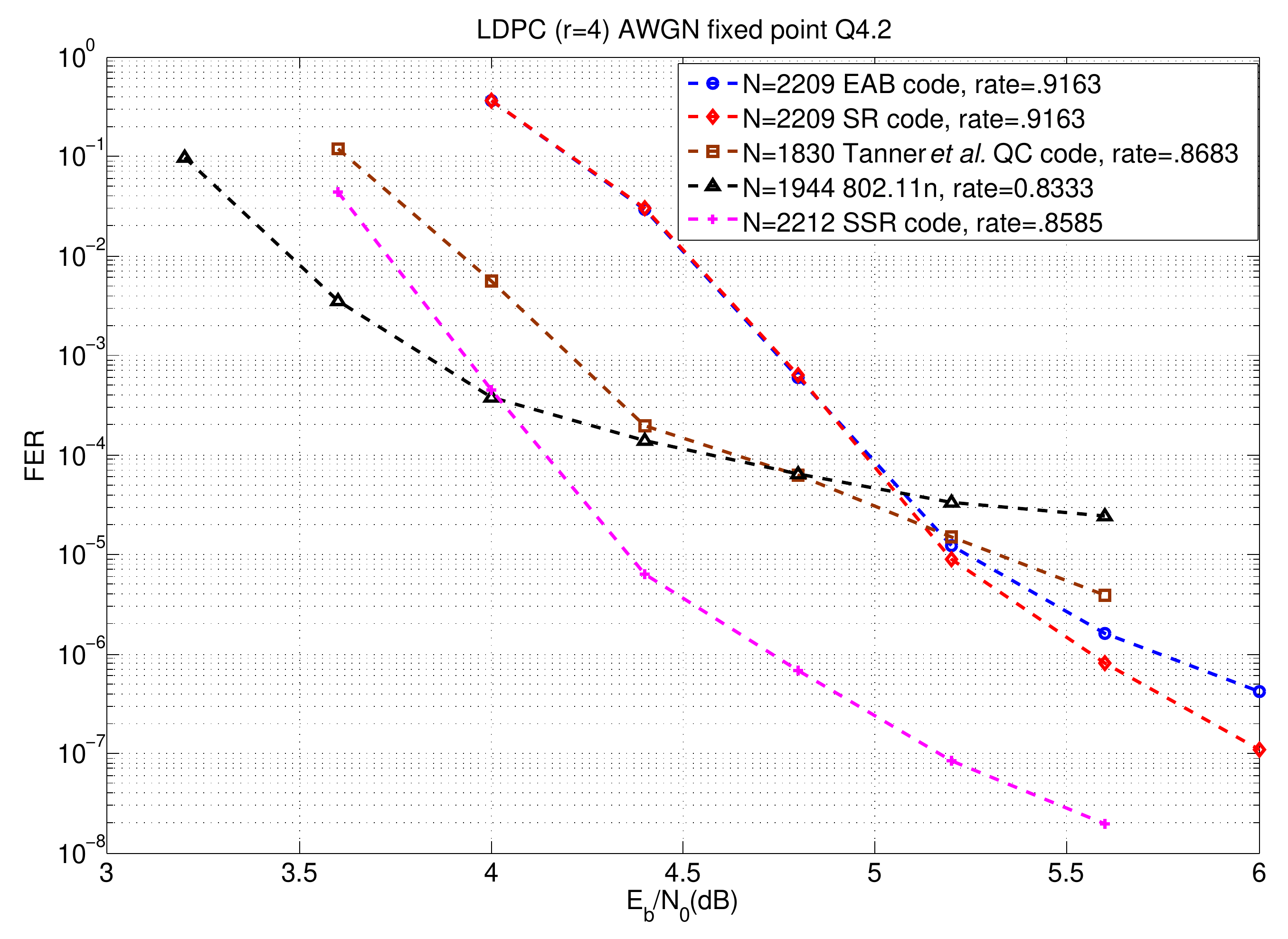}
\caption{Performance comparison of EAB, SR, SSR codes, a code from~\cite{tanner04}, and a code from the 802.11n standard with a decoder using 4.2 limited-precision {SXOR decoding}.}\label{figr4hard}
\includegraphics[width=0.35\textwidth]{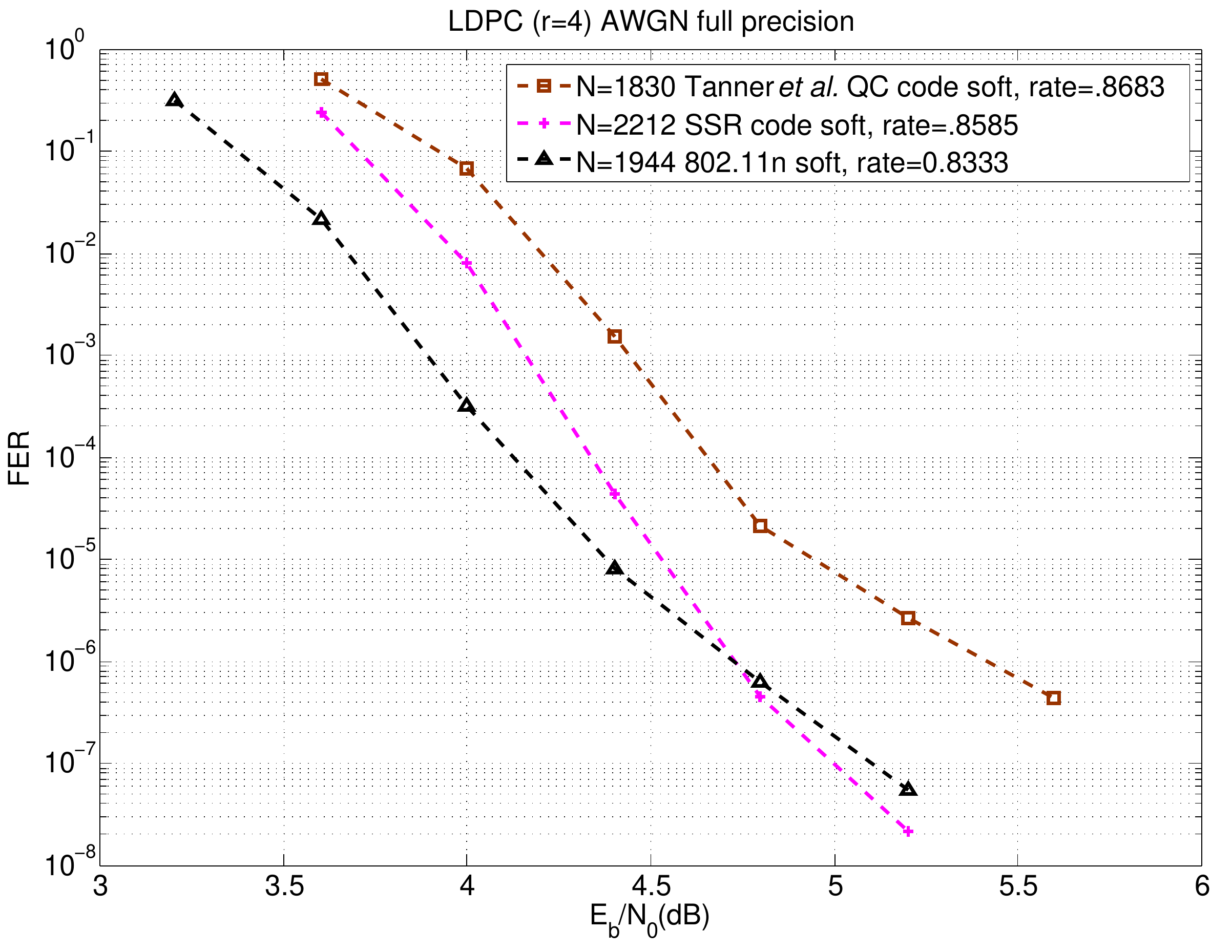}
\caption{Performance comparison of the SSR code, a code from~\cite{tanner04} and the code from the 802.11n standard as in Fig. \ref{figr4hard} but with a full-precision floating-point SPA decoder.}\label{figr4soft}
\vspace{-0.1in}
\end{figure}


\section{Conclusion}\label{conclusion}
We introduced a novel cycle consistency matrix (CCM) description of dominant absorbing sets to guide code design and analysis of circulant-based LDPC codes. Our approach is a deterministic method that can provably eliminate certain absorbing sets in a large family of  circulant-based codes, and can do so without changing code properties such as girth, rate and implementation complexity. This approach thus offers a class of codes with provably better performance than some known constructions.  Theoretical findings were substantiated by experimental results showing consistent performance improvement over a range of decoding algorithms and implementation platforms. An interesting problem for future investigation would be to apply the CCM method to other graphical structures of interest, e.g., trapping sets.

\section*{Ackowledgement}
The authors thank Prof. Zhengya Zhang for helpful discussions.






\bibliographystyle{unsrt}	
\bibliography{myrefs}		

\begin{thebibliography}{10}

\bibitem{globecom06}
Z.~Zhang, L.~Dolecek, B.~Nikolic, V.~Anantharam, and M.~Wainwright.
\newblock Investigation of error floors of structured low-density parity-check
  codes by hardware emulation.
\newblock In {\em Proc. IEEE Global Comm. Conf. (GLOBECOM)}, San Francisco, CA,
  Nov. 2006.

\bibitem{globecom08}
Z.~Zhang, L.~Dolecek, B.~Nikolic, V.~Anantharam, and M.~Wainwright.
\newblock {Lowering LDPC error floors by postprocessing}.
\newblock In {\em Proc. IEEE Global Comm. Conf. (GLOBECOM)}, New Orleans, LA,
  Nov. 2008.

\bibitem{zhangTCOM}
Z.~Zhang, L.~Dolecek, B.~Nikolic, V.~Anantharam, and M.~J. Wainwright.
\newblock {Design of LDPC decoders for improved low error rate performance:
  quantization and algorithm choices}.
\newblock {\em {IEEE Trans. Comm.}}, 57(11):3258--3268, Nov. 2009.

\bibitem{tanner04}
R.~M. Tanner, D.~Sridhara, A.~Sridharan, T.~E. Fuja, and D.~J. Costello.
\newblock {LDPC block and convolutional codes based on circulant matrices}.
\newblock {\em {IEEE Trans. Inform. Theory}}, 50(15):2966 --2984, Dec. 2004.

\bibitem{FossorierIT04}
M.~P.~C. Fossorier.
\newblock Quasi-cyclic low-density parity-check codes from circulant
  permutation matrices.
\newblock {\em {IEEE Trans. Inform. Theory}}, 58(8):1788--1793, Aug. 2004.

\bibitem{Fan00}
J.~L. Fan.
\newblock {Array codes as low-density parity-check codes}.
\newblock In {\em {Proc. 2nd Int. Symp. on Turbo Codes}}, Brest, France, Sep.
  2000.

\bibitem{AsvadiIT11}
R.~Asvadi, A.~H. Banihashemi, and M.~Ahmadian-Attari.
\newblock {Lowering the error floor of LDPC codes using cyclic liftings}.
\newblock {\em {IEEE Trans. Inform. Theory}}, 57(4):2213--2224, Apr. 2011.

\bibitem{AsvadiArxiv}
R.~Asvadi, A.~H. Banihashemi, and M.~Ahmadian-Attari.
\newblock {Design of finite-length irregular protograph codes with low error
  floors over the binary-input AWGN channel using cyclic liftings}.
\newblock {\em {IEEE Trans. Comm.}}, 60(4):902--907, April 2012.

\bibitem{dolecekITA10}
L.~Dolecek.
\newblock On absorbing sets of structured sparse graph codes.
\newblock In {\em Proc. Info. Theory and Apps. (ITA) Workshop}, San Diego, CA,
  Feb. 2010.

\bibitem{DOLECEKISTC10}
L.~Dolecek, J.~Wang, and Z.~Zhang.
\newblock {Towards improved LDPC code designs using absorbing set spectrum
  properties}.
\newblock In {\em {Proc. IEEE 6th Int. Symp. on Turbo Codes}}, Brest, France,
  Sep. 2010.

\bibitem{JWANGICC11}
J.~Wang, L.~Dolecek, and R.D. Wesel.
\newblock {Controlling LDPC absorbing sets via the null space of the cycle
  consistency matrix}.
\newblock In {\em {Proc. IEEE Int. Conf. on Comm. (ICC)}}, Kyoto, Japan, Jun.
  2011.

\bibitem{WangITA2011}
J.~Wang, L.~Dolecek, and R.~D Wesel.
\newblock {LDPC absorbing sets, the null space of the cycle consistency matrix,
  and Tanner's constructions}.
\newblock In {\em Proc. Info. Theory and Appl. (ITA) Workshop}, San Diego, CA,
  Feb. 2011.

\bibitem{WangISIT2011}
J.~Wang, L.~Dolecek, Z.~Zhang, and R.~D Wesel.
\newblock {Absorbing set spectrum approach for practical code design}.
\newblock In {\em Proc. IEEE Int. Symp. on Info. Theory (ISIT)}, Saint
  Petersburg, Russia, Jul. 2011.

\bibitem{richardson}
T.~Richardson.
\newblock Error-floors of {L}{D}{P}{C} codes.
\newblock In {\em Proc. 41st Annual Allerton Conf.}, Monticello, IL, Oct. 2003.

\bibitem{Bani07}
H.~Xiao and A.~H. Banihashemi.
\newblock {Estimation of bit and frame error rates of finite-length low-density
  parity-check codes on binary symmetric channels}.
\newblock {\em {IEEE Trans. Comm.}}, 55(12):2234 -- 2239, Dec. 2007.

\bibitem{dolecekIT10}
L.~Dolecek, Z.~Zhang, M.~J. Wainwright, V.~Anantharam, and B.~Nikolic.
\newblock {Analysis of absorbing sets and fully absorbing sets of array-based
  LDPC codes}.
\newblock {\em {IEEE Trans. Inform. Theory}}, 56(1):6261--6268, Jan. 2010.

\bibitem{CasadoTCOM2011}
A.~I.~Vila Casado, M.~Griot, and R.~D. Wesel.
\newblock {LDPC decoders with informed dynamic scheduling}.
\newblock {\em {IEEE Trans. Comm.}}, 58(12):3470--3479, Dec. 2010.

\bibitem{commletters09}
S.~Kim, K.~Ko, J.~Heo, and J.~Kim.
\newblock {Two-staged informed dynamic scheduling for sequential belief
  propagation decoding of LDPC codes}.
\newblock {\em {IEEE Comm. Letters}}, 13(3):193--195, Mar. 2009.

\bibitem{milenkovic}
S.~Laendner and O.~Milenkovic.
\newblock {Algorithmic and combinatorial analysis of trapping sets in
  structured {LDPC} codes}.
\newblock In {\em Proc. Wireless Comm.}, Honolulu, HI, Jun. 2005.

\bibitem{quant2}
Z.~Wang and Z.~Cui.
\newblock {Low-complexity high-speed decoder design for quasi-cyclic LDPC
  code}.
\newblock {\em {IEEE Trans. VLSI Systems}}, 15(1):104 -- 114, Jan. 2007.

\bibitem{TianTC2004}
T.~Tian, J.~D. Villasenor, and R.~D. Wesel.
\newblock {Selective avoidance of cycles in irregular LDPC code construction}.
\newblock {\em {IEEE Trans. Comm.}}, 52(8):1242--1247, Aug. 2004.

\bibitem{XiaoCL2004}
H.~Xiao and A.~H. Banihashemi.
\newblock {Improved progressive-edge-growth (PEG) construction of irregular
  LDPC codes}.
\newblock {\em {IEEE Comm. Letters}}, 8(12):715--718, Dec. 2004.

\bibitem{LanIT2007}
L.~Lan, L.~Zeng, Y.~Tai, L.~Chen, S.~Lin, and K.~Abdel-Ghaffar.
\newblock {Construction of quasi-cyclic LDPC codes for AWGN and binary erasure
  channels: a finite field approach}.
\newblock {\em {IEEE Trans. Inform. Theory}}, 53(7):2429--2458, Jul. 2007.

\bibitem{ChenTCOM04}
L.~Chen, J.~Xu, I.~Djurdjevic, and S.~Lin.
\newblock {Near-Shannon-limit quasi-cyclic low-density parity-check codes}.
\newblock {\em {IEEE Trans. Comm.}}, 52(7):1038--1042, Jul. 2004.

\bibitem{ZhangTCOM2010}
L.~Zhang, Q.~Huang, S.~Lin, K.~A. Ghaffar, and I.~Blake.
\newblock {Quasi-cyclic LDPC codes: an algebraic construction, rank analysis,
  and codes on Latin squares}.
\newblock {\em {IEEE Trans. Comm.}}, 58(11):3126--3139, Nov. 2010.

\bibitem{HuangIT2012}
Q.~Huang, S.~Lin, and K.~Abdel-Ghaffar.
\newblock {Cyclic and quasi-cyclic LDPC codes on row and column constrained
  parity-check matrices and their trapping sets}.
\newblock {\em {IEEE Trans. Inform. Theory}}, 58(5):2648--2671, May 2012.

\bibitem{NguyenITW10}
D.~V. Nguyen, B.~Vasic, M.~Marcellin, and S.K. Chilappagari.
\newblock {Structured LDPC codes from permutation matrices free of small
  trapping sets}.
\newblock In {\em Proc. IEEE Info. Theory Workshop (ITW)}, Dublin, Ireland,
  Sep. 2010.

\bibitem{NguyenIT11}
D.~V. Nguyen, S.K. Chilappagari, M.~Marcellin, and B.~Vasic.
\newblock {On the Construction of Structured LDPC Codes Free of Small Trapping
  Sets}.
\newblock {\em {IEEE Trans. Inform. Theory}}, 58(4):2280--2302, Apr. 2012.

\bibitem{milenkovicGlobecom06}
S.~Laendner, T.~Hehn, O.~Milenkovic, and J.~Huber.
\newblock When does one redundant parity-check equation matter?
\newblock In {\em Proc. IEEE Global Comm. Conf. (GLOBECOM)}, San Francisco, CA,
  Nov. 2006.

\bibitem{MilenkovicIT06}
O.~Milenkovic, N.~Kashyap, and D.~Leyba.
\newblock Shortened array codes of large girth.
\newblock {\em {IEEE Trans. Inform. Theory}}, 52(8):3707--3722, Aug. 2006.

\bibitem{Diestelgraphtheory}
{R. Diestel}.
\newblock {\em {Graph Theory}}.
\newblock {Springer}, 2006.

\bibitem{Dekhordi12}
M.~K. Dekhordi and A.~H. Banihashemi.
\newblock {An efficient algorithm for finding dominant trapping sets of LDPC
  codes}.
\newblock {\em {IEEE Trans. Inform. Theory}}, 58(11):6942 -- 6958, Nov. 2012.

\bibitem{Milenkovic07}
O.~Milenkovic, E.~Soljanin, and P.~Whiting.
\newblock {Asymptotic spectra of trapping sets in regular and irregular {LDPC}
  code ensembles}.
\newblock {\em {IEEE Trans. Inform. Theory}}, 53(1):39--55, Jan. 2007.

\bibitem{Mansour}
M.M. Mansour and N.R. Shanbhag.
\newblock {High-throughput LDPC decoders}.
\newblock {\em {IEEE Trans. VLSI Systems}}, 1(6):976 -- 996, Dec. 2003.

\bibitem{Bani12a}
S.~Khazraie, R.~Asvadi, and A.~H. Banihashemi.
\newblock {A PEG construction of finite-Length LDPC Codes with low error
  floor}.
\newblock {\em {IEEE Comm. Letters}}, 16(8):1288 -- 1291, Aug. 2012.

\end{thebibliography}

\end{document}